\documentclass[journal ,12]{IEEEtran}
\usepackage{colortbl}
\usepackage[table]{xcolor}
\usepackage[dvipsnames]{xcolor}

\usepackage{tikz}
\usepackage{lipsum}
\usepackage{algorithm}
\usepackage{algorithmic}
\usepackage{url}
\usepackage{float} 

\usepackage{amsmath,amsfonts,amssymb,mathrsfs,amsthm,amsbsy,graphicx,color,bm,balance,mathtools,xfrac,nccmath,siunitx,cite,xspace,soul}
\usepackage{multirow}

\usepackage[font=scriptsize,labelfont=bf]{caption}

\usepackage{hyperref}
\usepackage[shortlabels]{enumitem}

\usepackage{pgfplots}
\pgfplotsset{compat=newest}
\usetikzlibrary{plotmarks}
\usetikzlibrary{arrows.meta}
\usepgfplotslibrary{patchplots}
\usepackage{grffile}
\usetikzlibrary{plotmarks}

 \usepackage[utf8]{inputenc}
\usepackage{booktabs, caption, makecell}

\usepackage{threeparttable}

\newcommand{\q}[1]{{\bf{#1}}}
\newcommand{\mt}[1]{\mathtt{#1}}

\newcommand{\E}[1]{{\mathbb{E} \left\{ #1 \right\}}}

\newcommand{\tr}{\text{Tr}}

\renewcommand{\Re}{\text{Re}}

\theoremstyle{definition}
\makeatletter
\newcommand{\printfnsymbol}[1]{%
  \textsuperscript{\@fnsymbol{#1}}%
}
\makeatother

\begin{document}
\bstctlcite{IEEEexample:BSTcontrol}
\title{Cell-Free Integrated Sensing and Communication: Principles, Advances, and Future Directions}

\author{Diluka Galappaththige, \IEEEmembership{Member, IEEE},  Mohammadali Mohammadi, \IEEEmembership{Senior Member, IEEE,} Gayan Aruma Baduge, \IEEEmembership{Senior Member, IEEE} and Chintha Tellambura, \IEEEmembership{Fellow, IEEE}
\thanks{D. Galappaththige and C. Tellambura are with the Department of Electrical and Computer Engineering, University of Alberta, Edmonton, AB, T$6$G 1H9, Canada (e-mail: \{diluka.lg, ct4\}@ualberta.ca).} 
\thanks{M. Mohammadi is with the Centre for Wireless Innovation (CWI), Queen's University Belfast, BT3 9DT Belfast, U.K. (email: m.mohammadi@qub.ac.uk).}
\thanks{G. Aruma Baduge is with the School of Electrical, Computer, and Biomedical Engineering, Southern Illinois University, Carbondale, IL, USA (email: gayan.baduge@siu.edu). His work in part was supported by the U.S. NSF under Grant CCF-2326621.}
 \vspace{-0mm}}

\maketitle  
\begin{abstract}
Cell-free (CF) integrated sensing and communication (ISAC) combines CF architecture with ISAC. CF employs distributed access points, eliminates cell boundaries, and enhances coverage, spectral efficiency, and reliability. ISAC unifies radar sensing and communication, enabling simultaneous data transmission and environmental sensing within shared spectral and hardware resources. CF-ISAC leverages these strengths to improve spectral and energy efficiency while enhancing sensing in wireless networks. As a promising candidate for next-generation wireless systems, CF-ISAC supports robust multi-user communication, distributed multi-static sensing, and seamless resource optimization. However, a comprehensive survey on CF-ISAC has been lacking. This paper fills that gap by first revisiting CF and ISAC principles, covering cooperative transmission, radar cross-section, target parameter estimation, ISAC integration levels, sensing metrics, and applications. It then explores CF-ISAC systems, emphasizing their unique features and the benefits of multi-static sensing. State-of-the-art developments are categorized into performance analysis, resource allocation, security, and user/target-centric designs, offering a thorough literature review and case studies. Finally, the paper identifies key challenges such as synchronization, multi-target detection, interference management, and fronthaul capacity and latency. Emerging trends, including next-generation antenna technologies, network-assisted systems, near-field CF-ISAC, integration with other technologies, and machine learning approaches, are highlighted to outline the future trajectory of CF-ISAC research.
\end{abstract}

\begin{IEEEkeywords}
Cell-free architecture, massive multiple-input multiple-output communication, integrated sensing and communication, multi-static sensing, next-generation wireless networks.
\end{IEEEkeywords}

\thispagestyle{empty}

\section*{Abbreviations}
\addcontentsline{toc}{section}{Nomenclature}
\begin{IEEEdescription}[\IEEEusemathlabelsep\IEEEsetlabelwidth{$V_1,V_2,V_3,V_4$}]
\fontsize{0.33cm}{0.4cm}\selectfont
% \item[3GPP]     The 3rd Generation Partnership Project
\item[\num{5}G] Fifth generation
\item[\num{6}G] Sixth generation 
\item[AF]       Ambiguity function 
\item[AN]       Artificial noise 
\item[AO]       Alternating optimization 
\item[AoA]      Angle of arrival 
\item[AP]       Access point
\item[AWGN]     Additive white Gaussian noise 
\item[BackCom]  Backscatter communication
\item[BCD]      Block coordinate descent 
\item[BS]       Base station
\item[CF]       Cell-free
\item[CFMM]     Cell-free massive multiple-input multiple-output
\item[CoMP]     Coordinated multi-point
\item[CPU]      Central processing unit
\item[CRB]      Cram\'{e}r-Rao bound 
\item[CRLB]     Cram\'{e}r-Rao lower bound 
\item[CSI]      Channel state information
\item[CW]       Continuous-wave 
\item[DAS]      Distributed antenna system  
\item[DL]       Downlink 
\item[DLI]      Direct-link interference 
\item[DoF]      Degrees of freedom 
\item[EE]       Energy efficiency
\item[EM]       Electromagnetic 
\item[ESPRIT]   Estimation of signal parameters via rotational invariance
\item[FD]       Full-duplex 
\item[FFT]      Fast Fourier transform 
\item[FIM]      Fisher information matrix 
\item[GLRT]     Generalized likelihood ratio test 
\item[HMIMO]    Holographic multiple-input multiple-output  
\item[IoT]      Internet of Things 
\item[ISABC]    Integrated sensing and backscatter communication
\item[ISAC]     Integrated sensing and communication 
\item[LoS]      Line-of-sight 

\item[MA]       Movable antenna
\item[MI]       Mutual information 
\item[MIMO]     Multiple-input multiple-output
\item[ML]       Machine learning 
\item[MLE]      Maximum likelihood estimation 
\item[mMIMO]    Massive multiple-input multiple-output
\item[MRC]      Maximum ratio combining 
\item[MRT]      Maximum ratio transmission 
\item[MSE]      Mean-square error
\item[MUSIC]    Multiple signal classification 
\item[NLoS]     Non-line-of-sight  
\item[OFDM]     Orthogonal frequency division multiplexing 
\item[PRI]      Pulse repetition interval 
\item[QoS]      Quality-of-service 
\item[RCS]      Radar cross-section
\item[RIS]      Reconfigurable intelligent surface
\item[RSU]      Roadside units 
\item[SCA]      Successive convex approximation
\item[SCNR]     Signal-to-clutter-plus-noise-ratio 
\item[SDR]      Semidefinite relaxation 
\item[SE]       Spectral efficiency
\item[SI]       Self-interference 
\item[SIC]      Successive interference cancellation
\item[SINR]     Signal-to-interference-plus-noise ratio 
\item[SNR]      Signal-to-noise ratio 
\item[ToF]      Time-of-flight 
\item[UAV]      Unmanned aerial vehicle
\item[UC]       User-centric 
\item[UL]       Uplink 
\item[ULA]      Uniform linear array 
%\item[UMi]      Urban micro 
\item[URLLC]    Ultra-reliable low-latency communication 
\item[V2X]      Vehicle-to-everything 
\item[XL-MIMO]  Extremely large massive multiple-input multiple-output
\end{IEEEdescription}

\section{Introduction}
\IEEEPARstart{I}{ntegrated}  sensing and communication (ISAC) is poised to become a key component of future wireless standards \cite{Liu2022ISAC, Wang2022ISAC, Zhang2022, Azar2024, liu2023integratedbook}. Sensing enables networks to detect, localize, and track objects or environmental features within their coverage area by utilizing transmitted signals to gather information about targets, such as angle, range, velocity, shape, composition, or orientation \cite{Liu2022ISAC, Wang2022ISAC, Zhang2022, Azar2024, liu2023integratedbook}. The dual functions of communication and sensing can support diverse emerging applications, including the Internet of Things (IoT), vehicle-to-everything (V2X) communications, smart traffic control, virtual/augmented reality, smart homes/cities, unmanned aerial vehicles (UAVs), factory automation, and more \cite{Liu2022ISAC, Wang2022ISAC, Zhang2022, Azar2024, liu2023integratedbook}.

ISAC enables communication and sensing to efficiently share infrastructure, hardware, signal processing modules, and scarce spectrum and power resources \cite{Liu2022ISAC, Wang2022ISAC, Zhang2022, Azar2024, liu2023integratedbook}. For example, a base station (BS) with multiple antennas can direct beams toward users for communication while using others to monitor surrounding targets or environmental changes \cite{Liu2022ISAC, Wang2022ISAC, Zhang2022, Azar2024, liu2023integratedbook}. This dual functionality improves resource efficiency, reduces hardware and deployment costs, and facilitates seamless coordination between communication and sensing. Additionally, the BS can leverage sensing (e.g., by analyzing reflected signals) for channel estimation and communication beamforming optimization \cite{Liu2022ISAC}. 
{By jointly designing sensing and communication operations, ISAC systems can reinterpret part of the interference between users and targets as a source of additional information; for instance, echoes of communication signals can support sensing, and multi-user signals can enhance spatial diversity. This cooperative use of waveforms and resources introduces new design degrees of freedom (DoF), allowing flexible allocation of time, frequency, and spatial resources to improve both spectral efficiency (SE) and energy efficiency (EE) {\cite{Liu2022ISAC, Wang2022ISAC, Zhang2022, Azar2024, liu2023integratedbook}}.}

Most current ISAC studies focus on co-located multiple-input multiple-output (MIMO)/massive MIMO (mMIMO) cellular networks, i.e., one BS serves as an ISAC transceiver using mono-static sensing or two BSs act as the ISAC transmitter and sensing receiver utilizing bi-static sensing \cite{Liu2022ISAC, Wang2022ISAC, Zhang2022, Azar2024, liu2023integratedbook}. However,  multiple ISAC BSs will operate in the same geographical region, sharing the frequency and timing resources. Furthermore, mono-static and bi-static ISAC systems can only provide limited service coverage, communication, and sensing capabilities. In particular, when there are numerous obstacles in the environment and/or when the communication users and sensing targets are dispersed from the BS 
\cite{Mao2023, Demirhan2023, Huang2022Coordinated, Cao2023Design, Wang2023, Sakhnini2022Uplink, Silva2023, Behdad2022, Behdad2024Interplay}. This causes performance degradation, especially for cell-edge users/targets, frequent cell switching, and severe interference between communications and sensing \cite{Mao2023, Demirhan2023, Huang2022Coordinated, Cao2023Design, Wang2023, Sakhnini2022Uplink, Silva2023, Behdad2022, Behdad2024Interplay}. 

To overcome such challenges,  cooperation among multiple ISAC BSs or access points (APs)  can improve communication and sensing performance \cite{Mao2023, Demirhan2023, Huang2022Coordinated, Cao2023Design, Wang2023, Sakhnini2022Uplink, Silva2023, Behdad2022, Behdad2024Interplay}. On the other hand, recent advances in multi-BS/AP communication cooperation, such as coordinated multi-point (CoMP) transmission/reception, cloud-radio access networks, cell-free (CF) MIMO, distributed MIMO radar sensing, and networked ISAC, lay the groundwork for addressing the aforementioned challenges in conventional ISAC systems \cite{Gesbert2010, Dahrouj2010, Jun2015, Ngo2017, Fishler2004, Haimovich2008}. This ultimately leads to CF-ISAC systems, in which distributed ISAC APs jointly serve the same set of communication users and detect the same targets employing multi-static sensing \cite{Mao2023, Demirhan2023, Huang2022Coordinated, Cao2023Design, Wang2023, Sakhnini2022Uplink, Silva2023, Behdad2022, Behdad2024Interplay}. Multi-static sensing involves multiple spatially distributed transmitters and receivers to detect and track targets/objects, improving detection, localization, and tracking accuracy \cite{richards2005fundamentals}. However, it faces challenges such as synchronization, complex signal processing, and deployment costs.
{In CF-ISAC networks, multi-static sensing can offer a diversity gain by utilizing less correlated (approximately independent) sensing observations collected at spatially distributed receivers.}
Also, multi-static sensing can improve performance by increasing joint transmit/receive beamforming gain through the use of many transmitters/receivers in the network \cite{Mao2023, Demirhan2023, Huang2022Coordinated, Cao2023Design, Wang2023, Sakhnini2022Uplink, Silva2023, Behdad2022, Behdad2024Interplay}.  

\subsection{Existing Survey Papers and Contribution}
While CF-ISAC is attracting significant research attention, no survey papers have been published. In contrast,  separate surveys and tutorials exist for CF communication \cite{Demir2021book, Ammar2022, Giovanni2018, Zhang2020, Elhoushy2022, Zhang2019cellfree, Shuaifei2022, Kassam2023, Mohammadi2024} and co-located ISAC systems \cite{Liu2022ISAC, Wang2022ISAC, Zhang2022, Azar2024, liu2023integratedbook}. This study decisively addresses this critical gap by reviewing CF-ISAC advancements in depth, uncovering its capabilities, addressing challenges, and charting future directions for this technology.

The contributions are summarized as follows:
\begin{enumerate}
    \item A comprehensive discussion of CF-ISAC systems requires an examination of the core principles of CF-mMIMO (CFMM) and sensing. The discussion thus begins with an overview of CFMM, including cooperative transmission and reception, channel hardening, favorable propagation, and key performance metrics. Next, sensing fundamentals are explored, covering radar cross-section (RCS), clutter, signal transmission, radar types, sensing techniques, and methods for estimating target parameters.

    \item To establish the fundamentals of CF-ISAC networks, conventional ISAC systems and their properties are examined. Specifically, the discussion covers the levels of integration in ISAC systems, ISAC design philosophies, sensing metrics, and ISAC applications.

    \item With the foundational concepts established, the focus shifts to CF-ISAC systems and their unique characteristics. An in-depth analysis highlights the advantages of multi-static sensing in CF-ISAC over conventional ISAC, emphasizing its superior sensing capabilities. Key features are examined in detail, including distributed antenna systems (DAS), seamless handovers, user- and target-centric operations, advanced interference management, efficient resource allocation, AP cooperation, and robust synchronization mechanisms.

    \item State-of-the-art,  technical contributions are categorized into four key areas: performance analysis, resource allocation, secure CF-ISAC, and user/target-centric CF-ISAC. A detailed literature review is conducted for each category, highlighting methodologies, studies, and key findings. Additionally, case studies with extensive simulations evaluate the CF-ISAC system performance across these areas, offering more profound insights into their operational effectiveness and potential.
    
    \item Finally, the remaining challenges, open issues, and emerging trends in CF-ISAC systems are examined. Key challenges are addressed, including synchronization, multi-target detection, interference management, and fronthaul capacity and latency. Additionally, open research directions and future trends are highlighted, such as advancements in network-assisted CF-ISAC, the development of new antenna technologies, the integration of complementary technologies, and the application of machine learning (ML) techniques. These insights outline potential pathways for further innovation in the field.
\end{enumerate}

This paper is organized as follows: Section \ref{sec_basics_CF_radar} provides an overview of CF architecture and radar sensing fundamentals. Section \ref{sec_isac} covers the basics of conventional ISAC systems and recent advancements. Section \ref{sec_CF_isac} delves into CF-ISAC systems, highlighting their fundamentals, unique features, and advantages. Section \ref{sec_state_art} reviews current technical contributions and presents case studies. Finally, Section \ref{sec_future} discusses future research directions, opportunities, and challenges.

\textit{Notation:} 
Boldface lower and upper case letters denote vectors and matrices.   $\mathbb{C}^{M\times N}$ and ${\mathbb{R}^{M \times 1}}$ represent $M\times N$ dimensional complex matrices and $M\times 1$ dimensional real vectors, respectively. For a square matrix $\mathbf{A}$, $\mathbf{A}^{\rm{H}}$ and $\mathbf{A}^{\rm{T}}$ are the Hermitian conjugate transpose and transpose, respectively. $\mathbf{I}_M$ denotes the $M$-by-$M$ identity matrix. $\mathbf{0}_M$ is the $M$-dimensional all-zero vector. The Euclidean norm of a complex vector and the absolute value of a complex scalar are denoted by $\Vert\cdot\Vert$ and $\vert\cdot\vert$, respectively. Expectation, the real part of a complex number, and trace operation are denoted by $\E{\cdot}$, $\Re\{\cdot\}$, and $\tr(\cdot)$, respectively. A  circularly symmetric complex Gaussian (CSCG) random vector with zero mean and covariance matrix $\mathbf{C}$ is denoted by $\sim \mathcal{C}\mathcal{N}(\q{0},\,\mathbf{C})$.  Finally, $[\q{A}]_{ij}$ is the $\{i, j\}$-th element of  $\q{A}$.  

\section{Fundamentals of CFMM and Radar Sensing}\label{sec_basics_CF_radar} This section overviews two key concepts relevant to CF-ISAC: CFMM and radar sensing. CFMM networks enhance coverage, capacity, and user experience through collaboration among distributed APs without centralized BSs. Radar sensing, a core component of ISAC, enables real-time environmental awareness by detecting and tracking objects using electromagnetic (EM) waves. Together, these technologies underpin this study and offer innovative solutions for future ISAC systems.

%=======================================================================
\begin{figure*}[!t]\centering \vspace{0mm}
    \def\svgwidth{500pt} 
    \fontsize{8}{8}\selectfont 
    \graphicspath{{Figures/}}
    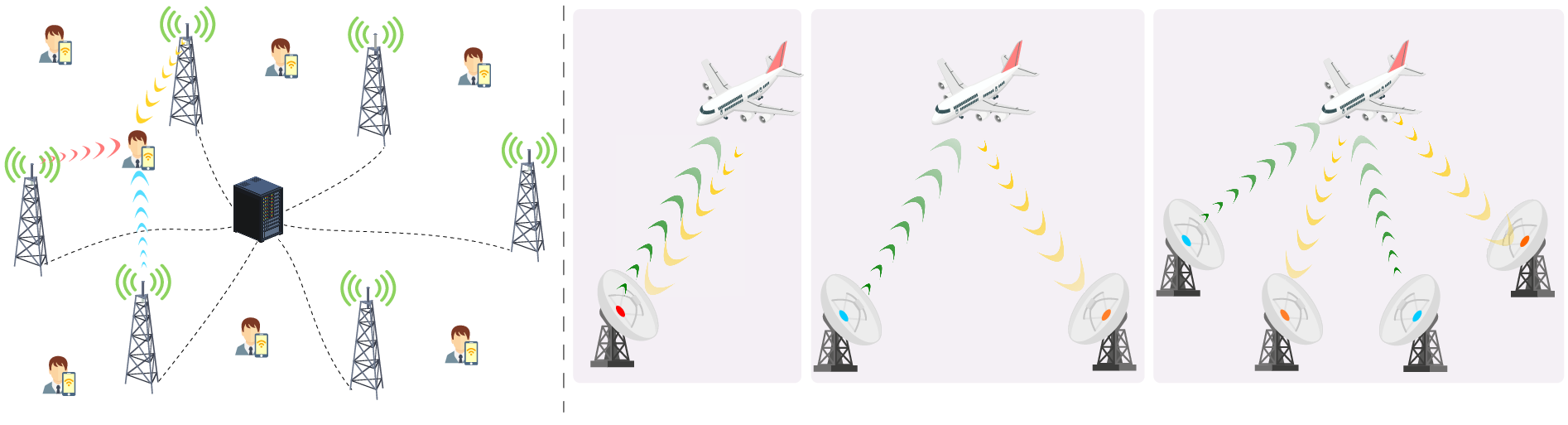 \vspace{0mm}
    \caption{{Conceptual illustration of CF-ISAC. (a) In a CF network, multiple distributed APs are connected to a CPU via fronthaul links and jointly serve users across all cell boundaries, enabling cooperative transmission and reception. (b) Conventional sensing configurations, including mono-static (co-located transmitter and receiver), bi-static (separated transmitter and receiver), and multi-static (multiple distributed transmitters and receivers) radar systems. By combining CF and multi-static sensing, CF-ISAC naturally enables distributed multi-static sensing using spatially separated APs, providing enhanced spatial diversity, improved target detection and localization, and seamless integration of communication and sensing functionalities.}}\vspace{0mm} \label{CF_and_RadarConfigurations}
\end{figure*}
%======================================================================

\subsection{Fundamentals of Cell-Free Massive MIMO}
CFMM is an emerging wireless communication paradigm that replaces the traditional cellular structure with a seamless, collaborative approach to user connectivity \cite{Demir2021book, Ngo2017, Zhang2019cellfree, Diluka2019, Galappaththige2021, Diluka2020, Galappaththige2024, Galappaththige2021Cellfree, Diluka2021}. To understand it, it can be compared with conventional co-located mMIMO, where the BS in each cell employs many antennas, typically 
\num{64} or more, to serve users within the cell \cite{Emil2017, Marzetta2016book}. In contrast, CFMM uses a distributed array of APs to collaboratively serve all users within a large geographical area (Fig.~\ref{CF_and_RadarConfigurations}a). This configuration offers several advantages,  including improved SE, EE, uniform service quality, and enhanced robustness to interference and shadowing effects~\cite{Ngo:TGCN:2018}. Another option to improve EE is the use of energy detection methods \cite{Wang2020,5429879,5208031,6987540}. 

\subsubsection{Elimination of Cell-Boundaries}
By deploying APs across a large region without forming discrete cells, CFMM networks overcome the performance limitations of co-located mMIMO. The network operates as a single cohesive unit, where multiple APs jointly serve each user, ensuring uniform service quality regardless of user location \cite{Demir2021book, Ngo2017, Zhang2019cellfree, Diluka2019, Galappaththige2021, Diluka2020, Galappaththige2024, Galappaththige2021Cellfree, Diluka2021}.

The absence of cell boundaries eliminates the requirement for cell handovers, which may cause latency and service outages. Instead, users experience seamless connectivity through cooperative transmission and reception facilitated by the distributed APs \cite{Demir2021book,  Ngo2017, Zhang2019cellfree, Diluka2024CFBiBC}. This approach enhances performance by exploiting spatial diversity, as multiple APs serve each user. The APs are closer to the user, reducing the path loss and improving reliability, particularly in challenging propagation environments \cite{Demir2021book,  Ngo2017, Zhang2019cellfree, Diluka2024CFBiBC}.

Nevertheless, CFMM  also introduces unique challenges. Distributed APs require substantial fronthaul signaling for acquiring channel state information (CSI) and data sharing, causing higher communication overhead. Also,  centralized or coordinated processing to handle distributed resources can result in high computational complexity. For instance, the fronthaul capacity, fronthaul signaling, and computational complexity linearly scale  (or faster) with the number of APs and users \cite{Emil:TCOM:2020, Parida2023}. 
Despite these challenges, CFMM offers significant gains over alternatives like DAS and CoMP, where static, disjoint cooperation clusters limit inter-cell collaboration \cite{Elhoshy2016, Irmer2011, Venkatesan2007, Simeone2008}. By dynamically and flexibly coordinating APs across the entire network, CFMM  achieves superior performance, especially in high user density and mobility scenarios \cite{Demir2021book,  Ngo2017, Zhang2019cellfree}.

\subsubsection{Cooperative Transmission and Reception} 
These involve the collaborative operation of distributed APs to jointly serve users, coordinated by a central processing unit (CPU) \cite{Demir2021book,  Ngo2017, Zhang2019cellfree}. This cooperation eliminates cell boundaries, enhances signal quality through spatial diversity, reduces interference, and improves SE and EE \cite{Demir2021book,  Ngo2017, Zhang2019cellfree}. It is necessary to ensure seamless connectivity, uniform service quality, and reliable communication, especially in dynamic and high-density user environments \cite{Demir2021book,  Ngo2017, Zhang2019cellfree}.
\begin{enumerate}
    \item \textit{CPU Coordination:}
    The CPU is the central intelligence of the CFMM system. It collects globally or locally estimated CSI from all APs to perform centralized beamforming, resource allocation, and other tasks. Unlike conventional cellular systems, where inter-cell interference limits performance, this coordinated framework minimizes interference while maximizing SE and EE and optimizing the network performance (Section~\ref{sec_performance_benifits}). Key functions of the CPU include:
\begin{itemize}
    \item {\textit{CSI acquisition:} This can be performed in two operational modes {\cite{Demir2021book, Ngo2017, Zhang2019cellfree}}: (i) Centralized mode, where each AP forwards the received pilot signals to the CPU, which then performs channel estimation; and (ii) Distributed mode, where each AP locally estimates its CSI without forwarding it to the CPU. The choice between centralized and distributed operation involves a trade-off between fronthaul load, computational complexity, and estimation accuracy {\cite{Demir2021book, Ngo2017, Zhang2019cellfree}}. In the centralized mode, the CPU can exploit the collected CSI to compute accurate multi-user beamforming and manage user-specific scheduling, enabling the APs to collaboratively transmit coherent signals that reduce interference and improve SE {\cite{Demir2021book, Ngo2017, Zhang2019cellfree}}.}

    \item \textit{Centralized processing:} With the acquired CSI, the CPU designs downlink (DL) beamforming weights/vectors for each AP to ensure constructive interference at the users. For the uplink (UL), the CPU performs joint decoding by coherently combining signals received by multiple APs, leveraging macro-diversity \cite{Demir2021book,  Ngo2017, Zhang2019cellfree}.

    \item \textit{Dynamic resource allocation:} The CPU can dynamically allocate system resources, such as power, spectrum resources, and user scheduling, based on the real-time requirements of users. This ensures uniform service quality across the network \cite{Demir2021book,  Ngo2017, Zhang2019cellfree}. Power control algorithms, such as max-min fairness or sum-rate maximization, can also balance SE and EE trade-offs.
\end{itemize}

However, this centralized coordination requires precise synchronization and fronthaul infrastructure.

\item \textit{Synchronization:}
APs must be synchronized to ensure coherent transmission and reception. Specifically, both time and phase synchronization are essential \cite{Demir2021book, Ngo2017, Zhang2019cellfree}. 
Time synchronization is typically achieved through high-precision network time protocols or synchronization signals distributed via the fronthaul or over-the-air pilots \cite{Demir2021book, Ngo2017, Zhang2019cellfree}. In practice, GPS-disciplined oscillators or reference signals transmitted from the CPU are used to align the clocks of all APs, ensuring their signals reach the intended users simultaneously and avoiding inter-symbol interference \cite{Demir2021book, Ngo2017, Zhang2019cellfree}. Phase synchronization is implemented by sharing reference oscillators or distributing phase-correction information over the fronthaul. The CPU or a reference AP can transmit calibration signals that allow each AP to estimate and compensate for phase offsets. This process ensures coherent beamforming by maintaining consistent carrier-phase alignment across distributed transmitters, which maximizes constructive interference and system performance \cite{Demir2021book, Ngo2017, Zhang2019cellfree}.

\item \textit{Fronthaul:} The fronthaul network consists of the control and communication links connecting distributed APs to the CPU. It enables the bidirectional exchange of user data, control information, and CSI, which are essential for cooperative transmission and reception \cite{Demir2021book, Ngo2017, Zhang2019cellfree}. As network density increases, this traffic grows rapidly, necessitating low-latency and high-bandwidth connections such as fiber, mmWave, or free-space optical links \cite{Demir2021book, Ngo2017, Zhang2019cellfree}.

Fronthaul architectures are broadly categorized as centralized or distributed \cite{intel2021fronthaul}. In centralized schemes (e.g., Cloud-RAN), the CP handles most baseband processing, which simplifies coordination but places stringent requirements on fronthaul bandwidth and latency \cite{Peng2015}. In contrast, distributed architectures (e.g., Fog-RAN or Edge-RAN) delegate part of the processing to APs, reducing fronthaul load at the cost of higher local complexity \cite{Mao2017}.

Optimizing the fronthaul  involves several strategies:
\begin{enumerate}
    \item \textbf{Compression:} Applying data and CSI compression algorithms can significantly reduce fronthaul traffic while maintaining acceptable performance levels.
    \item \textbf{Functional Splitting:} Dynamically adjusting the division of processing tasks between the APs and the CPU  can balance fronthaul load and processing latency based on network conditions.
    \item \textbf{Resource Allocation:} Efficient scheduling and allocation of fronthaul bandwidth can prioritize critical CSI and user data, reducing latency and congestion.
    \item \textbf{Hybrid Fronthaul Solutions:} Combining fiber and wireless (e.g., mmWave or Free Space Optics) links can offer flexibility, cost-efficiency, and redundancy to enhance network resilience.
\end{enumerate}
Furthermore, synchronization protocols across the fronthaul network, such as the Institute of Electrical and Electronics Engineers (IEEE) 1588 Precision Time Protocol (PTP), are critical to ensure the timely and coherent exchange of signals, especially for advanced applications like CoMP and CFMM \cite{Larsen2018}.

\end{enumerate}

\subsubsection{Multiple Antenna Effects}
When multiple-antenna APs communicate with user terminals, Channel hardening and Favorable propagation can occur \cite{Demir2021book}.

%=======================================================================
\begin{figure}[!t]\vspace{-0mm}
    \centering
    \includegraphics[width=0.47\textwidth]{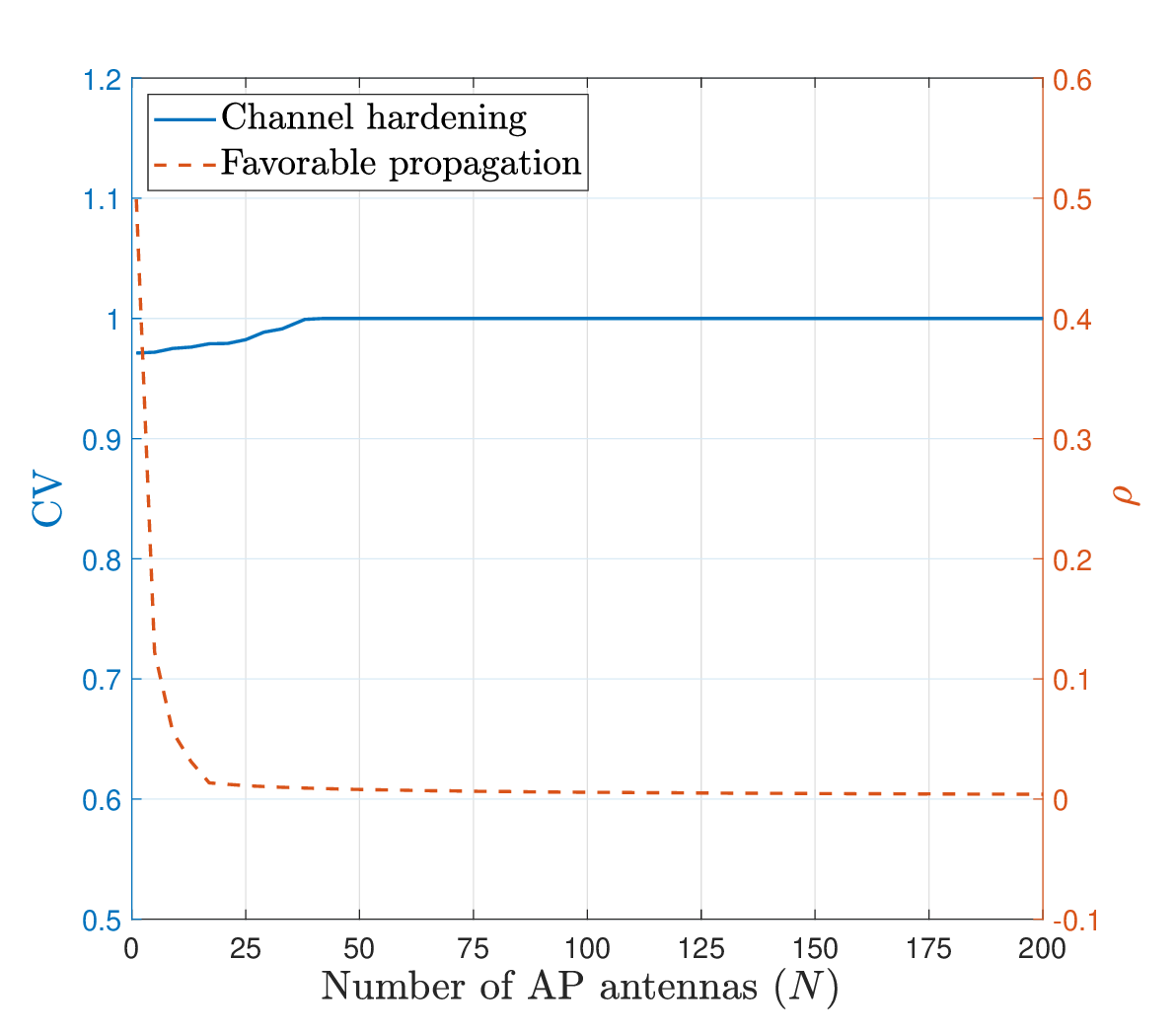}
    \caption{{Channel hardening (left $y$-axis) and favorable propagation (right $y$-axis) versus the number of AP antennas ($N$), under i.i.d. Rayleigh fading. The channel hardening curve shows that the normalized channel gain becomes a constant as $N$ increases, while the favorable propagation curve shows that inter-user channel correlation approaches zero, indicating increased orthogonality.}}
    \label{fig_CH_FP_numAntenna} \vspace{-0mm}
\end{figure}
%=======================================================================

\begin{enumerate}

\item \textit{Channel Hardening:}
This phenomenon describes the diminishing impact of small-scale fading as the number of AP antennas increases. Consequently, the effective channel gain becomes nearly deterministic, exhibiting minimal fluctuations around its mean value \cite{Demir2021book, Chen2018ChannelHardening, Zhang2019cellfree, Polegre2020}.

To illustrate, consider the channel between the $m$-th AP equipped with $N$ antennas and the $k$-th single-antenna user, denoted by $\q{h}_{mk} = [h_{mk,1}, \dots, h_{mk,N}]^{\mathrm{T}} \in \mathbb{C}^{N \times 1}$, where $h_{mk,n}$ represents the channel coefficient between the $n$-th antenna element and the user. For simplicity, the coefficients ${h_{mk,n}}$ are assumed to be independent and identically distributed (i.i.d.) Rayleigh fading variables, i.e., $h_{mk,n} \sim \mathcal{CN}(0, \beta_{mk})$, where $\beta_{mk}$ captures the effects of large-scale path loss and shadowing. The coefficient of variation of the  channel gain, i.e., $\Vert \q{h}_{mk} \Vert^2$,  a measure of the variability relative to the mean, is given as \cite{Demir2021book}
\begin{align}
    {\rm{CV}} = \frac{\Vert \q{h}_{mk} \Vert^2}{\E{\Vert \q{h}_{mk} \Vert^2} } \rightarrow 1 \quad \text{as} \quad N \rightarrow \infty.
\end{align}
As $N\rightarrow \infty$, the coefficient of variation approaches one,  indicating that the channel gain $\Vert\q{h}_{mk} \Vert^2$ becomes a deterministic constant (Fig.~\ref{fig_CH_FP_numAntenna}) \cite{Demir2021book}. In particular, since $\Vert\q{h}_{mk} \Vert^2$ is the sum of $N$ independent and i.i.d. random variables $\vert h_{mk,n} \vert^2$, each with finite means and variances, according to the law of large numbers, the sum converges to its expected value as $N$ increases \cite{papoulis2002probability}. Thus, the small-scale fluctuations diminish because the randomness of individual channel gains averages out over numerous antennas \cite{Demir2021book}. This effect stabilizes the channel gain around its mean.

It is important to emphasize that channel hardening in CF systems does not require each AP to be a mMIMO node. Even when individual APs are equipped with only a few antennas, the collective contribution of many distributed APs jointly serving the users produces a comparable hardening effect \cite{Demir2021book}. The macro-diversity and statistically independent fading across APs render the aggregate user-AP channel more deterministic, leading to network-level channel hardening. Hence, while the underlying principle resembles that of conventional mMIMO, CF architectures realize hardening both locally (per AP) and globally (across APs) through distributed cooperation \cite{Demir2021book}.

{Channel hardening benefits both communication and sensing in CF-ISAC systems. For communication, it simplifies linear processing (e.g., maximum ratio transmission (MRT) and maximum ratio combining (MRC)) and reduces the need for frequent CSI estimation {\cite{Demir2021book, liu2023integratedbook}}. From a sensing perspective, the reduced small-scale randomness stabilizes the propagation environment, making echo-based measurements more consistent over time. This helps separate target-induced variations from channel-induced fluctuations, improving the reliability of range, velocity, and localization estimates {\cite{Demir2021book, liu2023integratedbook}}. In multi-static sensing scenarios, where reflections from multiple APs are jointly processed, deterministic channel gains also simplify calibration and fusion across distributed nodes {\cite{Demir2021book, liu2023integratedbook}}. While the full exploitation of channel hardening for sensing remains an open research problem, its potential to enhance estimation stability and reduce recalibration overhead makes it an important physical property for future CF-ISAC design.}

\item \textit{Favorable Propagation:} This refers to the phenomenon of the different user-AP channels becoming virtually orthogonal as the number of AP antennas increases \cite{Demir2021book, Chen2018ChannelHardening, Zhang2019cellfree, Polegre2020}. This orthogonality reduces inter-user interference, enhancing system capacity and performance in multi-user communication. Mathematically, the inner product of the two-channel vectors between the $m$-th AP and the $k$-th and $l$-th users (i.e., $\q{h}_{mk}$ and $\q{h}_{ml}$) is $\q{h}_{mk}^{\rm{H}} \q{h}_{ml} = \sum_{j=1}^{N} h_{mk, j}^* h_{ml, j}$, which is a sum of $N$ i.i.d. random variables with zero mean and variance $\beta_{mk} \beta_{ml}$ \cite{Demir2021book}. As $N\rightarrow \infty$, according to the law of large numbers, $\q{h}_{mk}^{\rm{H}} \q{h}_{ml}$ converges to 0, implying that the two-channel vectors become orthogonal (Fig.~\ref{fig_CH_FP_numAntenna}), i.e.,
\begin{align}
   \rho =  \frac{\q{h}_{mk}^{\rm{H}} \q{h}_{ml}}{\sqrt{\E{\Vert \q{h}_{mk} \Vert^2} \E{\Vert \q{h}_{ml} \Vert^2} }} \rightarrow 0 \quad \text{as} \quad N \rightarrow \infty. 
\end{align}
{The correlation coefficient $\rho$ is a complex quantity. For illustration, Fig.~{\ref{fig_CH_FP_numAntenna}} shows its magnitude $|\rho|$, which is always non-negative and approaches zero as $N$ increases. This behavior indicates that the two channel vectors become nearly orthogonal, thereby reducing inter-user interference.}

Favorable propagation aligns with small-scale fading models like Rayleigh and Rician, especially when fading is uncorrelated across AP antennas and users \cite{Demir2021book, Chen2018ChannelHardening, Zhang2019cellfree, Polegre2020}. However, it weakens in highly correlated fading, severe shadowing, or densely deployed APs, leading to high spatial correlations. With a large number of AP antennas (\numrange{50}{100}), it can make user-AP channels nearly orthogonal, reducing inter-user interference by up to \qty{100}{\percent} \cite{Demir2021book, Chen2018ChannelHardening, Zhang2019cellfree, Polegre2020}.

{From a sensing perspective, favorable propagation also benefits multi-target parameter estimation. Nearly orthogonal channels across distributed APs provide distinct spatial signatures for echoes arriving from different directions or targets, reducing cross-target interference and improving angular and range resolution {\cite{Demir2021book, Chen2018ChannelHardening, Zhang2019cellfree, Polegre2020}}. This effect enables distributed APs to more effectively isolate weak reflections in cluttered environments and improves the accuracy of target localization and velocity estimation. Hence, favorable propagation serves as a unifying feature that supports both high-capacity communication and high-resolution sensing in CF-ISAC {\cite{Demir2021book, Chen2018ChannelHardening, Zhang2019cellfree, Polegre2020}}.}

\end{enumerate}

%=======================================================================
\begin{figure}[!t]\vspace{-0mm}
    \centering
    \includegraphics[width=0.47\textwidth]{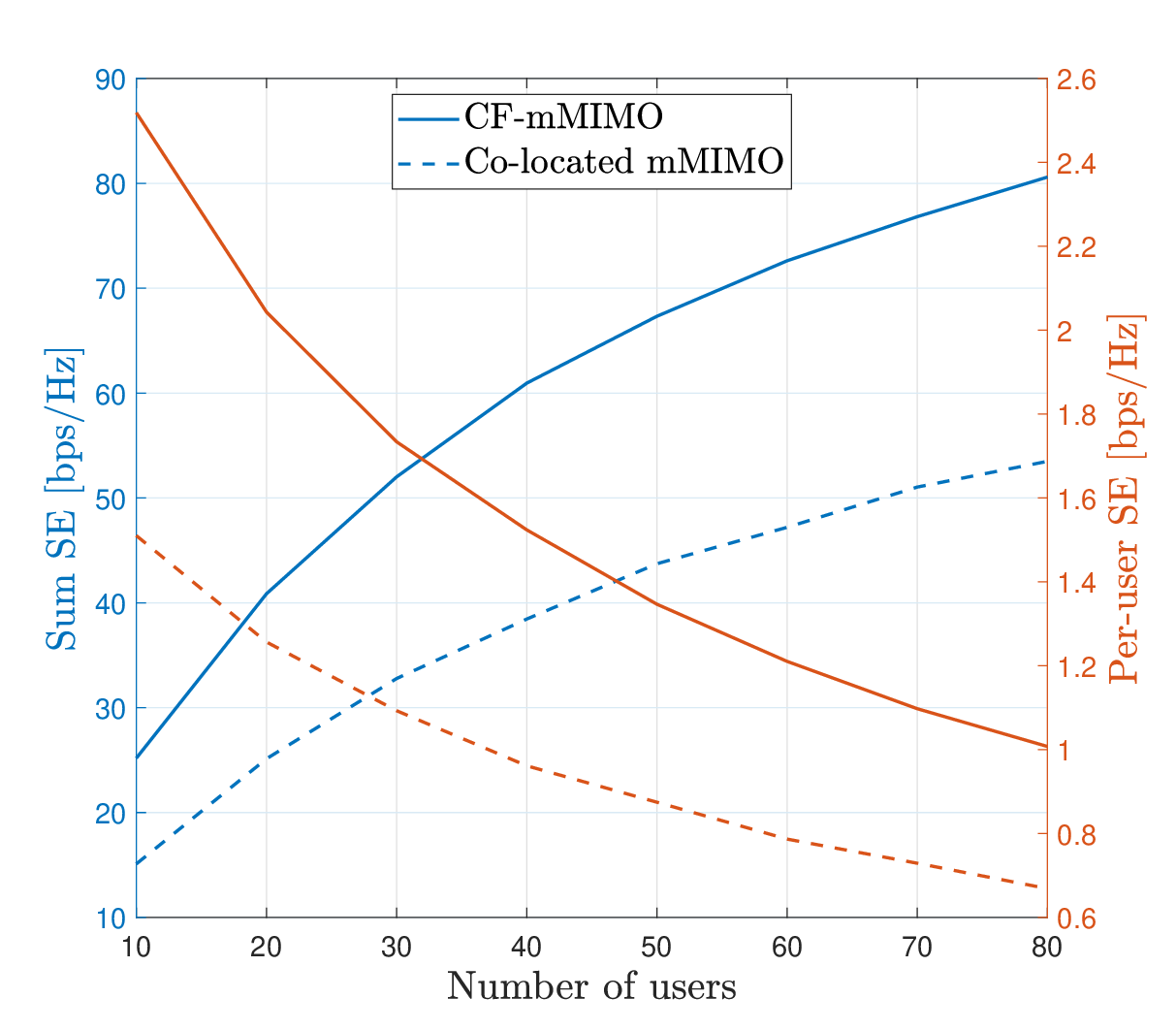}
    \caption{{Comparison of sum SE (left $y$-axis) and per-user SE (right $y$-axis) between CFMM and co-located mMIMO with {\num{100}} antennas in a {\qty{1}{\km^2}} area. In the CF system, {\num{100}} single-antenna APs are uniformly distributed, whereas in the co-located system, a {\num{100}}-antenna BS is located at the cell center. The DL SEs are evaluated using conjugate beamforming with statistical CSI at the users.}}
    \label{fig_SumAndPerUsereSE_Comp} \vspace{-0mm}
\end{figure}
%=======================================================================

\subsubsection{Performance and Benefits of CFMM over conventional cellular}\label{sec_performance_benifits} These include the following: 
\begin{itemize}
    \item \textit{Enhanced coverage:} Multiple distributed APs across a wide area provide users with consistent coverage, even in challenging environments like rural areas or dense urban centers. For example, CFMM systems can reduce outage probability by up to \qty{90}{\percent} compared to traditional cellular networks, particularly in poor line-of-sight (LoS) conditions \cite{Ngo2017EE, Papazafeiropoulos2020}.

    \item \textit{Improved SE:} CFMM enhances user data rates via spatial multiplexing, achieving \numrange{3}{5} times higher SE than traditional MIMO and exceeding \qty{100}{bps/\Hz} under ideal conditions \cite{Elhoushy2022}. As shown in Fig.~\ref{fig_SumAndPerUsereSE_Comp}, it outperforms co-located MIMO in both per-user and sum SE, even in dense user scenarios.

    \item \textit{Reduced interference:} AP cooperation minimizes inter-cell interference by eliminating cell boundaries, reducing interference by over \qty{50}{\percent} compared to conventional cellular systems, particularly for cell-edge users \cite{Yang2018}.
    
    \item \textit{Improved EE:} A large number of APs improves the likelihood of each user being close to an AP. This achieves substantial energy savings by reducing path loss and enabling lower transmit power levels. For example,  energy savings of \qtyrange{30}{50}{\percent} are achieved compared to co-located mMIMO \cite{Ngo2017EE, Yang2018}.
\end{itemize}

The fundamentals of CFMM are well known; interested readers are referred to \cite{Demir2021book} and the references therein for further information.

\subsection{Fundamentals of Radar Sensing}
{Radar (short for radio detection and ranging) systems detect and estimate object properties by transmitting signals toward targets and analyzing their reflected echoes {\cite{Mark2010RadarBook, richards2005fundamentals}}. Thus, a radar transmitter emits a narrow beam that scans the expected target area, such as aircraft, ships, spacecraft, vehicles, birds, and other targets. When the beam strikes a target, some energy scatters or reflects back to the radar receiver, which can be co-located with the transmitter (mono-static) or at a separate site (bi-static/multi-static) (Fig.~{\ref{CF_and_RadarConfigurations}b}). Some systems time-share a single antenna for transmission and reception {\cite{Mark2010RadarBook, richards2005fundamentals}}.}

{For  ISAC, the radar principles provide the analytical foundation for joint waveform design, parameter estimation, and performance analysis. Radar fundamentals, such as echo formation, RCS, clutter, and estimation accuracy, are critical for the sensing part of ISAC, where both sensing and communication occur under shared spectral and spatial resources. These parameters directly influence ISAC performance metrics such as sensing SE, Cram\'{e}r-Rao bound (CRB), and joint communication-sensing trade-offs, particularly in distributed or cooperative architectures {\cite{Mark2010RadarBook, richards2005fundamentals}}.}

{Reflections from targets form the signal of interest, while those from other sources, like the ground or rain, act as interference, degrading detection performance {\cite{Mark2010RadarBook, richards2005fundamentals}}. The radar receiver processes echo signals to estimate a target's presence, location, velocity, range, direction, size, and shape. By tracking its position over time, the target's trajectory and path can be predicted. Radar is widely used in defense, automotive, and weather forecasting due to its ability to operate under diverse conditions and to measure critical parameters such as range, velocity, and angle. From an ISAC perspective, these reflections can be interpreted as structured sensing channels that share the same spectral and spatial resources used for data transmission {\cite{liu2023integratedbook}}. The ability to distinguish target echoes from interference and multi-user signals underpins many ISAC performance metrics, such as sensing SE and CRB-based bounds.}

\subsubsection{Radar Cross-Section}
The RCS quantifies a target's radar-reflecting and scattering characteristics, i.e., its detectability by radar. It measures the intensity of the incident EM signal reflected back toward the radar by the target \cite{Mark2010RadarBook}. In particular, RCS is the effective area of a target that reflects radar signals back to the source/reader \cite{Mark2010RadarBook}. Mathematically, RCS ($\sigma$) is defined as the ratio of scattered power to incident power at a given distance, i.e., 
\begin{align}
    \sigma = \lim_{r\rightarrow \infty} 4 \pi r^2 \frac{P_{\rm{scatter}}}{P_{\rm{incident}}},
\end{align}
where $r$ is the distance between the radar and the target, $P_{\rm{scatter}}$ is the power scattered by the target back toward the radar, and $P_{\rm{incident}}$ is the power incident on the target \cite{Mark2010RadarBook}. RCS depends on the target's size, shape, material, radar frequency, incidence angle, and polarization. Larger, metallic objects typically have higher RCS, while smaller ones have lower values (e.g., an insect: \qty{e-5}{\m^2}, a large ship: \qty{e6}{\m^2}). Higher RCS aids detection at greater distances, whereas low RCS targets require more sensitive radar systems \cite{Mark2010RadarBook}. Table~\ref{tab_rcs} lists approximate RCS values for common objects \cite{knott2004radar, Rezende2002RCS, skolnik2001introduction}.

\begin{table}[t!]
\centering
\renewcommand{\arraystretch}{1.1}
\caption{Approximated RCS values of common objects.}\label{tab_rcs}
\begin{tabular}{|l|c|}
\hline
\textbf{Object} & \textbf{Approximated RCS (\qty{}{\m^2})} \\ \hline \hline
Insect &  $\num{e-6}-\num{e-5}$    \\ \hline
Bird (e.g., pigeon) &  \num{0.01}       \\ \hline
Human (standing, broadside)  &  \num{1} \\ \hline
Car (sedan, broadside)     &   $\num{10}-\num{100}$  \\ \hline
Large truck  &  $\num{100}-\num{200}$  \\ \hline
Commercial aircraft (e.g., Boeing 747)  &   $\num{30}-\num{1000}$  \\ \hline
Cargo aircraft   & Up to \num{100}   \\  \hline
Small combat aircraft   & $\num{2}-\num{3}$   \\  \hline
Large combat aircraft   & $\num{5}-\num{6}$   \\  \hline
Large ship (e.g., cargo ship, tanker)   & $\num{e5}-\num{e6}$   \\  \hline
\end{tabular}
\end{table}

{In ISAC systems, RCS diversity across angles and frequencies plays a key role in designing beamforming and waveform adaptation strategies {\cite{liu2023integratedbook}}. Particularly in distributed or cooperative architectures, variations in $\sigma$ across APs or viewpoints can be exploited for diversity gain in target detection and tracking  {\cite{Mao2023, Demirhan2023, Huang2022Coordinated}}.}

\subsubsection{Clutter}
Clutter refers to unwanted radar echoes/interference from objects that are not the intended targets, such as the ground, buildings, trees, or weather phenomena like rain and snow \cite{Mark2010RadarBook}. These reflections can mask or obscure the identification of intended targets, particularly small or low-RCS objects. Although the effect of clutter on detection can be comparable to that of noise, clutter is influenced by the environment, frequency, and transmitted signals, which may result in a distinct profile from noise \cite{Mark2010RadarBook}. To model the clutter, the Weibull, log-Weibull, log-normal, and $K$-distributions are commonly employed \cite{Sayama2001}. To reduce the impact of clutter, clutter-rejection techniques can be used \cite{Mark2010RadarBook}. These include moving target indication, Doppler filtering, constant false alarm rate, clutter maps, polarization diversity, and synthetic aperture radar \cite{Mark2010RadarBook}.

{For ISAC systems, clutter suppression must be addressed holistically alongside interference mitigation and communication resource allocation {\cite{liu2023integratedbook}}. To this end, adaptive beamforming and waveform optimization can jointly suppress clutter and multi-user interference, highlighting the synergy between classical radar processing and modern network coordination {\cite{Mark2010RadarBook}}.}

\subsubsection{Signal Transmission}
Radar signals are transmitted in the radio or microwave frequency bands \cite{Mark2010RadarBook, richards2005fundamentals, levanon2004radar, Blunt2016}. The frequency of the transmitted signal can vary depending on the application, but typical frequencies range from several megahertz (MHz) to gigahertz (GHz). The signal frequency determines both the range and resolution of the radar system. Two common types of radar are based on the transmitted signal: continuous-wave radar and pulsed radar \cite{levanon2004radar, Blunt2016}.
\begin{itemize}
    \item \textit{Continuous wave radar:} A continuous-wave (CW) radar transmits a constant, unmodulated radio frequency signal and continuously measures the reflected echoes from targets. Because the transmitted and received signals overlap in time, CW radar cannot directly measure range. Still, it can accurately determine a target's velocity by detecting the Doppler shift in its frequency. This principle forms the basis of Doppler radar, which is widely used for speed measurement and motion detection applications \cite{levanon2004radar, Blunt2016}.

    \item \textit{Pulsed radar:}  
    This system transmits short, high-power RF bursts and then switches to receive mode to capture echoes reflected from targets. By precisely measuring the time delay between the transmitted and received signals, the target's range can be determined, since this delay is proportional to the round-trip propagation distance. A series of such pulses is repeatedly transmitted for continuous detection. The pulse repetition interval (PRI) determines the maximum unambiguous range, while the pulse width governs the range resolution \cite{levanon2004radar, Blunt2016}.  

    For example, consider a pulsed radar operating with a PRI of $100~\mu\text{s}$ and a pulse width of $1~\mu\text{s}$.  
    \begin{itemize}
        \item The maximum unambiguous range is  
        \[
        R_{\max} = \frac{c \times \text{PRI}}{2} =  15~\text{km},
        \]
        where $c$ is the speed of light.  
        \item The range resolution, determined by the pulse duration, is  
        \[
        \Delta R = \frac{c \times \text{pulse width}}{2} = 150~\text{m}.
        \]
\end{itemize}
Thus, this radar can unambiguously detect targets up to $15~\text{km}$ away with a resolution of about $150~\text{m}$.

\end{itemize}

{In  ISAC, waveform selection (continuous vs. pulsed) affects not only sensing resolution but also the achievable communication rate and latency. Therefore, waveform design becomes a multi-objective problem balancing range-Doppler performance and information throughput {\cite{liu2023integratedbook}}.}

\subsubsection{General Radar Sensing Signal Model and Ambiguity Function}\label{sec_radarsignal}
{A general narrowband baseband radar sensing model can be expressed as the superposition of echoes reflected from multiple targets {\cite{Mark2010RadarBook}}. Considering $T$ point targets, the received complex baseband signal at the radar receiver is given as}
\begin{align}
    r(t) = \sum_{i=1}^{T} \alpha_i s(t - \tau_i) e^{j 2 \pi f_{D,i} t} + n(t),
\end{align}
{where $\alpha_i$ is the complex reflection coefficient (including propagation loss and target RCS), $\tau_i = 2R_i / c$ is the round-trip delay proportional to the target range $R_i$, $f_{D, i} = 2v_i f_c / c$ denotes the Doppler frequency shift caused by target radial velocity $v_i$, $f_c$ is the carrier frequency, and $n(t)$ represents additive white Gaussian noise (AWGN) {\cite{Mark2010RadarBook}}.}

{If the radar employs an antenna array with $N_r$ elements, the received signal vector can be written as}
\begin{align}
    \mathbf{r}(t) = \sum_{i=1}^{T} \alpha_i \mathbf{a}_r(\theta_i) s(t - \tau_i) e^{j 2 \pi f_{D,i} t} + \mathbf{n}(t),
\end{align}
{where $\mathbf{a}_r(\theta_i)$ is the receive array steering vector for a target at azimuth/elevation angle $\theta_i$. This model jointly captures range ($\tau_i$), velocity ($f_{D,i}$), and direction ($\theta_i$) dependencies, forming the basis for range-Doppler-angle estimation in modern radar and ISAC systems {\cite{Mark2010RadarBook}}.}

{To extract target parameters, a matched filter correlates the received signal with delayed and frequency-shifted versions of the transmitted waveform {\cite{Mark2010RadarBook, richards2005fundamentals, levanon2004radar, Blunt2016}}:}
\begin{align}
\chi(\tau, f_D) = \int_{-\infty}^{\infty} r(t) s^*(t - \tau) e^{-j 2 \pi f_{D} t} dt.
\end{align}
{where $\chi(\tau, f_D)$, the ambiguity function (AF), characterizes the radar's ability to resolve two targets separated in delay and Doppler~{\cite{Mark2010RadarBook, richards2005fundamentals, levanon2004radar, Blunt2016}}. The AF's main-lobe widths define the range and velocity resolutions as}
\begin{align}
\Delta R \approx \frac{c}{2B}, \qquad \text{and} \qquad \Delta v \approx \frac{\lambda}{2T_{\text{int}}},
\end{align}
{where $B$ and $T_{\text{int}}$ are the signal bandwidth and integration time, respectively. The zero-Doppler cut of $\chi(\tau, f_D)$ yields the waveform's autocorrelation function, governing range resolution and sidelobe behavior, while the zero-delay cut reveals Doppler sensitivity through the power spectrum {\cite{Mark2010RadarBook, richards2005fundamentals, levanon2004radar, Liu2023a, Ahmadipour2024}}.}

{In ISAC systems, the AF serves as a unifying metric linking sensing resolution, interference tolerance, and waveform design. Unlike traditional radar, ISAC waveforms must also satisfy communication constraints, introducing a deterministic-random trade-off in waveform structure~{\cite{Liu2023a, Ahmadipour2024}}. Recent studies have analyzed the statistical properties of random ISAC signals~{\cite{Liao2025, Liu2025, Vandendorpe2025}}, revealing that the expectation and variance of the AF can quantify the sensing-communication balance. Minimizing the AF sidelobes improves sensing accuracy but may reduce achievable data rates. This necessitates multi-objective optimization formulations of the form $\eta f_{\text{sense}}(\cdot) + (1-\eta) f_{\text{comm}}(\cdot)$, where $\eta$ governs task priority. Thus, AF-based analysis provides a powerful foundation for evaluating ISAC waveform efficiency and designing joint range-Doppler-angle-communication trade-offs in future distributed and CF-ISAC systems.}
{Moreover, this formulation establishes the analytical link between traditional radar signal models and ISAC performance evaluation. In ISAC, this model underlies joint design problems such as waveform optimization, power allocation, and sensing-communication trade-offs, in which $\tau_i$, $f_{D,i}$, and $\theta_i$ are jointly estimated using shared resources.}

\subsubsection{Types of Sensing}\label{sec_radartypes}
Radar sensing/systems can be classified into three groups based on the spatial relationship between the transmitter and receiver, i.e., mono-static, bi-static, and multi-static sensing/configurations (Fig.~\ref{CF_and_RadarConfigurations}b) \cite{Mark2010RadarBook, richards2005fundamentals, li2009mimoradar}. Each configuration presents unique advantages, limitations, and challenges in radar signal processing, hardware design, and system performance.
\begin{itemize}
    \item \textit{Mono-static radar:} This is characterized by a co-located transmitter and receiver pair, often sharing the same antenna (array) for transmission and reception (Fig.~\ref{CF_and_RadarConfigurations}b). It thus requires full-duplex (FD) operation \cite{Mark2010RadarBook}. However, the simultaneous transmission and reception can introduce strong self-interference (SI), necessitating SI cancellation techniques \cite{Mohammadi2023, Diluka2024CFFD}. Alternatively, the receiver should be isolated from the transmitter to protect it from the high SI. However, some modern systems use sophisticated duplexers to allow simultaneous transmission and reception at different frequencies or polarization states \cite{Mark2010RadarBook}. This configuration is widely used in most radar applications due to its simplicity in design and ease of signal processing \cite{Mark2010RadarBook}. Due to the co-located or shared antennas, this has less hardware complexity compared to other configurations. Mono-static radars benefit from well-developed signal processing algorithms, including pulse compression, Doppler processing, and clutter suppression \cite{Mark2010RadarBook}. The transmitter and receiver co-location allows for precise range measurements, with accuracy dependent on the pulse width or bandwidth \cite{Mark2010RadarBook}. Mono-static radar often fails to detect objects having a low RCS, which does not reflect much radar energy to the source \cite{Mark2010RadarBook}. 

    \item \textit{Bi-static radar:} The transmitter and receiver are separated, i.e., placed at different locations (Fig.~\ref{CF_and_RadarConfigurations}b) \cite{Mark2010RadarBook}. The separation can vary from short distances to several kilometers, making it suitable for various operational scenarios such as long-range surveillance and covert detection \cite{Mark2010RadarBook}. Accurate synchronization between the transmitter and receiver is critical because the receiver lacks direct access to the transmitted signal, thereby increasing system complexity \cite{Mark2010RadarBook}. External timing sources/techniques (e.g., direct signal reception, Global Positioning System (GPS) timing, or sophisticated signal tracking algorithms) are required to maintain accurate synchronization \cite{Mark2010RadarBook}. As the receiver is separated from the transmitter,  adversaries find it harder to detect and locate the radar system, making bi-static radar more suitable for military surveillance and stealth applications. The transmitter-receiver separation can reduce clutter caused by direct-path reflections (e.g., from the ground or other environmental features) \cite{Mark2010RadarBook}. On the other hand, the RCS of targets can vary substantially with bistatic angle, making them sensitive to target orientation and geometry \cite{Mark2010RadarBook}.  

    \item \textit{Multi-static radar:} This arrangement involves multiple transmitters and receivers distributed over different locations (Fig.~\ref{CF_and_RadarConfigurations}b) \cite{Mark2010RadarBook}. It leverages spatial diversity and offers significant advantages in target identification, tracking precision, and resistance to jamming or interference \cite{Mark2010RadarBook}. Multiple receivers enable more accurate triangulation of target positions than mono-static or bi-static radars. This is particularly useful in cluttered or complex environments \cite{Mark2010RadarBook}. Multi-static radar can be configured to produce high-resolution images of targets using techniques such as synthetic aperture radar and inverse synthetic aperture radar, incorporating multiple observations \cite{Mark2010RadarBook}. Jamming becomes substantially more difficult because an adversary must disrupt many radar links simultaneously. The spatial diversity of receivers also helps to reduce interference from environmental clutter and multi-path reflections \cite{Mark2010RadarBook}. Nevertheless, this configuration necessitates coordinating multiple radar nodes, thereby increasing the complexity of the hardware, signal processing, and communications infrastructure.  Additionally, precise synchronization between multiple transmitters and receivers is challenging, particularly in dynamic or mobile environments \cite{Mark2010RadarBook}.   
\end{itemize}

\subsubsection{Target Parameter Estimation Techniques}
Radar sensing detects targets and estimates parameters such as range, velocity, and angle of arrival (AoA) \cite{Mark2010RadarBook}. Various signal processing techniques enable accurate target parameter estimation:

\begin{itemize}
    \item \textit{Range estimation:} Matched filtering maximizes the signal-to-noise ratio (SNR) by correlating the received signal with a transmitted signal template. In pulse radar, this identifies target presence and range. The received signal is modeled as  
    \begin{align}
        y(t) = \alpha s(t-\tau) + n(t),
    \end{align}
    where $\alpha$ is the reflection coefficient, $\tau$ is the time delay, $s(t)$ is the transmitted signal,  and $n(t)$ is noise. The matched filter's peak at $t=\tau$ determines the target's range \cite{Mark2010RadarBook}.

    \item \textit{Velocity estimation:} Doppler processing determines the radial velocity of moving objects via the Doppler shift $\Delta f_{D}$, which is proportional to velocity. The fast Fourier transform (FFT) converts the received time-domain signal into the frequency domain, where velocity is inferred from spectral peaks \cite{Mark2010RadarBook}.  

    \item \textit{AoA estimation:} This is done by analyzing the direction of the reflected signal at the radar receiver \cite{Mark2010RadarBook}. Common methods include:  
    \begin{enumerate}
        \item \textbf{Monopulse radar} - this adds extra radio signal encoding to provide accurate directional information. It compares signal intensities across multiple beams or antennas to determine precise angular location.
        \item \textbf{Beamforming}, in phased-array radars, steers beams and processes received signals to determine target direction.
        \item \textbf{Subspace-based methods} -  techniques like multiple signal classification (MUSIC) and estimation of signal parameters via rotational invariance (ESPRIT) improve AoA accuracy.
    \end{enumerate}
\end{itemize}

{These techniques, ranging from matched filtering to subspace-based AoA estimation, serve as building blocks for ISAC signal processing {\cite{liu2023integratedbook}}. In distributed implementations, extensions of MUSIC and ESPRIT enable cooperative multi-AP parameter estimation, in which spatially diverse echoes are fused to achieve higher angular and range resolution. Such cross-disciplinary adaptations are key to realizing high-precision ISAC in next-generation wireless networks.}

\section{Integrated Sensing and Communication}\label{sec_isac}
ISAC is an emerging paradigm integrating sensing and communication into a unified framework \cite{liu2023integratedbook}. Unlike traditional systems, where different networks handle each function separately, ISAC leverages this synergy to enhance SE, reduce hardware costs, and enable new applications by sharing resources such as spectrum, power, and hardware.

\subsection{Fundamentals of ISAC}
Sensing and communication tasks are integrated into a single system using shared hardware, spectrum, and signal-processing algorithms \cite{liu2023integratedbook}. Since the BS can simultaneously transmit data and extract environmental information from echoes, this dual capability is exploited in ISAC \cite{liu2023integratedbook}. However, a key challenge arises: communication systems require highly random signals to achieve maximum channel capacity \cite{Xiong2022}, whereas radar systems rely on deterministic signals for accurate sensing \cite{Mark2010RadarBook}. This random-deterministic conflict is a fundamental challenge in ISAC, making it critical to research for the next-generation wireless networks \cite{liu2023integratedbook}.

Integration can be complete or partial, depending on the application. For instance, wireless sensor networks require hardware integration, surveillance radar systems need signaling integration, and cognitive radars rely on spectrum integration \cite{liu2023integratedbook}. ISAC systems are categorized into four types based on their integration level \cite{liu2023integratedbook}.
\begin{itemize}
    \item \textit{Level 1 - Spectral coexistence:} Spectral resources are shared between separate communication and sensing systems, minimizing mutual interference while allowing independent operation.

    \item \textit{Level 2 - Co-located hardware:} In addition to spectrum sharing, this level combines communication and sensing on a shared hardware platform, improving resource efficiency while maintaining distinct operations.

    \item {\textit{Level 3 - Joint waveform and processing:} At this level, communication and sensing are realized via a common waveform and a unified signal processing framework, achieving deep integration at the link level with performance gains through joint optimization.}

    \item {\textit{Level 4 - Perceptive networks:} This highest level goes beyond waveform sharing, embedding sensing as a native function of the wireless network. Multiple distributed nodes collaborate to provide network-wide communication, localization, and environmental awareness, enabling large-scale IoT, 5G, and 6G applications.}

\end{itemize}

Regardless of integration level, ISAC aims to (i) reduce hardware complexity by unifying communication and sensing, minimizing antennas, transceivers, and spectrum use; (ii) enhance SE by coordinating both functions in the same frequency band to minimize interference; and (iii) improve efficiency through joint optimization, reducing energy consumption \cite{liu2023integratedbook}. 

\subsection{ISAC Design Philosophy}
The approach is to jointly design sensing and communication functions to achieve direct trade-offs and mutual benefits \cite{liu2023integratedbook, Azar2024, Ma2020}. To briefly introduce this process, consider a general ISAC system with a BS equipped with an $M$-element uniform linear array (ULA), serving $K\geq 1$ single-antenna users and tracking targets. The BS transmitted signal at the $l$-th time slot can be expressed as
\begin{eqnarray}\label{eqn_tx_signal}
    \q{x}(l) = \sum\nolimits_{k =1}^{K} \sqrt{\rho} \q{w}_k q_k(l) + \sqrt{(1-\rho)} \q{s}(l) \in \mathbb{C}^{M\times 1},
\end{eqnarray}
where $\q{w}_k \in \mathbb{C}^{M \times 1}$  and $q_k(l)$ denote the communication beamforming vector and the $l$-th time slot data  for the $k$-th user. The sensing signal, $\q{s}(l) \sim \mathcal{CN}(\boldsymbol{0}, \q{R}_s)$, follows a complex Gaussian distribution with covariance matrix $\q{R}_s = \E{\q{s} \q{s}^{\rm{H}}} \succeq 0$. This covariance matrix is designed to extend the DoF of the BS transmit signal, enhancing the system's sensing performance \cite{liu2023integratedbook}. Finally, $\rho \in [0,1]$ denotes the power allocation or priority factor between communication and sensing, determining the balance and priority between these two tasks~\cite{liu2023integratedbook}.

Based on the priority level, there are three ISAC design philosophies: (i) Communication-centric design with high $\rho$ ($\rho \rightarrow 1$), (ii) Sensing-centric design with low $\rho$ ($\rho \rightarrow 0$), and (iii) Joint design with moderate $\rho$ \cite{Ma2020}.
\begin{enumerate}
    \item \textit{Communication-centric design:} 
This design adapts existing communication systems for dual communication and sensing \cite{Ma2021}. Since standard communication signals are not optimized for sensing, advanced processing techniques are utilized to enhance performance. The approach prioritizes communication while extracting target information from signal echoes with minimal modifications.

    \item \textit{Sensing-centric design:} Sensing takes priority over communication, with radar as the primary function and communication as a secondary task \cite{Chen2022}. For example, communication symbols can be embedded in radar waveforms, ensuring accurate sensing while meeting communication needs.  However, the embedded symbols in the sensing signal must not compromise the accuracy or reliability of the primary sensing function, ensuring dependable detection and measurement while meeting communication requirements \cite{Chen2022}.

    \item \textit{Joint design:} Here, the signal is designed to assign equal or well-balanced priorities to both sensing and communication, enabling improved trade-offs between the two functionalities. This approach enables a more flexible resource-allocation framework for sensing and communication tasks. As a result, the joint signal design provides greater flexibility and higher DoF, effectively balancing the requirements of both sensing and communication \cite{Liu2020}.
\end{enumerate}

These ISAC design concepts push conventional communication-only wireless networks to a new dimension  \cite{liu2023integratedbook}.  Sensing capabilities, in particular, may become a standard component of future wireless networks, serving as auxiliary or essential for many users and applications. Conversely, sensing data can enhance communication performance. For example, sensing-aided vehicle beamforming and resource management can improve signal quality and optimize network resources, leading to more efficient and reliable communication in dynamic environments \cite{liu2023integratedbook}. Additionally, sensing-enabled mobile networks can continuously sense their environment and provide traffic monitoring, weather forecasting, and human activity recognition services. This sensing data develops intelligence for the ISAC network and its associated applications, such as smart homes, transportation, and cities \cite{liu2023integratedbook}.

%=======================================================================
\begin{figure*}[!t]\centering \vspace{0mm}
    \def\svgwidth{480pt} 
    \fontsize{9}{9}\selectfont 
    \graphicspath{{Figures/}}
    %% Creator: Inkscape 1.3 (0e150ed6c4, 2023-07-21), www.inkscape.org
%% Accompanies image file 'ISACApplications.eps' (pdf, eps, ps)
\begingroup%
  \makeatletter%
  \providecommand\color[2][]{%
    \errmessage{(Inkscape) Color is used for the text in Inkscape, but the package 'color.sty' is not loaded}%
    \renewcommand\color[2][]{}%
  }%
  \providecommand\transparent[1]{%
    \errmessage{(Inkscape) Transparency is used (non-zero) for the text in Inkscape, but the package 'transparent.sty' is not loaded}%
    \renewcommand\transparent[1]{}%
  }%
  \providecommand\rotatebox[2]{#2}%
  \newcommand*\fsize{\dimexpr\f@size pt\relax}%
  \newcommand*\lineheight[1]{\fontsize{\fsize}{#1\fsize}\selectfont}%
  \ifx\svgwidth\undefined%
    \setlength{\unitlength}{283.60140991bp}%
    \ifx\svgscale\undefined%
      \relax%
    \else%
      \setlength{\unitlength}{\unitlength * \real{\svgscale}}%
    \fi%
  \else%
    \setlength{\unitlength}{\svgwidth}%
  \fi%
  \global\let\svgwidth\undefined%
  \global\let\svgscale\undefined%
  \makeatother%
  \begin{picture}(1,0.45831563)%
    \lineheight{1}%
    \setlength\tabcolsep{0pt}%
    \put(0,0){\includegraphics[width=\unitlength]{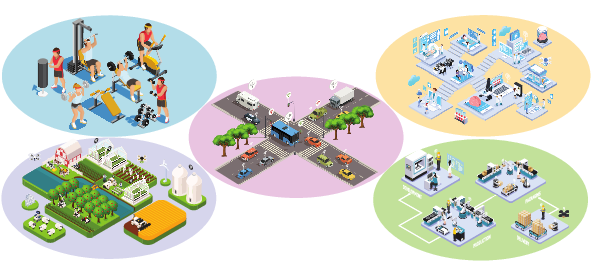}}%
    \put(0.08359202,0.43471243){\makebox(0,0)[lt]{\lineheight{1.25}\smash{\begin{tabular}[t]{l}Human activity recognition\end{tabular}}}}%
    \put(0.69713094,0.43471243){\makebox(0,0)[lt]{\lineheight{1.25}\smash{\begin{tabular}[t]{l}Healthcare and assisted living\end{tabular}}}}%
    \put(0.05316943,0.00883063){\makebox(0,0)[lt]{\lineheight{1.25}\smash{\begin{tabular}[t]{l}Agriculture and precision farming\end{tabular}}}}%
    \put(0.73017771,0.0035415){\makebox(0,0)[lt]{\lineheight{1.25}\smash{\begin{tabular}[t]{l}Smart manufacturing\end{tabular}}}}%
    \put(0.42359237,0.10850363){\makebox(0,0)[lt]{\lineheight{1.25}\smash{\begin{tabular}[t]{l}Autonomous vehicles\end{tabular}}}}%
  \end{picture}%
\endgroup%
 \vspace{0mm}
    \caption{{ISAC supports a wide range of applications, including autonomous vehicles, smart manufacturing, precision agriculture, healthcare and assisted living, and human activity recognition. By enabling simultaneous data communication and environmental sensing, ISAC facilitates real-time monitoring, localization, and intelligent decision-making.}}\vspace{0mm} \label{fig_ISACApplications}
\end{figure*}
%======================================================================

\subsection{Sensing Metrics in ISAC} 
In ISAC, sensing metrics, such as sensing rate/SE, CRB, and transmit/receiver beampattern gain, are critical in evaluating the performance of the sensing functionality \cite{Bell1993, Tang2010, Zhang2021Radar, Mark2010RadarBook, Kay1998, He2022, Stoica2007, Cui2014, Hua2023}. These metrics assess the system's ability to precisely detect and estimate parameters of interest while reflecting the inherent trade-offs between sensing and communication operations \cite{Bell1993, Tang2010, Zhang2021Radar}.

\subsubsection{Sensing SE}\label{sec_sensing_SE}
{The sensing SE, measured in {\qty{}{bps/\Hz}}, quantifies the amount of target-related information that an ISAC system can extract about its environment, such as range, velocity, or angle, within a given time and bandwidth {\cite{Bell1993, Tang2010, Zhang2021Radar}}. The sensing rate is defined as the mutual information (MI) between the transmitted signal and the received reflections/echoes {\cite{Bell1993, Tang2010, Zhang2021Radar}}. For example, consider an ISAC BS with $M$ transmit and $M$ receiver antennas. The received echo signal over $L$ symbols can be expressed as}
\begin{align}\label{eqn_obs_sig}
    \q{Y} = \q{G} \q{X} + \q{Z} \in \mathbb{C}^{M \times L},
\end{align}
{where $\q{X} \in \mathbb{C}^{M \times L}$ is the transmitted signal, $\q{G} \in \mathbb{C}^{M\times M}$ is sensing channel or the target response matrix {\cite{Mark2010RadarBook}}, and $\q{Z}$ is the AWGN matrix at the ISAC BS with independently distributed elements, i.e., $\sim \mathcal{CN}(0, \sigma^2)$. Thus, the sensing MI at the receiver can be given as {\cite{Zhang2021Radar}}}
\begin{align}
    I(\q{Y}; \q{G}\vert \q{X}) = M \log_2 \left(\det \left(\frac{1}{\sigma^2} \q{X}^{\rm{H}} \q{R}_G \q{X} + \q{I}_L\right)  \right),
\end{align}
{where $I(X;Y\vert Z)$ is the MI between $X$ and $Y$ conditioned on $Z$, $\q{I}_L$ is an identity matrix of size $L\times L$, and $\q{R}_G = \E{\q{G} \q{G}}/M$. Thus, the achievable sensing SE in {\qty{}{bps/\Hz}} is defined as {\cite{Tang2019, Zhenyao2023, Cui2014, Ouyang2022}}}
\begin{align}
    \mathcal{S}^{\rm{Sen}} = \max_{\tr(\q{X} \q{X}^{\rm{H}})\leq p_{\max}}  \frac{M}{L} \log_2 \left(\det \left(\frac{1}{\sigma^2} \q{X}^{\rm{H}} \q{R}_G \q{X} + \q{I}_L\right)  \right),
\end{align}
{where $p_{\max}$ is the maximum allowable transmit power at the ISAC BS. Alternatively, the sensing SE can be approximated as $\mathcal{S}^{\rm{Sen}} \approx \log_2 \left(1+ \mt{SINR}^{\rm{Sen}} \right)$, where $\mt{SINR}^{\rm{Sen}}$ is the sensing signal-to-interference-plus-noise ratio (SINR) {\cite{Tang2019, Zhenyao2023, Cui2014, Ouyang2022}}.}

Unlike the communication rate, which has an operational interpretation in terms of coding strategies, the sensing rate quantifies the information-extraction capability of the radar component in ISAC systems. Despite lacking a coding-based interpretation, it serves as a stand-alone and practically meaningful metric for sensing performance {\cite{Bell1993, Tang2010, Zhang2021Radar}}. On the other hand, maximizing the sensing rate yields optimal waveform and resource allocation strategies and has been shown to achieve estimation performance equivalent to minimizing mean-square error (MSE) or to the CRB {\cite{Bell1993, Tang2010, Zhang2021Radar}}.

A higher sensing rate indicates greater information extraction per unit time, which is critical for accurate and rapid estimation of target parameters {\cite{Tang2019, Zhenyao2023, Cui2014, Ouyang2022}}. Consequently, a high sensing rate is essential for many real-time applications, including autonomous vehicles, surveillance radar systems, healthcare monitoring, and environmental monitoring. In these applications, the volume and timeliness of sensed information are often more crucial for decision-making than individual measurement precision. For instance, autonomous vehicles require frequent updates to avoid collisions and navigate dynamic environments, while surveillance systems and healthcare monitoring rely on high sensing rates for real-time tracking and continuous observation {\cite{Tang2019, Zhenyao2023, Cui2014, Ouyang2022}}.

{In addition, the quality of target parameter estimation is proportional to the sensing SE or corresponding sensing SINR {\cite{Tang2019, Zhenyao2023, Cui2014}}. Improved sensing SE enhances parameter estimation by proper echo-signal processing {\cite{Tang2019, Zhenyao2023, Cui2014, Diluka2023}} and enables estimation using both transmit and receive beampatterns, thereby reducing interference among targets. However, achieving a high sensing rate necessitates careful allocation of shared resources, such as power and bandwidth, between sensing and communication functions {\cite{Tang2019, Zhenyao2023, Cui2014, Ouyang2022}}.}

In summary, the sensing rate provides a rigorous and practically relevant benchmark for the efficiency of information extraction in ISAC, while the CRB offers a complementary measure of estimation accuracy. Together, they form a comprehensive framework for evaluating and optimizing sensing performance in future CF-ISAC networks.

\subsubsection{CRB}
The CRB is the lower bound for the variance of an unbiased estimator designed to estimate a certain parameter and serves as a theoretical limit on the precision of parameter estimation in statistical signal processing \cite{Kay1998}. Hence, it offers insights into the maximum achievable accuracy for estimating parameters such as target range, velocity, and angle in ISAC systems \cite{Mark2010RadarBook, Kay1998}. Minimizing the CRB thus results in improved target parameter estimation \cite{Mark2010RadarBook, Kay1998}.

The CRB is derived from the Fisher information matrix (FIM), which quantifies the sensitivity of the likelihood function to changes in the estimated parameters. Mathematically, the CRB for estimating the parameter $\boldsymbol{\theta} = [\theta_1,\ldots, \theta_N]^{\rm{T}}$ is given by \cite{Kay1998}
\begin{align}
    \mt{CRB}(\theta_n) = \left[\q{F}^{-1}(\boldsymbol{\theta})\right]_{nn},
\end{align}
where $\q{F}(\boldsymbol{\theta})$ is the FIM. In particular, the CRB is a lower bound for the minimum variance unbiased estimation, i.e., ${\rm{Var}}(\hat{\theta}_n) \geq \mt{CRB}(\theta_n)$, where ${\rm{Var}}(\hat{\theta}_n)$ is the estimation variance and $\hat{\theta}_n$ is the estimated value of $\theta_n$ \cite{Mark2010RadarBook, Kay1998}. The elements of the FIM are computed as the expected value of the second derivatives of the log-likelihood function $p(\q{Y}; \boldsymbol{\theta})$ with respect to $\boldsymbol{\theta}$ and is given as $[\q{F}(\boldsymbol{\theta})]_{n,m} =  - \E{\frac{\partial^2 \ln(p(\q{Y}; \boldsymbol{\theta}))}{\partial \theta_n  \partial \theta_m } }$,  where $\q{Y}$ is the received echo signal \eqref{eqn_obs_sig}. From \eqref{eqn_obs_sig}, $\q{Y} \sim \mathcal{CN}(\q{G} \q{X}, \sigma^2 \q{I}_M)$, and hence the $\{n, m\}$-th element of $\q{F}(\boldsymbol{\theta})$ can be expressed as \cite{Bekkerman:TSP:2006, dan2024}
\begin{align}\label{eqn_F_nm}
    [\q{F}(\boldsymbol{\theta})]_{n,m} = \frac{2M}{\sigma^2} \Re \left\{\tr \left(\frac{\partial (\q{G} \q{X})}{\partial \theta_n} \frac{\partial (\q{G} \q{X})^{\rm{H}}}{\partial \theta_m} \right) \right\}.
\end{align}
This indicates how strongly the parameters $\boldsymbol{\theta}$ influence the observations $\q{Y}$ \cite{Mark2010RadarBook, Kay1998}. A higher Fisher information corresponds to more precise parameter estimation and a lower CRB \cite{Mark2010RadarBook, Kay1998}. 

In sum,  the CRB is a key metric for evaluating sensor performance, designing signal processing schemes, and optimizing ISAC systems \cite{Li2007, Liu2020Radar, Rivetti:WCNC:2024}. It sets theoretical benchmarks for estimation algorithms, resource allocation, and waveform design. For instance, minimizing the CRB for sensing parameters while preserving communication quality helps balance communication-sensing trade-offs \cite{Li2007, Liu2020Radar, Rivetti:WCNC:2024}.

\subsubsection{Beampattern Gain}Sensing beampattern gains are a key ISAC metric, determining how energy is directed (transmit), and signals are analyzed (receive) \cite{He2022, Stoica2007, Cui2014, Hua2023}. The transmit beampattern shapes energy radiation for efficient target illumination, while the receive beampattern, optimized via sensing combiners, enhances echo reception \cite{He2022, Stoica2007, Cui2014, Hua2023}. These patterns are crucial for optimizing sensing, target detection, and communication efficiency in ISAC systems \cite{He2022, Stoica2007, Cui2014, Hua2023, Zhenyao2023, galappaththige2024RSMA, zargari2024ISABC, Diluka2026, Diluka2026WBNF}. Thus, the three pivotal beampattern gains are  
\begin{subequations}\label{eqn_beamgain}
\begin{align} 
p_t(\theta) &= \left| \q{a}^{\rm{H}}(\theta) \q{x} \right|^2, \label{eqn_tx_beamgain} \\ 
p_r(\theta) &= \left| \q{u}^{\rm{H}} \q{b}(\theta) \right|^2, \label{eqn_rx_beamgain} \\
p_c(\theta) &= \left| \q{u}^{\rm{H}} \q{b}(\theta) \q{a}^{\rm{H}}(\theta) \q{x} \right|^2, \label{eqn_com_beamgain} 
\end{align}
\end{subequations}
where $\q{x} \in \mathbb{C}^{M \times 1}$ is the transmitted signal, $\q{a}(\theta) \in \mathbb{C}^{M \times 1}$ and $\q{b}(\theta) \in \mathbb{C}^{M \times 1}$ are the transmit and receiver steering vectors that capture the array responses from direction $\theta$, and $\q{u} \in \mathbb{C}^{M \times 1}$ is the combiner at the receiver. First, \eqref{eqn_tx_beamgain} is the transmit beampattern gain and illustrates how the transmitted energy disperses as a function of angle $\theta$. Second, \eqref{eqn_rx_beamgain} is the receiver beampattern gain and encapsulates the sensitivity of the ISAC system across different angles during the reception of reflected energy. Finally, \eqref{eqn_com_beamgain} is the combined beampattern gain and provides a single representation that integrates the effects of transmission and subsequent reflection processing.  

Note that in CF-ISAC,  the beampattern gains, eq. \eqref{eqn_beamgain},  must be computed at multiple APs. These distributed gains from the APs provide a significant advantage over traditional co-located ISAC \cite{He2022, zargari2024CFISAC}. Unlike co-located ISAC, which relies on a single viewpoint (e.g., a BS) and primarily estimates target angles, CF-ISAC leverages spatial diversity to refine both angular and distance localization (Fig.~\ref{fig_BeamGani_Comp}). By coherently combining beampattern gains, whether from transmission, reception, or both,  CF-ISAC mitigates directional ambiguities and improves localization accuracy \cite{zargari2024CFISAC}. In a multi-static scenario, optimizing transmit beampatterns at transmitting APs and enhancing received beampatterns at sensing APs provide reliable data for CPU-level processing. This spatial diversity enhances sensing accuracy, robustness, and detection reliability in dynamic and multi-target environments \cite{zargari2024CFISAC}.

Since each AP must evaluate local beampattern gains and exchange information with the CPU for precise target localization, this advantage comes at the cost of increased computational complexity and coordination overhead \cite{He2022, zargari2024CFISAC}. However, computing beampattern gains at all APs may be unnecessary as a well-placed subset can achieve near-optimal localization while reducing system overhead \cite{He2022, zargari2024CFISAC}. The required number of APs depends on factors such as target count, visibility, desired localization precision, and AP distribution. Generally, three or more APs with sufficient angular separation are needed for 2D localization, while additional APs enhance robustness in multi-target scenarios and mitigate uncertainties from noise and obstructions.

\subsection{ISAC Applications}
{Standardization bodies, such as 3GPP and ITU-R, have identified over 30 use cases for ISAC across multiple verticals, including transportation, healthcare, industrial automation, and smart cities {\cite{IMT2023, 3GPPISAC2024, liu2023integratedbook, Wang2022ISAC}}. The examples discussed in this section, i.e., autonomous vehicles, healthcare, smart manufacturing, and precision agriculture (Fig.~{\ref{fig_ISACApplications}}), represent broad categories under which many of these standardized use cases can be grouped. These applications highlight the practical relevance of ISAC and its potential to enable intelligent, interconnected environments across future wireless networks.}

\subsubsection{ISAC for Autonomous Vehicles} 
Self-driving vehicles need both high-resolution radar sensing for obstacle detection and reliable communication for V2X connectivity. This fusion can transform the transportation industry by enhancing navigation safety, highway capacity, and traffic flow while reducing fuel consumption, pollution, and accident rates \cite{Zeng2019, Milanes2012}. In particular,  autonomous vehicles acquire ambient information while exchanging data with roadside units (RSUs), other vehicles, and pedestrians \cite{Zeng2019, Milanes2012}. Conversely, ISAC-aided V2X can offer environmental information for rapid vehicle platooning, secure and seamless access, and simultaneous localization and mapping,  addressing EM compatibility and spectrum congestion challenges \cite{Zeng2019, Milanes2012}. 

\subsubsection{Human Activity Recognition}
Computing systems can monitor, evaluate, and assist people daily by recording their behaviors, making activity recognition essential \cite{Yongsen2019}. Wireless signal fluctuations, influenced by static and moving objects and human actions, can be used to detect activities such as presence, proximity, falls, sleep, breathing, and more. These capabilities have significant applications in healthcare and transportation \cite{Yongsen2019}. For instance, measuring a driver's blink rate with high-resolution sensors can help detect drowsy driving, thereby enhancing road safety. Additionally, integrating sensing capabilities into commercial wireless devices, such as Wi-Fi, can identify occupant behaviors, thereby creating innovative, human-centric environments \cite{Yongsen2019}.

\subsubsection{Smart Manufacturing (Industry 4.0)} Communication and sensing enable the automation of production lines in smart manufacturing \cite{Popovski2019}. Modern factories feature interconnected machinery, robotic arms, and autonomous systems that collaborate in real time to ensure efficiency and precision. Machines communicate wirelessly and sense their environment, detecting factors like vibrations, temperature changes, and product quality. This dual functionality supports predictive maintenance, allowing machines to identify potential failures proactively, reducing downtime, and minimizing repair costs \cite{Popovski2019}.

\subsubsection{Healthcare and Assisted Living}Integrated solutions for continuous monitoring and real-time communication are vital for improving patient outcomes and elderly care \cite{Philip2021}. These systems enable remote monitoring by collecting health data, such as heart rate, blood pressure, and oxygen saturation, and transferring them to healthcare providers for real-time evaluation. This allows personalized care and timely interventions without requiring hospital visits \cite{Philip2021}. In assisted living, sensors track daily activities, detect falls, and identify abnormal behavior, such as prolonged inactivity, alerting caregivers or medical professionals during emergencies. Hospitals can also benefit from optimizing patient flow, medical equipment utilization, and staff coordination, thereby enhancing healthcare delivery \cite{Philip2021}.

\subsubsection{Agriculture and Precision Farming}Critical environmental conditions, such as soil moisture, nutrient levels, weather, and crop health, play a key role in agriculture \cite{Chander2023}. For instance, soil sensors can monitor soil moisture in real time, enabling irrigation systems to adjust water use automatically. Drones equipped with ISAC  can survey large fields, detect plant diseases and pest infestations, analyze growth patterns, and use this data to apply pesticides or fertilizers precisely. This integration of communication and sensing reduces resource waste, such as water and chemicals, while boosting crop yields and sustainability \cite{Chander2023}.

%%%%%%%%%%%%%%%%%%%%%%%%%%%%%%%%%%%%%%%%%%
\subsection{Evolution of the ISAC Networks}\label{subsec_CF_isac}
%%%%%%%%%%%%%%%%%%%%%%%%%%%%%%%%%%%%%%%%%%
{Early research in ISAC has predominantly focused on \textit{system-/link-level designs}, demonstrating the feasibility of combining sensing and communication functionalities within a single BS~{\cite{liu2023integratedbook, Liu2022ISAC, zargari2024riemannian, Rahman2020, Liu2020, Diluka2023, Diluka2024NF, Diluka2025NF, Zhenyao2023, Ouyang2022Uplink, Ouyang2022}.}}\footnote{Throughout this paper, ``system-/link-level ISAC" is used interchangeably with ``conventional ISAC".} {While these efforts laid the groundwork for ISAC technologies, they also exposed several critical limitations: \textit{i)} severe inter-cell interference, \textit{ii)} restricted coverage (especially in millimeter-wave bands) and \textit{iii)} limited sensing resolution, accuracy, and detection probability. To overcome these challenges, the field is now shifting toward \textit{network-level ISAC}, also referred to as multi-cell cooperation, cooperative ISAC, or distributed ISAC~{\cite{Zhiqing:Net:2024,Xu:COMMG:2024,Guo:WC:2025,Meng:WC:2025,Wang:WC:2025}}. In this paradigm, multiple BSs or APs collaborate to enhance both communication and sensing capabilities. This cooperation enables advanced techniques such as coordinated beamforming, resource allocation, and joint transmission/reception via CoMP, while also supporting distributed MIMO radar architectures that significantly improve target detection and localization. Such a transition is particularly vital for emerging applications such as smart cities, autonomous driving, and industrial automation, where long-range, high-accuracy, and continuous sensing of dynamic targets is essential. In complex environments, cooperative sensing across multiple BSs enables multi-angle observations, significantly enhancing spatial diversity. This leads to improved detection probability, more accurate parameter estimation, and robust tracking of high-mobility targets such as intelligent vehicles, UAVs, and automated guided vehicles. Therefore, network-level ISAC unlocks spatial diversity and multi-static sensing capabilities that are unattainable in isolated link-level designs. It enables coordinated sensing over wider areas, mitigates inter-cell interference, and supports high-resolution sensing alongside reliable communication coverage.}

{A mathematical framework for analyzing the performance of networked ISAC systems was developed in~{\cite{Meng:WC:2024}} using stochastic geometry. The proposed cooperative ISAC scheme integrates CoMP-based joint transmission with distributed radar sensing and provides analytical expressions for both sensing and communication performance, including a closed-form Cram\'{e}r-Rao lower bound (CRLB) for localization accuracy. A key insight is that deploying $N$ ISAC transceivers improves average cooperative sensing performance across the network, following a $\ln^2N$ scaling law. However, this gain is less pronounced than the $N^2$ scaling achieved when all transceivers are equidistant from the target, due to the increased path loss from distant BSs, which limits their contribution to sensing. In~{\cite{Kaitao:TWC:2025}}, the authors further investigated how antenna-to-BS allocation strategies affect cooperative ISAC performance. They examined the trade-off between centralized (mMIMO) and distributed (CF) antenna deployments: the former enhances beamforming and coherent processing, while the latter improves spatial diversity and reduces access distances. For sensing, three localization methods were considered: AoA, time-of-flight (ToF), and a hybrid AoA-ToF approach. In a network with $N$ ISAC transceivers, each equipped with $M_t$ transmit and $M_r$ receive antennas, these methods are implemented as follows:}

\begin{itemize}
    \item {To estimate the AoA at receiver $j$, the covariance matrix of the received signal is first computed. This matrix is then decomposed via eigendecomposition to separate the signal and noise subspaces. The MUSIC algorithm is subsequently applied to compute a pseudospectrum, whose peak indicates the estimated AoA. By measuring the AoAs of both mono-static and bi-static links, the target location can be estimated using maximum likelihood estimation (MLE)~{\cite{Li:TSP:1993}}. Specifically, the AoA measurement of the bi-static link from the $j$-th BS to the target and then to the $i$th BS can be expressed by}
%---------------
\begin{align}
    \hat\theta_{i,j} = \tan^{-1}\left(\frac{y_t-y_i}{x_t-x_i}\right) + n_{i,j}^{\mathtt{AoA}},
\end{align}
%--------------
{where $(x_i,y_i)$ and $(x_t,y_t)$ represent the location of
BS $i$ and target, respectively. Moreover, $n_{i,j}^{\mathtt{AoA}}~\mathcal{CN}(0,\rho_{i,j}^2)$ denotes the AoA measurement error with $\rho_{i,j}^2=6{[\pi^2 \cos^2\theta_i M_r(M_r^2-1)G_t\gamma_{i,j}]^{-1}}$, where $\theta_i$ denotes the angle of bearing for the $i$-th BS to the target with respect to the horizontal
axis, $G_t$ is the transmit beamforming gain, and $\gamma_{i,j}$ was given in~{\cite{Kaitao:TWC:2025}}}.

\item 
{For ToF-based range estimation, matched filtering is applied to the received signal by correlating it with a known replica of the transmitted waveform. This process accentuates peaks corresponding to the time delays introduced by reflections from targets. These delays are then converted into range estimates using the speed of light. Specifically, the estimate of the distance from transmitter $j$ to the target and then to receiver $i$ is given by}
%---------------
\begin{align}
    \hat d_{i,j} &= \sqrt{ (x_i-x_t)^2 + (y_i-y_t)^2}
    \nonumber\\
    &+\sqrt{ (x_j-x_t)^2 + (y_j-y_t)^2}+n_{i,j}^{\mathtt{ToF}},
\end{align}
%--------------
{where $n_{i,j}^{\mathtt{ToF}}~\mathcal{CN}(0,\eta_{i,j}^2)$ denotes the TOD measurement error with $\eta_{i,j}^2=\frac{3 c^3 \sigma_s^2}{2\pi^2 M_rG_tB^2\gamma_{i,j}}$, where $c$ denotes the speed of light, $B^2$ represents the squared effective bandwidth, showing that the larger bandwidth offers the more accurate ToF estimation~{\cite{Kaitao:TWC:2025}}}.

\item  {The hybrid AoA-ToF approach combines both angle-of-arrival and time-of-flight measurements, rather than relying solely on one, significantly enhancing the accuracy and robustness of the localization system.}
\end{itemize}

{The CRLB analysis based on AoA and ToF estimates reveals that, in networks with $N$ ISAC nodes modeled as a Poisson point process, the localization accuracy of ToF-based methods scales as $\ln^2N$,  while AoA-based methods exhibit a scaling of $\ln N$. The hybrid AoA-ToF approach achieves a combined scaling of $a\ln^2N+b \ln N$, where $a$ and $b$ reflect the respective contributions of ToF and AoA. These differences arise from the distinct error-propagation behaviors of the methods, with AoA-based localization being more sensitive to BS-target distances. On the communication side, a tractable expression for the achievable data rate was derived, showing that higher path-loss exponents favor distributed antenna deployments to reduce access distances, whereas lower exponents benefit from centralized configurations that enhance beamforming gain.}

While network-level ISAC systems effectively mitigate inter-cell interference and enhance both communication and sensing performance, realizing their full potential introduces several technical challenges. These challenges include~{\cite{Li:TWC:2024, Meng:TWC:2024}}
\begin{itemize}
    \item  \textit{Phase synchronization for coherent ISAC cooperation:} Phase coherence is essential for maximizing cooperative gains in distributed ISAC systems. However, achieving and maintaining phase synchronization is highly challenging due to the need for precise time and frequency alignment and the presence of phase noise from independent oscillators at each node. These factors lead to time-varying phase offsets that undermine long-term coherent operation. Despite these difficulties, coherent processing offers notable performance benefits, making phase synchronization a key research focus for scalable and high-performance ISAC networks.
    \item \textit{Scalability limitations:}
    One of the major challenges in network-level ISAC design is the high complexity and backhaul demands of centralized coordination. While centralized signal processing can, in principle, optimize performance, it often incurs increased latency and reduced system efficiency due to extensive data exchange among distributed nodes. To enable practical and scalable deployment, future ISAC systems must adopt decentralized signaling and local processing architectures, allowing transmitters and receivers to operate with minimal reliance on centralized control.    
    \item \textit{EE:} CoMP operations often involve high-power transmission and centralized coordination across BSs, which can significantly increase energy consumption. A major contributor to this inefficiency is the extensive fronthaul communication required to support centralized processing. The exchange of large volumes of raw or partially processed data between BSs and central units places a heavy burden on fronthaul links, leading to substantial power consumption, particularly in high-throughput, low-latency ISAC scenarios.     
    \item \textit{Security and Privacy:} Networked ISAC systems introduce distinct security and privacy risks compared to conventional communication-only networks. As more APs or BSs participate in sensing, the coverage area expands, along with the potential for the leakage of sensitive information. Each sensing-enabled node can passively collect environmental data, which may be intercepted by unauthorized entities. Traditional encryption methods are ineffective here, as there is no data link to secure. This opens new attack vectors that could compromise system integrity or cause critical failures. Moreover, the pervasive nature of ISAC sensing raises privacy concerns, as users may be monitored without consent, even in private spaces, exposing behavioral patterns and other sensitive data. These challenges necessitate a rethinking of security and privacy frameworks for ISAC deployment.

\end{itemize}

\subsection{Integration of ISAC with Other Technologies}
{The future evolution of ISAC relies on synergy with other key wireless technologies that expand spatial coverage, EE, and environmental awareness. These integrations aim to realize a perceptive, intelligent, and adaptive wireless network capable of jointly sensing and communicating across heterogeneous environments.}

\subsubsection{{RIS-Assisted ISAC}}
{Reconfigurable Intelligent Surfaces (RISs) introduce programmable reflections that can dynamically manipulate the wireless environment by adjusting the phase, amplitude, and polarization of incident signals~{\cite{Ismail2024, Yang2024, Luo2023, Chen2024}}. When integrated with ISAC, RISs can passively enhance both communication and sensing by steering or focusing signals without the need for additional power-hungry RF chains. They effectively act as passive sensing apertures or virtual array extensions, improving angular resolution, extending sensing coverage, and mitigating blockages in non-LoS (NLoS) scenarios~{\cite{Ismail2024, Yang2024, Luo2023, Chen2024}}.}

However, the benefits of RIS-assisted ISAC depend heavily on accurate channel knowledge and real-time control. Practical challenges include optimizing large numbers of reflection elements in the presence of channel estimation errors, feedback delays, and hardware constraints. Future work should explore learning-based and model-driven approaches for adaptive RIS configuration and joint waveform-beamforming optimization to fully harness RIS's potential in next-generation ISAC networks.

\subsubsection{{BackCom and ISAC}}
{Backscatter communication (BackCom) enables ultra-low-power connectivity by allowing passive tags to modulate and reflect ambient RF signals instead of generating their own~{\cite{Rezaei2023Coding, Diluka2022, rezaei2023timespread, Galappaththige2023SR, Galappaththige2023RIS, Rezaei2024NOMA, Galappaththige2024passive}}. Integrating BackCom with ISAC yields the concept of integrated sensing and BackCom (ISABC)~{\cite{Diluka2025ISABC, Diluka2023, galappaththige2024RSMA, zargari2024ISABC}}, in which structured backscattered signals from tags are jointly exploited for communication, sensing, and device-state estimation. ISABC enables battery-free IoT devices to act as both communication nodes and sensing reflectors, improving environmental awareness without additional transmission power or hardware~{\cite{Diluka2025ISABC, Diluka2023, galappaththige2024RSMA, zargari2024ISABC}}.} 

{However, the weak and highly reflective nature of backscattered signals introduces challenges such as interference management, waveform design for joint sensing-decoding, and scalability with large tag populations. Future ISAC research may focus on cooperative processing, multi-tag detection, and energy-efficient optimization to fully realize the potential of ISABC in large-scale, perceptive IoT networks.}

\subsubsection{{UAV-Assisted ISAC}}
{UAVs add a three-dimensional spatial layer to ISAC systems, offering flexible and adaptive coverage~{\cite{Jiang2024, Orikumhi2022, Song2025, Zhao:TWc:2025,Cheng:TCOM:2025,Shaoqiang:TWC:2025}}. By adjusting their trajectories, altitudes, and beam orientations, UAV-mounted transceivers can dynamically balance sensing and communication demands, making them highly suitable for disaster response, surveillance, and temporary or infrastructure-free network deployments~{\cite{Jiang2024, Orikumhi2022, Song2025, Zhao:TWc:2025,Cheng:TCOM:2025,Shaoqiang:TWC:2025}}. The inherent mobility of UAVs provides geometric diversity, enhancing spatial resolution, target tracking accuracy, and connectivity in obstructed or hard-to-reach areas~{\cite{Jiang2024, Orikumhi2022, Song2025, Zhao:TWc:2025,Cheng:TCOM:2025,Shaoqiang:TWC:2025}}.}

{However, UAV-assisted ISAC also faces several challenges~{\cite{Zhao:TWc:2025,Cheng:TCOM:2025,Shaoqiang:TWC:2025}}. Continuous motion induces rapid channel variations, Doppler shifts, and time-varying propagation delays, thereby complicating joint waveform design and synchronization. Energy constraints on UAV platforms further limit the sensing duration and transmit power, necessitating energy-efficient trajectory planning and adaptive resource allocation. Future research directions include cooperative multi-UAV sensing, learning-based trajectory optimization, and synchronization protocols to ensure robust, scalable airborne ISAC operations.}

\subsubsection{{Holographic MIMO for ISAC}}
{Holographic MIMO extends conventional mMIMO into the continuous-aperture domain, where densely packed sub-wavelength elements form an almost continuous EM surface~{\cite{Zhang2022Holographic, Adhikary2024, Gavras2023}}. This architecture enables precise control of spatial wavefronts, supporting extremely fine angular resolution and seamless operation across near- and far-field regimes. When applied to ISAC, holographic MIMO allows simultaneous high-resolution sensing and high-capacity communication, facilitating detailed environmental mapping and target localization~{\cite{Zhang2022Holographic, Adhikary2024, Gavras2023}}.}

{Despite its promise, practical realization remains challenging. Accurate surface calibration and EM modeling are required to account for mutual coupling and phase inconsistencies across the dense array. Moreover, the large number of radiating elements introduces significant processing and hardware complexity, motivating research into scalable beamforming architectures, low-overhead channel estimation, and efficient baseband design for real-time holographic ISAC operation.}

\subsubsection{{ISAC and New Antenna Technologies}}
{Recent advancements in antenna technologies have been increasingly integrated into modern wireless communication systems to better exploit spatial resources and enhance system performance. Among these innovations are movable antennas (MAs)~{\cite{Zhu:COMMG:2024}}, which allow dynamic adjustment of antenna positions to optimize channel conditions; fluid antennas{\cite{Wong:TWC:2021}}, which leverage reconfigurable liquid-based structures to achieve flexible radiation patterns and adaptability;  rotatable antennas~{\cite{Shao:Tut:2025}}, which allow orientation adjustments to improve directional gain and coverage; and pinching antennas~{\cite{Yuanwei:MCOM:2025}}, which enable precise control over EM fields through localized structural manipulation, offering unique beamforming capabilities. Beyond conventional communication scenarios, integrating these technologies into ISAC  opens new opportunities for joint data transmission and environmental sensing, thereby enabling improved SE, enhanced localization accuracy, and more intelligent wireless networks.}

{While research on the applications of these emerging antenna technologies in ISAC systems is still in its early stages, significant interest has already been devoted by various research groups. Specifically, Li~\textit{et al.}~{\cite{Zhendong:WC:2025}} introduced application scenarios for MA-enabled ISAC systems, including the industrial IoT, low altitude economy, and Internet of vehicles. They emphasized the benefits of MA-enabled ISAC systems, including improved SE, highly flexible and precise beamforming, and the ability to dynamically adjust the signal coverage range. Recent research has investigated flexible beamforming designs for diverse MA-enabled ISAC architectures, including one-dimensional MA-aided ISAC for bistatic radar systems~{\cite{Lyu:TWC:2025}}, two-dimensional MA-assisted ISAC for bistatic configurations~{\cite{Jiang:TWC:2025}}, two-dimensional MA-enabled networked ISAC with distributed BSs~{\cite{Guo:TWC:2025}}, two-dimensional monostatic systems employing a FD BS equipped with multiple movable transmit and receive MAs across large regions~{\cite{Ding:TWC:2025}}, and six-dimensional monostatic systems featuring FD BSs with rotatable MAs~{\cite{Sun:IoT:2025}}.}

{Pinching antenna-assisted ISAC systems have attracted attention due to their adaptability, cost-effectiveness, and capability to enable LoS transmission. The authors in~{\cite{Yunhui:WCL:2025}} proposed an ISAC design for a two-waveguide pinching antenna system, in which one waveguide emits information-bearing signals for ISAC transmission while the other receives reflected echo signals. Building on this framework, a penalty-based alternating optimization algorithm was introduced to maximize illumination power while guaranteeing communication quality-of-service requirements. In~{\cite{Weihao:TWC:2025}}, a multi-waveguide pinching antenna ISAC architecture was developed, where transmit and receive pinching antennas cooperate to simultaneously detect a potential target and serve a DL user.}

%%%%%%%%%%%%%%%%%%%%%%%%%%%%%%%%%%%%%%%%%%
\section{Cell-Free Integrated Sensing and Communication}\label{sec_CF_isac}
%%%%%%%%%%%%%%%%%%%%%%%%%%%%%%%%%%%%%%%%%%

CFMM provides a robust platform for implementing various sensing systems, including monostatic and multistatic sensing (Section~\ref{sec_radartypes}). Unlike mono-static, multi-static sensing in CF-ISAC systems eliminates the requirement for FD nodes while also providing diversity gain via distributed/non-colocated transmitters and receivers, 
{i.e., it can achieve improved sensing performance by leveraging observations that are less correlated across spatially separated APs {\cite{Mao2023, Demirhan2023, Huang2022Coordinated, Cao2023Design, Wang2023, Sakhnini2022Uplink, Silva2023, Behdad2022, Behdad2024Interplay, Diluka2025CFBook}}.}
To implement multi-static CF-ISAC, the concept of network-assisted CFMM~\cite{Mohammadi:JSAC:2023} can be leveraged to identify the ISAC transmitters and receivers. Furthermore, CF-ISAC can use orthogonal waveforms to exploit the intrinsic spatial diversity of target RCS, enhancing sensing accuracy in estimating target parameters and detection probability \cite{Haimovich2008, Fishler2006}.

\subsection{Key Features of Cell-Free ISAC}
The key features of CF architecture, such as DASs, seamless handovers, enhanced interference management, AP cooperation, and scalability, make CF-ISAC an appealing solution for future wireless communication and sensing \cite{Mao2023, Yuanyuan2024, Demirhan2023, demirhan2024cellfree, Elfiatoure2023, Mao2024, Liu2024, Huang2022Coordinated, Cao2023Design, Behdad2024Interplay, elfiatoure2024multiple, Diluka2025CFBook}. These characteristics not only allow for joint optimization of communication and sensing but also improve communication coverage and sensing accuracy \cite{Mao2023, Yuanyuan2024, Demirhan2023, demirhan2024cellfree, Elfiatoure2023, Mao2024, Liu2024, Huang2022Coordinated, Cao2023Design, Behdad2024Interplay, Diluka2025CFBook}. In addition, CF-ISAC systems utilize wireless resources more efficiently than traditional co-located ISAC architectures, making them crucial for next-generation applications such as $6$G networks, smart cities, self-driving cars, healthcare, and industrial IoT.

\subsubsection{Distributed Antenna Systems}
In CF-ISAC, APs are geographically distributed across the coverage region (i.e., DAS) rather than concentrated at a single BS, offering distinct advantages for both communication and sensing \cite{Moerman2022, You2010, Haimovich2008}. This distribution provides significant spatial diversity, improving communication by mitigating spatially correlated fading, shadowing from obstacles, and reducing end-to-end transmission distances \cite{Demir2021book, Ngo2017, Zhang2019cellfree}. For sensing, it enhances environmental perception by enabling multi-static sensing (Section~\ref{sec_sensing_CFvsCL}), which improves resolution and accuracy \cite{Mao2023, Yuanyuan2024, Demirhan2023, demirhan2024cellfree, Elfiatoure2023, Mao2024, Liu2024, Huang2022Coordinated, Cao2023Design}.

Distributed APs also provide wider and more uniform coverage, thereby mitigating the cell-edge issue encountered in traditional co-located systems \cite{Demir2021book, Ngo2017, Zhang2019cellfree}. In CF-ISAC, the users and targets are simultaneously served by multiple APs, enhancing both communication quality and sensing precision. In addition, because APs are distributed across the network, users/targets are always in close proximity to multiple APs \cite{Mao2023, Yuanyuan2024, Demirhan2023, demirhan2024cellfree, Elfiatoure2023, Mao2024, Liu2024, Huang2022Coordinated, Cao2023Design, Behdad2024Interplay}. This effectively lowers latency and provides more reliable connections. This is especially important in time-sensitive ISAC applications such as autonomous driving, where communication and sensing must occur with minimal latency \cite{liu2023integratedbook}.

\subsubsection{Seamless Handover and User/Target-Centric Operation}
In traditional cellular networks, each user is served by a single BS. When a user moves from one cell (or BS) to another, a handover is necessary \cite{Demir2021book,  Ngo2017, Zhang2019cellfree}. This can result in delays, signal overhead, and even connection interruptions \cite{Demir2021book,  Ngo2017, Zhang2019cellfree}. However, in CF-ISAC, the concept of no cell boundaries assures that a user/target is not tied to a specific AP. Instead, numerous APs cooperate to serve the user/target concurrently. When the user/target moves, nearby APs take over, eliminating the requirement for a hard handover \cite{Demir2021book,  Ngo2017, Zhang2019cellfree}. This improves reliability, reduces latency, and boosts overall network performance. Furthermore, unlike cellular networks, the system does not require frequent switching between BSs (APs), making it appropriate for applications with significant mobility, such as vehicular networks or drones \cite{liu2023integratedbook}. In a CF-ISAC network, seamless handover ensures uninterrupted communication while performing environmental sensing. It eliminates signal degradation or interruption, even when users or targets move between various AP regions. 

User/target-centric operations optimize network resources based on the specific location and requirements of users (for communication) and targets (for sensing) \cite{Demir2021book, Ammar2022}. Unlike traditional cell-centric networks, in which each cell serves a user independently, CF-ISAC networks assign each user/target to a nearby set of APs, effectively placing them at the center of a dedicated serving cluster. This cluster, formed based on selected criteria such as service performance and network efficiency, comprises APs that contribute useful signals/data. A two-stage process can further refine clustering, first using large-scale fading statistics to form a base cluster, then optimizing it with scheduling or power allocation algorithms per time slot \cite{Demir2021book, Ammar2022}. This approach ensures optimal resource allocation, enhancing communication quality and sensing coverage.

\subsubsection{Interference Management and Resource Allocation} Managing interference between sensing and communication tasks, as well as between different users and targets, is one of the most significant issues in co-located ISAC \cite{liu2023integratedbook}. In particular, the concurrent operation of these tasks in the same frequency spectrum often leads to conflicting requirements, as optimizing for communication performance might reduce sensing accuracy, and vice versa \cite{liu2023integratedbook}. Furthermore, interference among multiple users and sensing targets exacerbates the issue, underscoring the need for sophisticated beamforming and resource allocation algorithms to balance these competing objectives while minimizing cross-task and inter-user interference.    

Nevertheless, in CF-ISAC, the distributed nature and cooperative architecture of APs can substantially reduce interference. By enabling APs to collaborate and share information, the system achieves greater spatial diversity and flexibility in resource management, thereby improving interference suppression \cite{Demir2021book, Ngo2017, Zhang2019cellfree}. Advanced techniques such as CoMP processing, interference alignment, and beamforming can be used to reduce cross-user/target interference by ensuring that signals from various users or sensing tasks do not overlap \cite{Demir2021book,  Ngo2017, Zhang2019cellfree}. The distributed design also enables more precise control of beamforming and power allocation, thereby reducing the likelihood of unintentional interference between sensing and communication operations \cite{Demir2021book,  Ngo2017, Zhang2019cellfree}.   

\subsubsection{AP Cooperation and Synchronization}
In a CF network, numerous distributed APs collaborate to serve users and perform sensing tasks in a coordinated manner \cite{Demir2021book, Ngo2017, Zhang2019cellfree}. This collaboration is key to realizing the benefits of CF-ISAC, which uses the same infrastructure for both communication and environmental sensing. The advantages of AP cooperation include increased communication and sensing coverage, reduced interference, and efficient use of network resources. 
   
AP cooperation can be achieved through cooperative beamforming, in which APs coordinate their transmissions to improve signal strength while minimizing interference \cite{Demir2021book,  Ngo2017, Zhang2019cellfree}. In sensing, this improves spatial resolution and coverage by combining signals from multiple APs to create a more accurate representation of the environment. This is especially critical in situations where targets may be hidden from certain APs or where multipath reflections make sensing more difficult \cite{Mao2023, Yuanyuan2024, Demirhan2023, demirhan2024cellfree, Elfiatoure2023, Mao2024, Liu2024, Huang2022Coordinated, Cao2023Design, Behdad2024Interplay}. In communication, coordinated APs improve data throughput and reduce the risk of signal dropouts during transmission. Furthermore, the CPU is crucial in AP cooperation, acquiring data from APs, processing it, and coordinating transmissions to improve both communication and sensing \cite{Demir2021book,  Ngo2017, Zhang2019cellfree}. The CPU can dynamically assign AP resources (e.g., power, spectrum, and time slots) in response to real-time demands.

\subsection{Cell-Free ISAC Versus Link-Level ISAC Design}\label{sec_sensing_CFvsCL}
The CF-ISAC designs with multi-static sensing provide significant advantages over link-level mono-static or bi-static ISAC designs. These advantages stem primarily from their distributed architecture and the ability of multiple APs to participate in both communication and sensing~\cite{Mao2023, Demirhan2023, Huang2022Coordinated, Cao2023Design, Wang2023, Sakhnini2022Uplink, Silva2023, Behdad2022, Behdad2024Interplay}.

\begin{table*}[t]
\centering
\renewcommand{\arraystretch}{1.2}
\caption{A compression of sensing in CF-ISAC and conventional ISAC.}\label{tab_features}
% \rowcolors{1}{ForestGreen!15}{white}
\begin{tabular}{|p{3cm}|p{6cm}|p{6cm}|}
\hline
\textbf{Feature} & \textbf{CF-ISAC sensing (Multi-static)} & \textbf{Conventional ISAC sensing (Mono/bi-static)} \\ \hline \hline
Sensing coverage   &   Broader with fewer blind spots  & Limited to single/double viewpoint  \\ \hline
Sensing accuracy  &  High due to triangulation from multiple points &  Lower and limited by fewer viewpoints  \\ \hline
Robustness  &  High resilience to node failure & Vulnerable to single-point failure \\ \hline
SNR and sensitivity  & Improved through signal diversity &  Limited SNR with sensitive to noise \\ \hline
Scalability  & Easily scalable across APs  &  Difficult to scale \\ \hline
Multi-path exploitation   & Can utilize multi-path signals for extra information  &  Often treats multi-path as interference \\ \hline
Multi-target tracking   & 	More effective and can track multiple targets & Limited in tracking multiple targets simultaneously \\ \hline
Interference management   &  Better due to spatial diversity &  More susceptible to interference  \\ \hline
Resource optimization    & Highly flexible for joint communication-sensing  & Limited flexibility     \\ \hline
NLoS handling    &  More effective due to multiple angles &  Often requires LoS or favorable geometry \\ \hline
\end{tabular}
\end{table*}

\subsubsection{Improved Sensing Coverage} CF-ISAC with multi-static sensing employs multiple distributed transmitter-receiver pairs (APs) that cover a wide area. This provides a more extensive coverage than link-level ISAC design, which normally relies on a single or two locations for sensing. Furthermore, link-level ISAC design is frequently restricted by LoS conditions. CF-ISAC avoids blind spots and provides reliable coverage across broader regions. 

\subsubsection{Improved Sensing Accuracy} CF-ISAC leverages measurements from multiple distributed nodes, thereby improving triangulation and enabling more accurate localization or target detection. Furthermore, cross-correlating observations from multiple sources can also greatly improve resolution, precision, and detection. On the other hand, because link-level ISAC design relies on one or two points of view, its precision is limited, and it is more likely to overlook fine-grained details of the target, e.g., position, movement, or characteristics. 

\subsubsection{Enhanced Robustness and Resilience} 
Multi-static sensing in CF-ISAC improves system robustness by enabling continued operation even in the event of node failures or interference, ensuring the system remains functional under challenging conditions. The distributed structure provides robustness and reliability when mono-static/bi-static systems fail due to obstruction, occlusion, or physical limitations. In link-level ISAC design, mono-static/bi-static sensing is more susceptible to single-point failures, i.e., if the BS fails to perform sensing, the system loses all sensing capabilities. 

\subsubsection{Improved SNR and Detection Sensitivity} CF-ISAC systems employ coherent combining and diversity gain to improve sensing SNR, leveraging multiple observations through received signals. This can assist in detecting weak or distant targets that may have been overlooked in mono-static or bi-static systems. Mono-static/bi-static systems, on the other hand, are limited by the signal SNR or dual channels, i.e., transmitting and receiving, reducing detection sensitivity, particularly in challenging environments (e.g., dense urban areas and cluttered environments).

\subsubsection{Distributed and Scalable System Architecture} Multiple APs in CF-ISAC perform joint communication and sensing in a distributed and scalable manner, i.e., each AP contributes to both tasks, optimizing network resources and providing seamless scalability. In co-located ISAC, communication and sensing tasks are centralized and less flexible, i.e., they lack inherent scalability compared to CF networks when expanding coverage or adding new sensing nodes. 

\subsubsection{Multi-Path (Macro-Diversity) Exploitation} CF-ISAC can leverage multi-path propagation or macro-diversity, which involves signals reflected from multiple points and surfaces, providing more information about the environment and targets. This is especially beneficial in indoor or cluttered environments with plenty of reflections.  However, link-level ISAC design may fail to effectively utilize multi-path signals, often treating them as interference rather than meaningful information. 

\subsubsection{Simultaneous Multi-Target Tracking} By leveraging the user-centric (UC) nature of the CFMM architecture, CF-ISAC allows different APs to focus on distinct targets and combine the gathered data, enabling simultaneous monitoring of multiple targets. This approach enhances the efficiency and accuracy of multi-target tracking, particularly in dynamic environments with moving objects. In contrast, the link-level ISAC design with a single BS has limited capacity to manage multiple targets simultaneously, leading to performance degradation due to signal overlap or occlusion.

\subsubsection{Interference Management} CF-ISAC can use the spatial diversity of the distributed architecture to reduce interference. In particular, observations/measurements from several angles and distances aid in differentiating targets from noise. Due to the limited number of observations in the link-level ISAC design, interference from other communication channels or ambient clutter is more difficult to mitigate. 

\subsubsection{Joint Communication-Sensing Optimization} In CF-ISAC systems, the network may optimize resources (e.g., power, bandwidth) among scattered APs to balance communication and sensing demands. The distributed design provides greater flexibility to allocate resources in response to dynamic demands. For example, sensing in a specific region while maintaining communication quality elsewhere. Nonetheless, co-located ISAC systems have fewer DoF for joint resource allocation, making it more difficult to accomplish effective communication-sensing trade-offs, particularly in dynamic environments.

\subsubsection{Non-LoS Capability} In CF-ISAC, multi-static sensing can successfully manage NLoS conditions as APs provide alternate viewpoints of the target, potentially circumventing obstructions or blockages.  In contrast, mono-static/bi-static sensing in the link-level ISAC design relies on LoS or favorable positioning. Thus, obstacles or structures that conceal the target are significant limitations.  

In conclusion, multi-static sensing in CF-ISAC systems offers significant advantages in sensing coverage, accuracy, robustness, and flexibility compared to the link-level mono-static or bi-static ISAC design. These benefits make CF-ISAC particularly well-suited for complex, dense, or large-scale future-generation wireless networks. A summary of these aspects is provided in Table~\ref{tab_features}.

\section{State of the Art in Cell-Free ISAC}\label{sec_state_art}

\subsection{Performance Analysis in Cell-Free ISAC}Analytical performance evaluation provides insights into a wireless system's behavior, potential, and limitations under various conditions \cite{1199275,1556835,545899,4694096}. It may predict performance without requiring extensive simulations or real-world testing. In CF-ISAC systems, it is crucial to optimize communication and sensing interactions, guide resource allocation, balance trade-offs, ensure scalability, and assess real-world feasibility.

Nevertheless, comprehensive CF-ISAC performance analyses have not yet been developed, except for \cite{Sakhnini2022, Behdad2024,Kulathunga2025}. In particular, reference \cite{Sakhnini2022} presents a communication and sensing protocol for a single-target CF-ISAC system, in which an AP is allocated to participate in the UL alongside users and to deliver a radar signal. This approach models the radar system as a distributed bistatic radar to recover radar echoes in the presence of multi-user interference. The cost imposed on the communication system is the loss of an AP, since it is treated as a virtual user. Two radar sensing modes are proposed, i.e., sensing during either the UL training period or the data payload segment of the communication frame. A subspace signal model and a corresponding generalized likelihood ratio test (GLRT) are developed to assess detection performance, with expressions for the probability of detection and false alarm presented.

Reference \cite{Behdad2024} studies a centralized CF-ISAC mMIMO system for single-target detection, where transmit APs serve DL users while steering a beam toward the target in a multi-static sensing setup. A maximum a posteriori ratio test detector is developed to detect the target amid clutter, with sensing SE as a key performance metric. The study also explores two ISAC signal models-leveraging communication beams for sensing and adding dedicated sensing beams-and proposes a power allocation algorithm to maximize sensing SINR while ensuring minimum communication requirements.

In \cite{Kulathunga2025}, the performance of multi-user single-target CF-ISAC with transmit APs using MRT precoding is analyzed. Each AP uses locally estimated CSI to design and transmit a superimposed ISAC waveform for user communication, while designated sensing APs process reflected echoes for sensing. A max-min algorithm optimizes transmit power to maximize the weakest user rate while maintaining a predefined signal-to-clutter-plus-noise-ratio (SCNR) threshold at each sensing AP. Achievable user rates and the two-dimensional MUSIC spectrum for target localization are derived, considering CSI errors, spatially correlated Rician fading, and clutter interference. Numerical results demonstrate the potential of CF-ISAC with efficient MRT precoders.

\subsubsection{Case Study and Discussion}\label{sec_CSD_performance}
\begin{figure}[!t]\vspace{4mm}
    \centering 
    \def\svgwidth{240pt} 
    \fontsize{8}{8}\selectfont 
    \graphicspath{{Figures/}}
    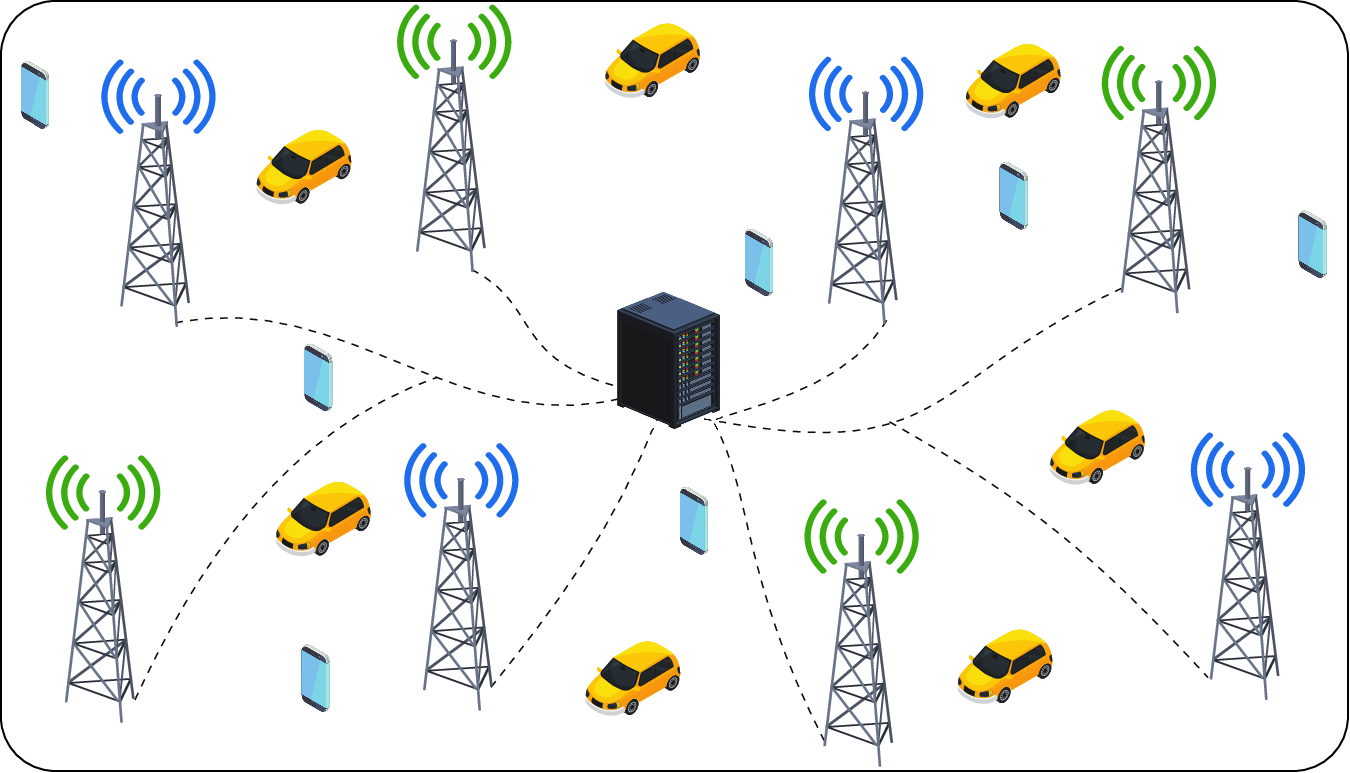 \vspace{-0mm} 
    \caption{{A CF-ISAC system with distributed UL and DL APs connected to a CPU. The APs jointly serve users and sense targets, enabling cooperative communication and multi-static sensing.} }  \label{fig_SystemModelPerfomance}\vspace{-0mm}
\end{figure}

This case study evaluates a generalized CF-ISAC system with multiple targets and users, extending beyond existing single-target scenarios (Fig.~\ref{fig_SystemModelPerfomance}). It examines the interplay between communication and sensing, highlighting key trade-offs.

%=======================================================================
%=======================================================================

It considers the following model. 
A CF-ISAC mMIMO system with $M$ DL APs and $N$ UL APs, each with $L$ antennas, serves $K$ single-antenna DL users and detects $T$ targets (Fig.~\ref{fig_SystemModelPerfomance}). All APs connect to a CPU via fronthaul/backhaul links. DL APs serve users while steering sensing beams toward targets using the same time-frequency resources. UL APs capture reflected echoes for sensing. Analytical expressions for communication SE at users and sensing SE at UL APs for each target are derived, with channel and transmission models and performance evaluation detailed in Appendix~\ref{apx_perfomance}.

\textit{Simulation Example:} 

This employs the 3GPP Urban micro (UMi) model for large-scale fading $\zeta_{\q{a}}$, where $\q{a} \in \{ \q{h}_{mk}, \q{g}_{mt}^{\mt{d}}, \q{g}_{mt}^{\mt{u}}\}$, with an operating frequency of $f_c = \qty{3}{\GHz}$ \cite[Table B.1.2.1]{3GPP2010}. The AWGN variance is modeled as  $\sigma^2 = 10 \log_{10}{(N_0 B N_f)}$\,\qty{}{\dB m}, where $N_0=\qty{-174}{\dB m/\Hz}$, $B=\qty{10}{\MHz}$ is the bandwidth, and $N_f=\qty{10}{\dB}$ is the noise figure. Additionally, the UL and DL APs are uniformly distributed, while the users and targets are randomly distributed over a $\num{200} \times \qty{200}{\m^2}$ area. {The RCS of targets are set to $\alpha_t = {\num{e-2}}$.}

%=======================================================================
\begin{figure}[!t]\vspace{-0mm}
    \centering
    \includegraphics[width=0.47\textwidth]{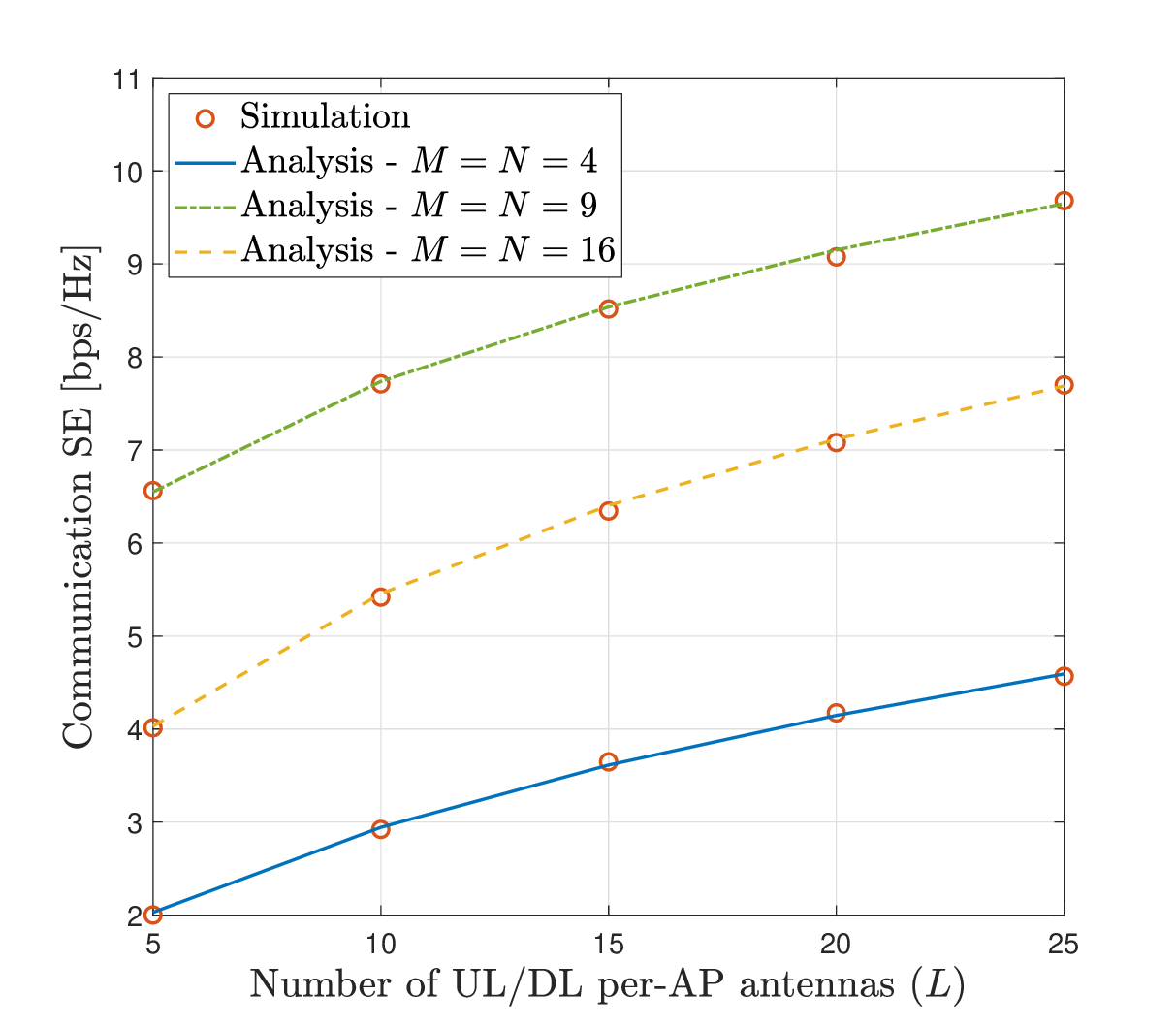}
    \vspace{-0mm}
    \caption{{Communication SE versus the number of UL/DL antennas per AP ($L$). This figure compares simulation and analytical results for several system configurations ($M = N = \{4, 9, 16\}$).}}
    \label{fig_ComSumRate_perf} \vspace{-0mm}
\end{figure}
%=======================================================================

%=======================================================================
\begin{figure}[!t]\vspace{-0mm}
    \centering
    \includegraphics[width=0.47\textwidth]{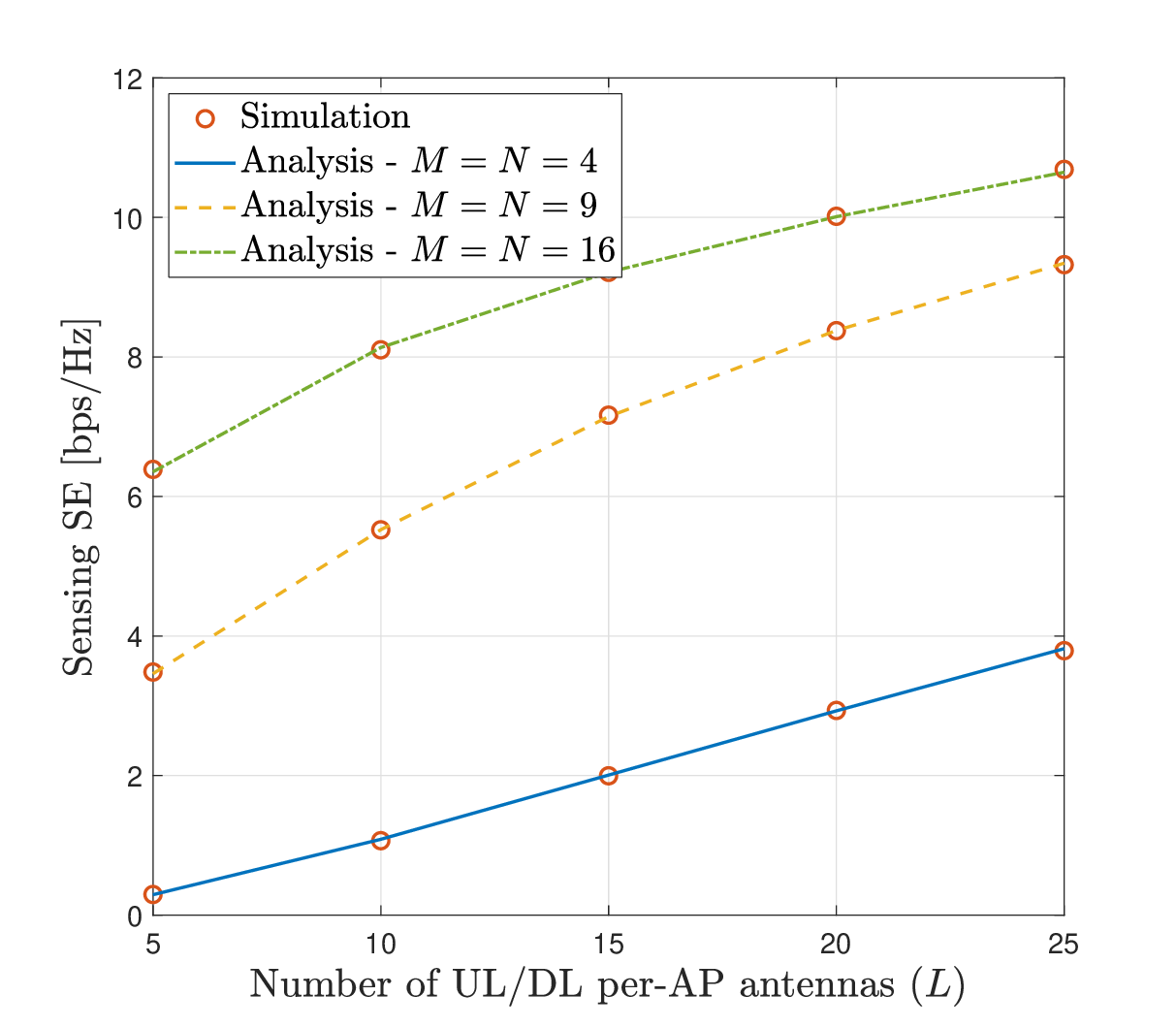}
    \vspace{-0mm}
    \caption{{Sensing SE versus the number of UL/DL antennas per AP ($L$). This figure compares simulation and analytical results for several system configurations ($M = N = \{4, 9, 16\}$).}}
    \label{fig_SensSumRate_perf} \vspace{-0mm}
\end{figure}
%=======================================================================

{Fig.~{\ref{fig_ComSumRate_perf}} and Fig.~{\ref{fig_SensSumRate_perf}} show the communication and sensing SE as functions of the number of UL/DL AP antennas ($L$) for different AP counts $M=N=\{4, 9, 16\}$. The analytical SE expressions are verified via Monte Carlo simulations, in which the analytical and simulated results agree closely across all configurations, confirming the accuracy of the derived models.}

{Both figures demonstrate that increasing the number of antennas per AP significantly enhances the communication and sensing SE. This improvement stems from stronger spatial diversity, higher beamforming gain, and reduced multi-user interference achieved by larger antenna arrays. For example, when $M=N=9$, increasing the number of antennas from {\num{10}} to {\num{20}} yields approximately {\qty{31.3}{\percent}} and {\qty{51.7}{\percent}} improvements in communication and sensing SE, respectively. Furthermore, increasing the number of APs provides additional distributed spatial diversity and broader coverage, thereby improving SE even in interference-limited regimes. The gains from both antenna and AP scaling confirm the inherent advantage of CF-ISAC architectures, where distributed antennas simultaneously enhance both communication reliability and sensing precision through coherent combining across multiple spatially separated APs.}

{These results also highlight an important system-level implication: while higher AP density and larger antenna arrays enhance performance, they simultaneously impose stricter requirements on synchronization accuracy, channel estimation quality, fronthaul capacity, user/target mobility, and cooperative processing. Consequently, the observed trends motivate future CF-ISAC research on advanced distributed signal processing, synchronization, and coordination algorithms to sustain these performance gains under realistic network conditions.}

\subsection{Resource Allocation in Cell-Free ISAC}\label{sec_resource_allocation}
Resource allocation optimizes bandwidth, power, time, and antennas to meet data demands, support multiple users, ensure fairness, and enhance performance. However, varying requirements across wireless applications complicate allocation. Different networks (e.g., Wi-Fi, LTE, sensor, and energy-harvesting) require tailored solutions rather than a one-size-fits-all approach. While standard algorithms address some challenges, others demand customized strategies.

Conversely, CF-ISAC resource allocation faces additional challenges in balancing sensing and communication, and in managing resources across distributed APs. Their complex, distributed nature and joint functionality requirements demand tailored solutions. To address these challenges, various resource allocation schemes have been developed  \cite{Mao2023, Yuanyuan2024, Demirhan2023, demirhan2024cellfree, Elfiatoure2023, Mao2024, Liu2024, Huang2022Coordinated, Cao2023Design, Behdad2024Interplay, behdad2024Joint, Silva2023, Behdad2022, Wang2023, Sakhnini2022Uplink, zargari2024CFISAC,Mobini:SPAWC:2025}.
In particular, many studies focus on designing AP beamforming approaches to enable efficient resource allocation \cite{Mao2023, Demirhan2023, demirhan2024cellfree, Cao2023Design, Wang2023, Silva2023, Mao2024, Liu2024, zargari2024CFISAC}. Although traditionally considered a signal processing technique, beamforming in CF-ISAC inherently involves resource management as the AP transmit power is embedded within the beamforming vectors. This integration directly affects the distribution of system power across APs to optimize communication and sensing performance. Thus, beamforming plays a critical role in CF-ISAC resource allocation frameworks.

\subsubsection*{{Beamforming Optimization}}
{Reference~{~\cite{Mao2023}} presents a beamforming design for a multi-user, single-target CF-ISAC system under CSI errors. The beamforming problem is formulated as an MSE minimization problem for sensing beampattern matching, subject to power budget constraints at the APs and user rate requirements. Given lower-bound user rates under imperfect CSI, a successive convex approximation (SCA)-based algorithm is proposed. For a single-target CF-ISAC system, references~{\cite{Demirhan2023, demirhan2024cellfree}}  propose a max-min fairness joint beamforming design. The approach utilizes two benchmarks: communication-prioritized sensing beamforming and sensing-prioritized communication beamforming, to balance performance between communication and sensing tasks. In~{\cite{Wang2023}}, a hybrid beamforming architecture for a multi-user, multi-target cooperative CF dual-function radar-communication network is investigated. The hybrid beamforming at the APs is designed to maximize the weighted communication sum rate while satisfying the MSE constraint on the beampattern gain for radar sensing. This is achieved through a semi-distributed approach using an alternating optimization (AO) algorithm.}

{For a multi-user, multi-target CF-ISAC system, reference~{\cite{Mao2024}} provides transmit beamforming designs to accommodate ISAC, communication-only, and sensing-only scenarios. Three different solutions are presented via the Lagrangian dual transform,  the quadratic fractional transform technique, the block coordinate descent (BCD) method, and the SCA method. In~{\cite{Silva2023}}, a transmit beamforming design for a single-target, multi-user CF-ISAC system is presented, where a user acts as an adversary attempting to infer the target's position. To overcome this, an expectation-maximization algorithm is employed to estimate the transmitted signal, which is subsequently used to generate replicas of the transmit beampattern for each AP. In~{\cite{zargari2024CFISAC}}, a new efficient and low-complexity beamforming design for CF-ISAC networks is proposed. The proposed algorithm leverages the augmented Lagrangian model-based Riemannian manifold optimization technique to maximize the communication sum rate while satisfying sensing beampattern gains of multiple targets, and per-AP transmit power constraints.}

\subsubsection*{{Power Allocation}}
{In~{\cite{Huang2022Coordinated}}, a coordinated power control scheme is investigated for CF-ISAC transmitters. The objective is to minimize the total transmit power of the APs while meeting the minimum communication SINR for users and the maximum CRLB for target location estimation. Two efficient algorithms based on semidefinite relaxation (SDR) and CRLB approximation are proposed. A power allocation algorithm is offered to maximize the sensing SNR while meeting the minimum SINR requirements for users in~{\cite{Behdad2022}}. Thereby, a maximum a posteriori ratio test detector is derived to detect the target using signals received at distributed APs. In~{\cite{Sakhnini2022Uplink}}, a CFMM system with integrated virtual UL radar is proposed. The radar system operates in the UL, with a subset of APs dedicated to radar transmission in addition to user communications. Power control is employed at the APs to manage the interference from the communication system on the radar, incorporating linear interference constraints into the sum SE and sum-log-SNR policies. Additionally, a low-complexity solution based on the largest large-scale fading heuristic is presented.}

{The studies~{\cite{Behdad2024Interplay, behdad2024Joint}} examine DL ultra-reliable low-latency communication (URLLC) in a CF-ISAC system. It proposes an SCA-based power-allocation algorithm to maximize the network EE while meeting the sensing-SINR and communication-decoding-error-probability requirements. Reference~{\cite{Yuanyuan2024}} considers a multi-user, single-target CF-ISAC mmWave mMIMO system with capacity-limited fronthaul links. A power allocation scheme with a fronthaul compression design is proposed that employs hybrid analog and digital precoders, with digital precoders implemented in the CPU and analog precoders designed at each transmitting AP. Two BCD-based algorithms were presented. In~{\cite{Mobini:SPAWC:2025}}, a beam scanning protocol was introduced for practical three-dimensional (3D) target localization. To improve sensing performance under SE constraints for communication users, a power optimization problem was formulated, yielding a significant reduction in the CRLB for target estimation.}

\subsubsection*{{Miscellaneous}}
{Reference~{\cite{Cao2023Design}} examines a multi-user, multi-target CF-ISAC system and proposes vector orthogonal frequency division multiplexing (OFDM) signals to enhance SE, detection resolution, and latency/Doppler estimation accuracy. To estimate the AoAs for multiple targets, a low-complexity grid-searching approach is introduced. Additionally, AP beamforming and power allocation are designed using an alternative Lagrange multiplier algorithm. In~{\cite{Elfiatoure2023}}, an AP operating-mode selection approach is presented for a multi-user, single-target CF-ISAC system. In particular, some APs are dedicated to DL communication, while the remaining APs are utilized for sensing. With closed-form SE and mainlobe-to-average-sidelobe ratio, a max-min fairness problem is formulated to maximize the minimum user SE. In~{\cite{Liu2024}}, the joint AP mode selection, transmit beamforming, and receive filter designs are investigated for cooperative CF-ISAC networks, in which the APs jointly serve multiple communication users and detect targets. Three heuristic AP mode selection techniques and an efficient joint beamforming design method are presented.}

\subsubsection{Case Study and Discussion}\label{sec_CSD_resource}
\begin{figure}[!t]\vspace{4mm}
    \centering 
    \def\svgwidth{240pt} 
    \fontsize{8}{8}\selectfont 
    \graphicspath{{Figures/}}
    %% Creator: Inkscape 1.3 (0e150ed6c4, 2023-07-21), www.inkscape.org
%% Accompanies image file 'SystemModelResourceAllocation.eps' (pdf, eps, ps)
\begingroup%
  \makeatletter%
  \providecommand\color[2][]{%
    \errmessage{(Inkscape) Color is used for the text in Inkscape, but the package 'color.sty' is not loaded}%
    \renewcommand\color[2][]{}%
  }%
  \providecommand\transparent[1]{%
    \errmessage{(Inkscape) Transparency is used (non-zero) for the text in Inkscape, but the package 'transparent.sty' is not loaded}%
    \renewcommand\transparent[1]{}%
  }%
  \providecommand\rotatebox[2]{#2}%
  \newcommand*\fsize{\dimexpr\f@size pt\relax}%
  \newcommand*\lineheight[1]{\fontsize{\fsize}{#1\fsize}\selectfont}%
  \ifx\svgwidth\undefined%
    \setlength{\unitlength}{598.53051758bp}%
    \ifx\svgscale\undefined%
      \relax%
    \else%
      \setlength{\unitlength}{\unitlength * \real{\svgscale}}%
    \fi%
  \else%
    \setlength{\unitlength}{\svgwidth}%
  \fi%
  \global\let\svgwidth\undefined%
  \global\let\svgscale\undefined%
  \makeatother%
  \begin{picture}(1,0.5870527)%
    \lineheight{1}%
    \setlength\tabcolsep{0pt}%
    \put(0,0){\includegraphics[width=\unitlength]{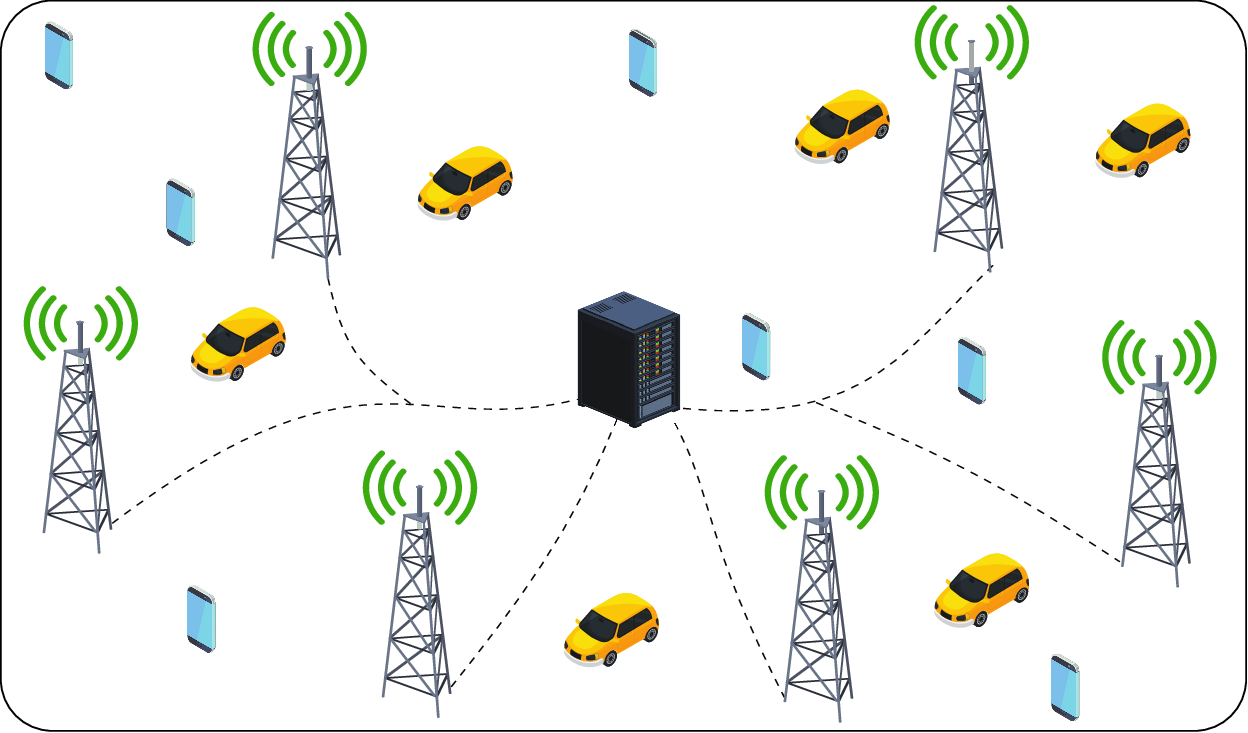}}%
    \put(0.02818211,0.11428925){\color[rgb]{0,0,0}\makebox(0,0)[lt]{\lineheight{1.25}\smash{\begin{tabular}[t]{l}AP-$1$\end{tabular}}}}%
    \put(0.87992321,0.08691825){\color[rgb]{0,0,0}\makebox(0,0)[lt]{\lineheight{1.25}\smash{\begin{tabular}[t]{l}AP-$M$\end{tabular}}}}%
    \put(0.46271266,0.36277318){\color[rgb]{0,0,0}\makebox(0,0)[lt]{\lineheight{1.25}\smash{\begin{tabular}[t]{l}CPU\end{tabular}}}}%
    \put(0.12378976,0.02999772){\color[rgb]{0,0,0}\makebox(0,0)[lt]{\lineheight{1.25}\smash{\begin{tabular}[t]{l}User-$1$\end{tabular}}}}%
    \put(0.47663936,0.47632577){\color[rgb]{0,0,0}\makebox(0,0)[lt]{\lineheight{1.25}\smash{\begin{tabular}[t]{l}User-$K$\end{tabular}}}}%
    \put(0.43438669,0.01753471){\color[rgb]{0,0,0}\makebox(0,0)[lt]{\lineheight{1.25}\smash{\begin{tabular}[t]{l}Target-$1$\end{tabular}}}}%
    \put(0.85714148,0.41345724){\color[rgb]{0,0,0}\makebox(0,0)[lt]{\lineheight{1.25}\smash{\begin{tabular}[t]{l}Target-$T$\end{tabular}}}}%
  \end{picture}%
\endgroup%
 \vspace{-0mm} 
    \caption{{A CF-ISAC system with distributed APs connected to a CPU, jointly serving users and sensing targets in the network.}}  \label{fig_SystemModelResourceAllocation}\vspace{-0mm}
\end{figure}

Herein, a beamforming design for a generalized CF-ISAC system (Fig.~\ref{fig_SystemModelResourceAllocation}) is provided and evaluated with numerical examples.  

%=======================================================================
%=======================================================================

Fig.~\ref{fig_SystemModelResourceAllocation}  investigates a CF-ISAC system comprising $M$ APs, each equipped with $L$ ULA antennas, $K$ single-antenna users, and $T$ potential targets. The CPU connects all APs and coordinates the joint communication and sensing, ensuring time synchronization across all APs \cite{Ngo2017}. The communication SE at the users and the transmit beampattern gains at the AP towards the targets are utilized to evaluate the communication and sensing performance, respectively. The objective is to maximize the communication SE for the users while satisfying sensing beampattern gain requirements for each target and per-AP transmit power constraints. Interested readers are referred to \cite{zargari2024CFISAC} for comprehensive details on the system, channel, and transmission models, as well as the analytical performance evaluation and the associated algorithm.

{It is worth noting that this case study uses beampattern gain as the sensing metric, as it offers an intuitive and visual illustration of target localization and resolution, aligning well with the tutorial nature of this paper. While more advanced metrics, such as sensing SE or the CRB, could provide deeper quantitative analysis, their inclusion would require a different system model (e.g., echo reception or FD APs). Such evaluations, therefore, provide ample opportunities for future CF-ISAC research {\cite{Wenrui:TWC:2025, Zhenyao2023}}.}

\textit{Simulation Example:}
The simulation setup is the same as in Section~\ref{sec_CSD_performance}, excluding the UL APs. The maximum allowable per-AP transmit power is set to $p_{\rm{max}} = \qty{30}{\dB m}$, and the required sensing beampattern gain for each target is set to $\Gamma_{t}^{\rm{th}} = \qty{10}{\dB m}$.

%=======================================================================
\begin{figure}[!t]\vspace{-0mm}
    \centering
    \includegraphics[width=0.47\textwidth]{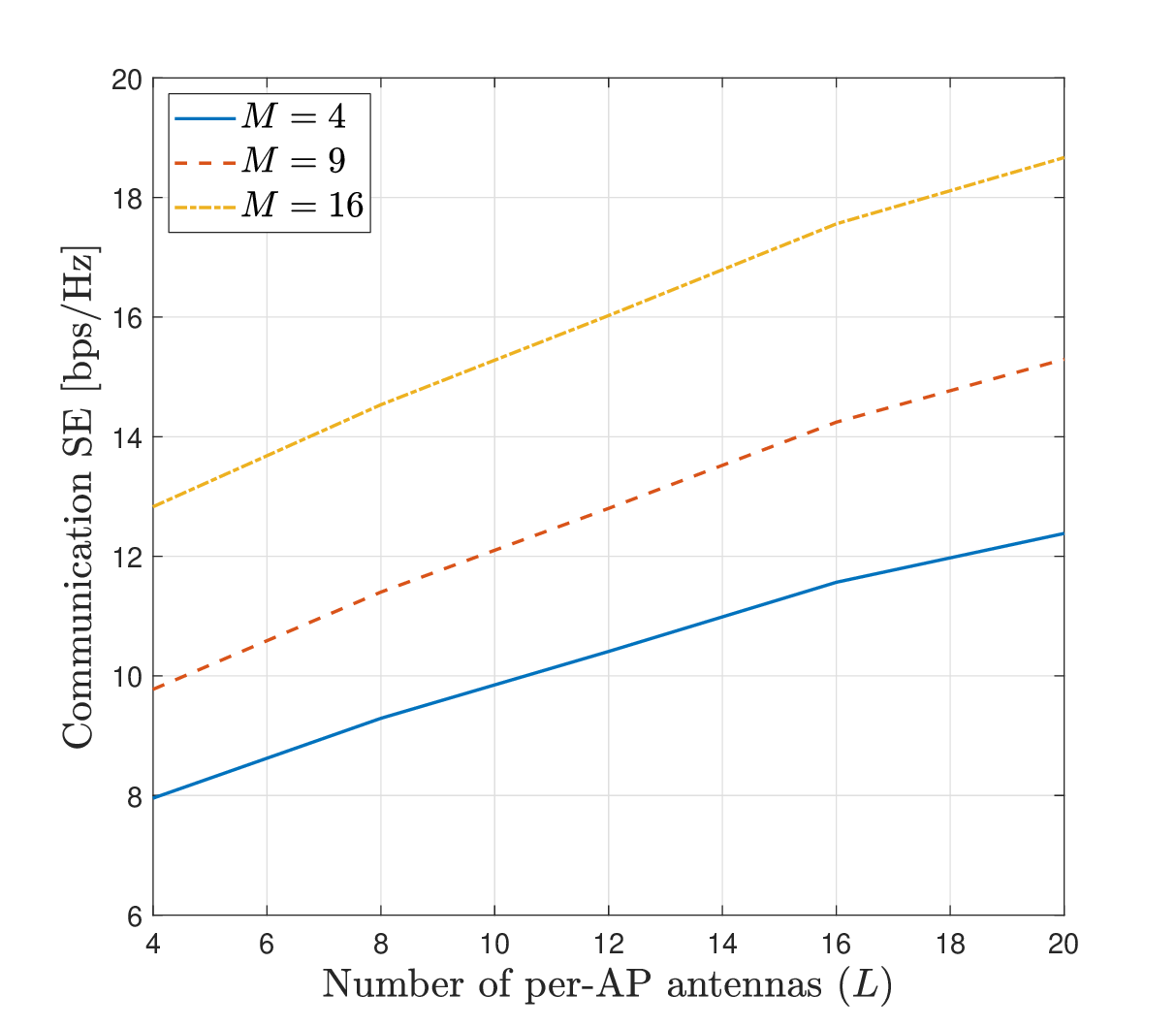}
    \vspace{-0mm}
    \caption{{Communication SE versus the number of per-AP antennas ($L$) for $K = 2$ users and $T = 3$ targets, with different numbers of APs ($M = \{4, 9, 16\}$).}}
    \label{fig_Sumrate_antennas_resource} \vspace{-0mm}
\end{figure}
%=======================================================================
%=======================================================================
\begin{figure*}[!t]\vspace{-0mm}
    \centering
    \includegraphics[width=1.0\textwidth]{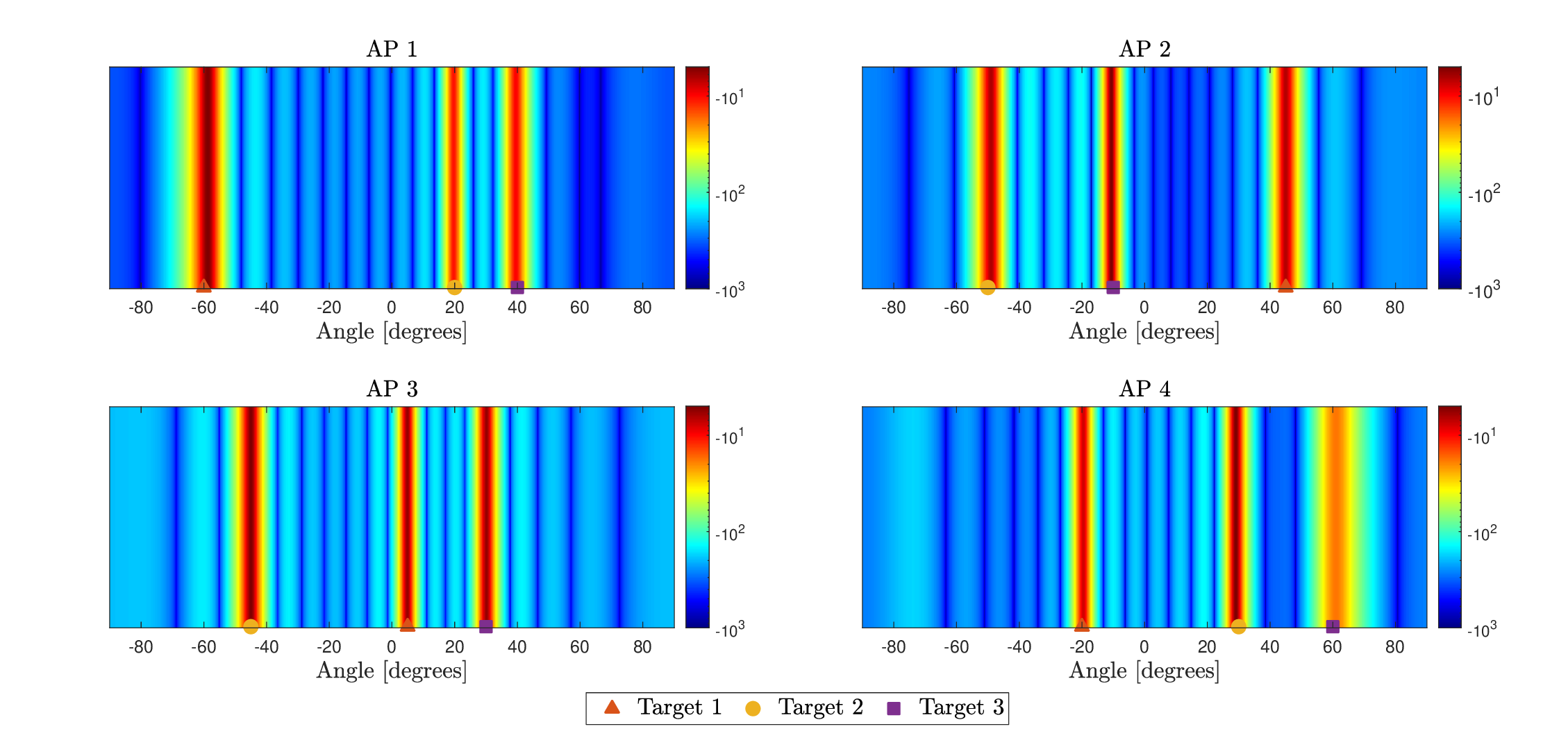}
    \vspace{-0mm}
    \caption{{Beampattern gain profiles/heatmaps over a {\qty{\pm 90}{\degree}} angular spread at different APs, illustrating the gain variations and directivity in a color-coded scale for $L=8$, $M=4$, $K=2$, and $T=3$. The plots highlight the variation in beamforming gain and spatial directivity across APs.}}
    \label{fig_HeatMap_resource} \vspace{-0mm}
\end{figure*}
%=======================================================================

{Fig.~{\ref{fig_Sumrate_antennas_resource}} presents the communication SE as a function of the number of AP antennas ($L$) for different AP counts, $M=\{4, 9, 16\}$, with $K=2$ and $T=3$. The results show that, for all values of $M$, the communication SE increases with $L$, and for any given $L$, higher $M$ yields improved SE. For example, with $L={\num{12}}$, $M={\num{16}}$ achieves {\qty{53.9}{\percent}} and {\qty{25.2}{\percent}} higher communication SE than the $M={\num{4}}$ and $M={\num{9}}$ cases, respectively. These results confirm that increasing both the number of antennas and APs enhances SE by improving spatial diversity, interference suppression, and cooperative gain.}

{This improvement highlights the scalability and architectural flexibility of CF-ISAC systems. Larger antenna arrays yield greater beamforming gains, whereas denser AP deployments provide finer-grained spatial coverage and enable cooperative processing. Unlike conventional co-located ISAC, where performance is bounded by a single node's view, distributed CF-ISAC configurations exploit multiple independent channels to manage interference and shadowing. However, such scalability also increases synchronization and fronthaul coordination demands, emphasizing the need for lightweight distributed optimization in future CF-ISAC implementations.}

Fig.~\ref{fig_HeatMap_resource} presents the effects of beamforming gains utilizing $L = \num{8}$ AP antennas with $M=\num{4}$ APs for $K=2$ and $T=3$. In particular, Fig.~{\ref{fig_HeatMap_resource}} plots directional gain profiles in a color-coded scale to evaluate the performance of beamforming gains at each AP. The direction angles for sensing targets from AP 1, AP 2, AP 3, and AP 4, are set to $\{-60, 20, 40\}\qty{}{\degree}$, $\{25, -70, -10\}\qty{}{\degree}$, $\{25, -45, 75\}\qty{}{\degree}$, and $\{-20, 30, 60\}\qty{}{\degree}$, respectively, ensuring coverage of a broad angular range.

\begin{figure}[!t]\vspace{-0mm}
    \centering
    \includegraphics[width=0.5\textwidth]{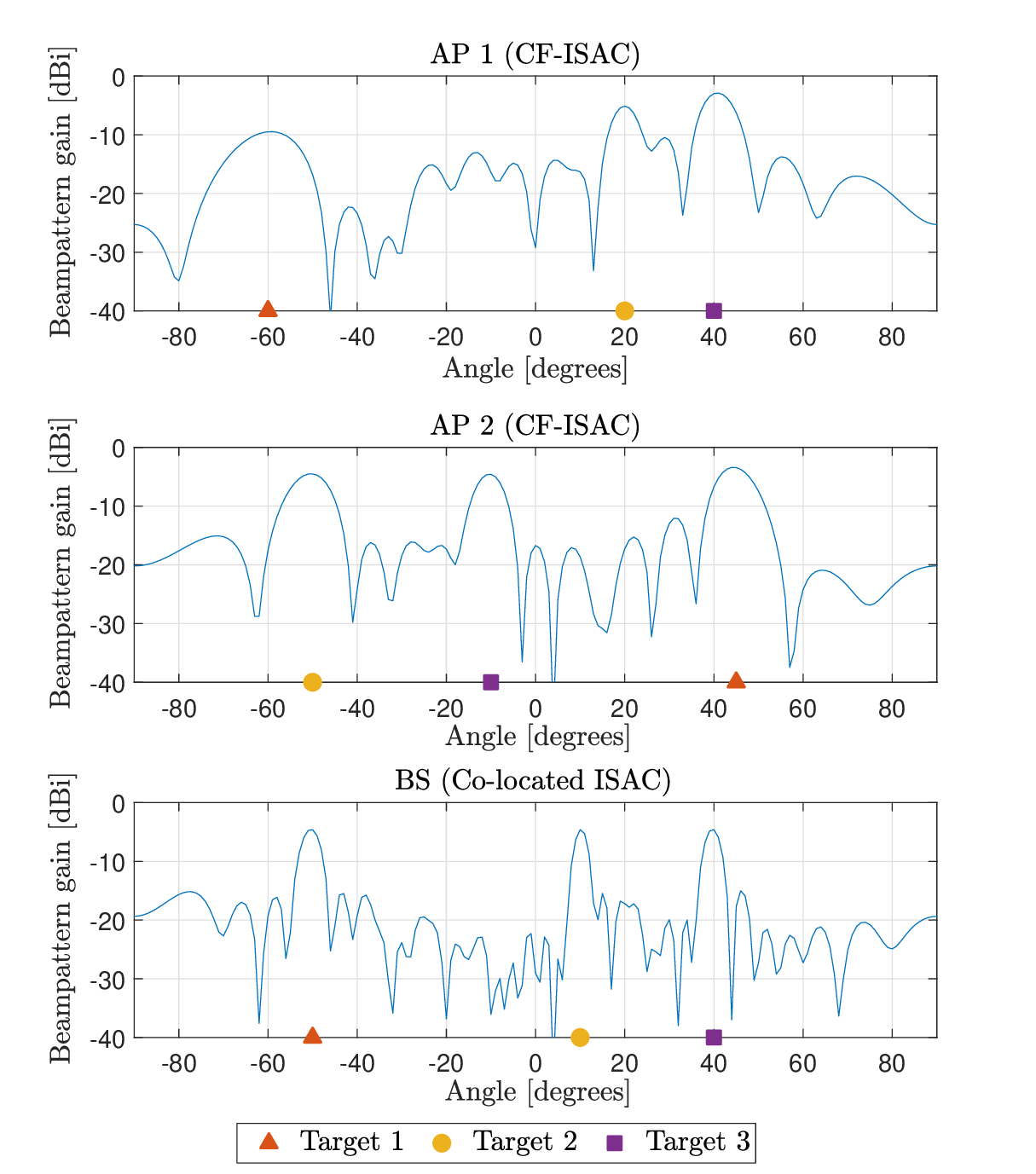}
    \vspace{-0mm}
    \caption{{A comparison of directional beampattern gain profiles over a {\qty{\pm 90}{\degree}} angular range for CF-ISAC (AP 1 and AP 2) and co-located ISAC (BS).}}
    \label{fig_BeamGani_Comp} \vspace{-0mm}
\end{figure}

The color gradient represents beamforming gain magnitude, with red indicating strong gains from focused energy and blue showing minimal radiated power. Intersecting beampattern gain directions across APs enables precise target localization, a key advantage of CF-ISAC's multi-static sensing over mono-static or bi-static approaches in co-located ISAC (Fig.~\ref{fig_BeamGani_Comp}). The distributed APs provide multiple viewpoints, enhancing spatial resolution and mitigating multi-path fading. Overlapping beampatterns enable accurate triangulation using AoA information from multiple perspectives, improving localization accuracy and resilience to blockages and interference. Additionally, if one AP experiences poor signal quality, others compensate, ensuring consistent sensing performance across the coverage area.

%=======================================================================
\begin{figure}[!t]\vspace{-0mm}
    \centering
    \includegraphics[width=0.47\textwidth]{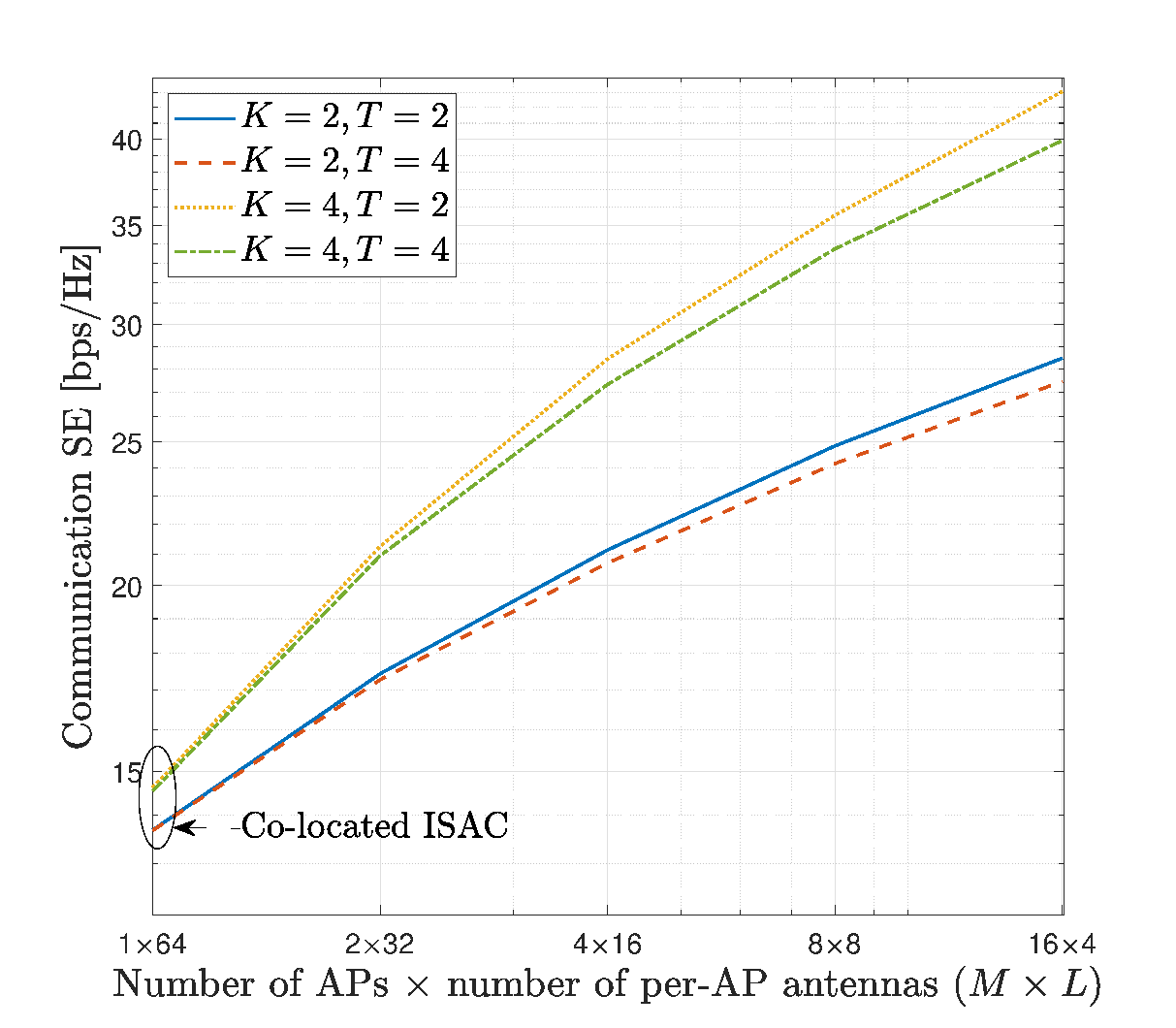}
    \vspace{-0mm}
    \caption{{Communication SE comparison between CF-ISAC and co-located ISAC systems for a fixed total number of antennas ($M\times L =\num{64}$), under different numbers of users ($K$) and targets ($T$).}}
    \label{fig_Comm_SE_Comp} \vspace{-0mm}
\end{figure}
%=======================================================================

{Fig.~{\ref{fig_Comm_SE_Comp}} compares the communication SE between CF-ISAC and co-located ISAC systems for equal total antenna counts ($M\times L = {\num{64}}$). The co-located ISAC corresponds to $M={\num{1}}$ and $L={\num{64}}$, while CF-ISAC configurations have $M> {\num{1}}$. The CF-ISAC consistently outperforms co-located systems due to micro-diversity gains and reduced path loss achieved through spatially distributed APs. These distributed configurations provide more uniform coverage and superior edge-user performance, validating one of CF-ISAC's key advantages: uniform quality-of-service (QoS) through cooperation rather than centralized beam concentration. The figure also reveals a clear communication-sensing trade-off: as the number of sensing targets increases for a fixed number of users, communication SE decreases because more transmit power and spatial DoF are devoted to target illumination and detection. CF-ISAC's distributed structure provides flexibility to dynamically balance this trade-off through adaptive power allocation and coordinated beamforming, maintaining robust performance under varying sensing and communication demands {\cite{zargari2024CFISAC}}.}

%=======================================================================
%=======================================================================

{Fig.~{\ref{fig_BeamGani_Comp}} compares the beampattern gains of CF-ISAC and co-located ISAC systems for $T=3$. The CF-ISAC configuration employs two distributed APs (AP1 and AP2), each equipped with 16 antennas, whereas the co-located ISAC setup comprises a single {\num{32}}-antenna BS located at the center of the coverage area. The target directions for AP1 and AP2 are set to ${\{-60, 20, 40\}\qty{}{\degree}}$ and ${\{45, -50, -10\}\qty{}{\degree}}$, respectively, while the co-located ISAC system targets are at ${\{-50, 10, 40\}\qty{}{\degree}}$ from the BS.}

{As observed in Fig.~{\ref{fig_BeamGani_Comp}}, both configurations form distinct main lobes toward their respective target directions, accurately indicating target locations. However, CF-ISAC achieves notably higher spatial resolution and more uniform coverage due to its multi-view diversity. By coherently combining beampattern gains from multiple distributed APs, CF-ISAC provides a richer set of angular and range observations, enabling joint angle-distance localization that co-located ISAC cannot achieve. In contrast, a single BS, limited to one viewpoint, can primarily resolve angular directions, leading to ambiguities in range estimation. This distributed gain pattern also demonstrates CF-ISAC's resilience to blockages and fading: if one AP's link to a target is degraded, other APs with different geometrical perspectives can maintain detection and tracking reliability. The overlapping main lobes across APs enhance localization accuracy through geometric diversity, effectively reducing sidelobe interference and improving overall sensing SINR.}

%=======================================================================
\begin{figure}[!t]\vspace{-0mm}
    \centering
    \includegraphics[width=0.47\textwidth]{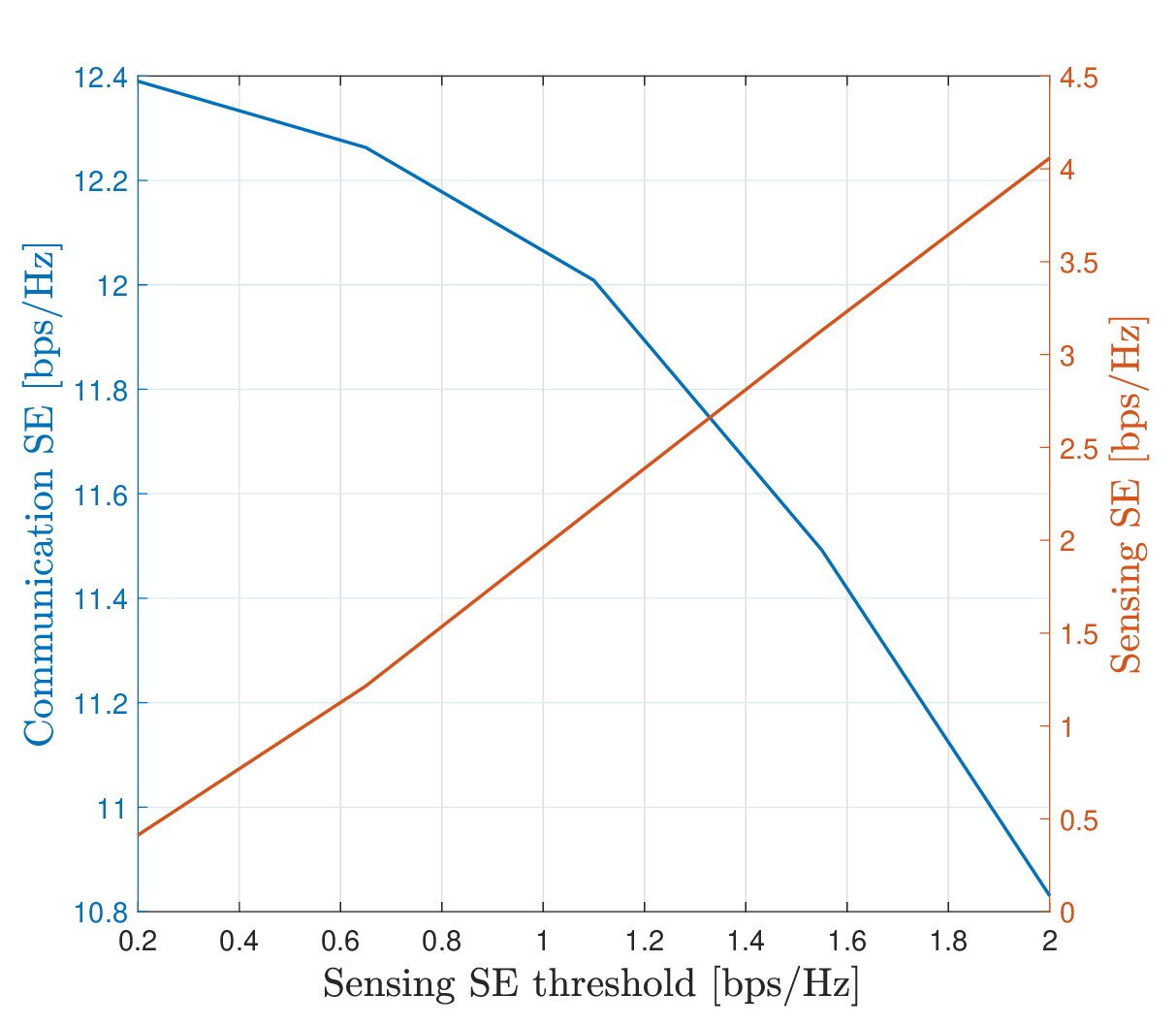}
    \vspace{-0mm}
    \caption{{Communication SE and sensing SE versus the per-target sensing SE threshold, illustrating the trade-off between communication performance and sensing requirements.}}
    \label{fig_ComSenTradeOff} \vspace{-0mm}
\end{figure}
%=======================================================================

{Fig.~{\ref{fig_ComSenTradeOff}} illustrates the trade-off between communication SE (left $y$-axis) and sensing SE (right $y$-axis) as a function of the per-target sensing SE threshold. The simulation considers a CF-ISAC mMIMO system with $M=9$ DL APs and $N=9$ UL APs, each equipped with $L=8$ antennas, serving $K=3$ single-antenna users and sensing $T=2$ targets (see Fig.~{\ref{fig_SystemModelPerfomance}}). The optimization objective is to maximize the users' communication SE subject to individual sensing SE thresholds and per-AP transmit power constraints.}

As shown in Fig.~{\ref{fig_ComSenTradeOff}}, increasing the sensing SE threshold improves the achievable sensing performance but degrades the communication SE. This inverse relationship stems from the coupling between joint communication and sensing operations: as the system prioritizes sensing, more transmit power and spatial DoF are allocated to the sensing beams, thereby reducing the power and beamforming gain available for communication. Moreover, stronger sensing signals generate higher interference in the communication band, further lowering communication SE. This trade-off underscores the fundamental resource coupling in CF-ISAC. The optimal operating point depends on application requirements, e.g., high sensing SE is preferred for precise localization or tracking, while communication-centric services favor higher throughput. Therefore, dynamic resource allocation strategies (e.g., adaptive beamforming or power splitting based on network context) are essential to balance sensing accuracy and communication capacity in real time.

Moreover, the results illustrate that cooperative beamforming across distributed APs enables flexible power allocation and target-specific focusing, which is particularly valuable in dynamic sensing environments with mobile or closely spaced targets. Although these performance gains depend on accurate synchronization and phase alignment among APs, reliable channel estimation, and sufficient fronthaul capacity, the distributed CF-ISAC architecture establishes a strong foundation for scalable, multi-perspective sensing. This framework can support future network-level ISAC deployments that leverage adaptive spatial coordination for enhanced situational awareness and robustness.

%-------------------------------
\subsection{Security Challenges of  CF-ISAC}
%-------------------------------
CF-ISAC and ISAC networks face more significant security risks than traditional wireless systems. This is because using information beams to enhance sensing can lead to information leakage, particularly when targets include adversarial entities such as eavesdropping UAVs \cite{Qu2024}. They not only intercept transmitted data but may also exploit sensing information to disrupt system performance. Conventional countermeasures such as beamforming, sensing covariance matrix design, and artificial noise (AN) transmission can enhance secrecy~\cite{Zhu:TuT:2024}. However, these solutions must be adapted for CF networks. While deploying more APs and receivers improves spatial diversity, it also increases vulnerabilities to active eavesdropping and pilot signal manipulation, potentially disrupting legitimate sensing.

These CF-ISAC security issues have been studied recently~\cite{Rivetti:WCNC:2024, Nasir2024,  Ren2024}. In~\cite{Rivetti:WCNC:2024}, a secure CF-ISAC system with multiple users and a single target, assumed to be an eavesdropper, is considered. The system aims to ensure reliable communication for the users while simultaneously detecting the target using mono-static sensing and degrading the eavesdropper's channel quality. To achieve this, an ISAC waveform embedded with AN is designed to minimize the CRB when estimating the target's direction relative to the APs. Using the SDR method, the CPU optimizes the precoding vectors and AN covariance matrices for each AP. This optimization is constrained to maintain the required SINR for the users while limiting the eavesdropper's maximum SNR. The study reveals an inverse proportionality between the optimal CRB and the user-eavesdropper distance, highlighting the impact of spatial positioning on ISAC performance.
In~\cite{Nasir2024}, a multi-static sensing model is examined in which multiple eavesdroppers try to intercept confidential information intended for communication users. The authors propose a joint communication and sensing beamforming design that maximizes the sensing SNR while ensuring that the secrecy rate for all communication users remains above a specified threshold. In~\cite{Ren2024}, the security of both communication and sensing is examined in the presence of information and sensing eavesdroppers, who seek to intercept confidential communications and extract target information, respectively. The study formulates a transmit beamforming optimization problem to maximize the detection probability while adhering to several constraints: SINR constraints for communication users, SNR constraints for information eavesdroppers, detection-probability constraints for sensing eavesdroppers, and transmit-power limitations for each transmitter. The global optimal solution was obtained using an SDR-based approach.

\subsubsection{Case Study and Discussion}\label{sec_CSD_secure}
Here, the resource allocation framework of the CF-ISAC system (Fig.~\ref{fig_SystemModelResourceAllocation}) is extended to examine its security aspects.
In particular, one or more targets not only serve as objects of interest for the ISAC system but also act as eavesdroppers, attempting to intercept confidential information intended for communication users. It is assumed that a malicious target attempts to decode any user's information from the received signal. If successful, it can then use the successive interference cancellation (SIC) technique to decode the information of all other users. Therefore, the leakage SEs at the targets for decoding user data are considered for evaluating the security performance. Our goal is to maximize the communication SE for users while meeting the sensing beampattern gain requirements and minimizing the leakage SE to targets/eavesdroppers for any user. This approach ensures that eavesdroppers cannot decode any user information, thereby securing communication data against eavesdropping.

\textit{Simulation Example:} The simulation setup is the same as the case study in Section~\ref{sec_resource_allocation}. Moreover, the maximum allowable leakage SE at all targets/eavesdroppers, $\delta_{\rm{max}}$, is set to \qty{0.5}{bps/\Hz}. 

%=======================================================================
\begin{figure}[!t]\vspace{-0mm}
    \centering
    \includegraphics[width=0.47\textwidth]{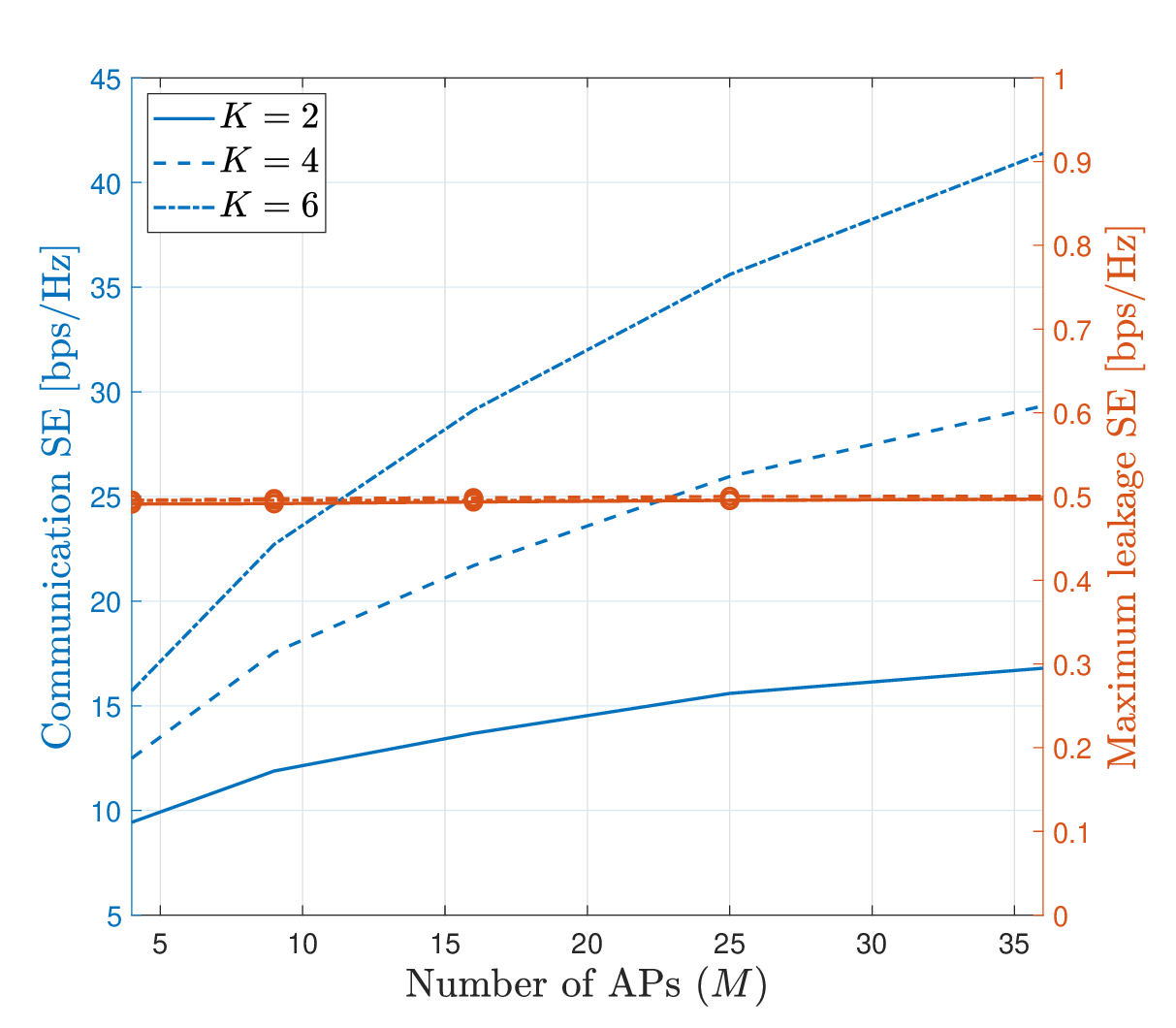}
    \vspace{-0mm}
    \caption{{Communication SE and maximum leakage SE as functions of the number of APs $M$ for different numbers of users $K$.}}
    \label{fig_ComRate_LeakRate_M_security} \vspace{-0mm}
\end{figure}
%=======================================================================

{Fig.~{\ref{fig_ComRate_LeakRate_M_security}} illustrates the impact of the number of APs ($M$) on both communication SE (left $y$-axis) and maximum leakage SE at potential eavesdroppers or sensing targets (right $y$-axis) for various user counts $K=\{2, 4, 6\}$. As $M$ increases, the communication SE improves markedly due to enhanced spatial diversity, stronger beamforming gains, and lower path-loss and shadowing effects. The CF-ISAC system also scales effectively with user density by exploiting distributed resources and spatial multiplexing. For instance, at $M={\num{16}}$, the configuration with $K={\num{6}}$ achieves communication SE gains of {\qty{112.8}{\percent}} and {\qty{34.2}{\percent}} over the $K={\num{2}}$ and $K={\num{4}}$ cases, respectively, demonstrating the system's ability to accommodate multiple users while sustaining high throughput.}

{Importantly, the secrecy performance, quantified by the maximum leakage SE, remains nearly constant across different AP densities. This stability underscores the effectiveness of the secure beamforming design, which shapes the transmit signals to enhance intended-user links while minimizing energy leakage toward potential eavesdroppers. The distributed nature of CF-ISAC inherently enhances secrecy: because APs operate from diverse spatial locations, the probability of coherent information leakage to unintended receivers is substantially reduced. Nonetheless, as the network densifies, maintaining low leakage SE requires careful coordination of beamforming phases and power allocation, indicating a key trade-off between scalability and security complexity.}

%=======================================================================
\begin{figure*}[!t]\vspace{-0mm}
    \centering
    \includegraphics[width=1.0\textwidth]{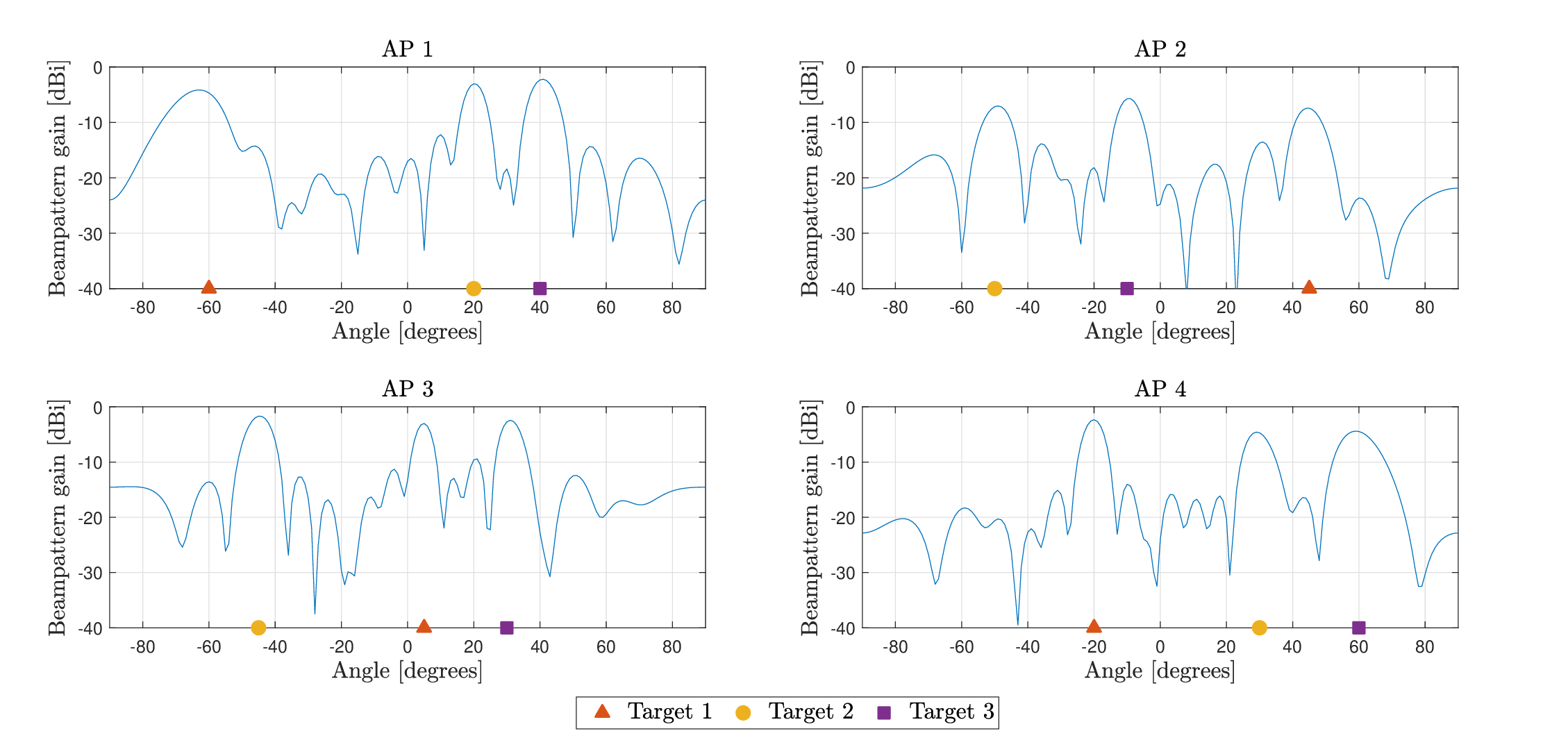}
    \vspace{-0mm}
    \caption{{Directional beampattern gain profiles over a {\qty{\pm 90}{\degree}} angular range at different APs, illustrating the per-AP beam responses toward multiple targets.}}
    \label{fig_BeamGain_Secure} \vspace{-0mm}
\end{figure*}
%=======================================================================

{Fig.~{\ref{fig_BeamGain_Secure}} depicts the directional beampattern gain profiles for secure CF-ISAC with $L=\num{8}$, $M=\num{4}$, $K=\num{2}$, and $T=\num{3}$. The subfigures correspond to APs 1-4, with target directions set as in Fig.~{\ref{fig_HeatMap_resource}}. Each AP's beampattern shows concentrated main lobes toward its assigned targets, confirming effective beam steering and precise power focusing. This localized power control is achieved by jointly optimizing sensing and communication beamforming, thereby suppressing mutual interference and minimizing radiation leakage toward unauthorized directions.}

The sharp main lobes reflect accurate target illumination and strong sensing accuracy, while deep troughs and low sidelobe levels signify robust leakage suppression. These sidelobe characteristics directly translate into lower interception probability at potential eavesdroppers. The differences among the AP-specific beampatterns further illustrate CF-ISAC's cooperative diversity; each AP adapts its beam pattern based on local geometry and global coordination, jointly maximizing communication reliability, sensing precision, and physical-layer security.

Overall, these results confirm that CF-ISAC's distributed architecture inherently supports secure and efficient joint operation. By combining multi-AP diversity with cooperative beamforming, CF-ISAC achieves high SE and sensing accuracy while maintaining strong confidentiality, key attributes for future mission-critical, IoT, defense, and surveillance applications that demand simultaneous situational awareness and secure connectivity. Nevertheless, these evaluations are based on idealized conditions to clearly demonstrate fundamental trends. More realistic analyses that incorporate channel estimation errors, synchronization imperfections, fronthaul capacity constraints, and user/target mobility represent important future research directions for assessing CF-ISAC performance.

%----------------------------------------
\subsection{User/Target-Centric Cell-Free ISAC}
%-----------------------------------
Scalability is a key challenge for CF networks, as the computational complexity and fronthaul capacity required by each AP to process and share data signals for all users grow linearly (and sometimes even faster) with the number of users. As an alternative, UC-CF has been introduced~\cite{Emil:TCOM:2020}. In this approach, each user is served by a subset of APs (an AP cluster) offering the most favorable channel conditions, leading to the formation of overlapping cooperation clusters across the network. As a result, each AP must dynamically collaborate with other APs to serve different users, all within the same time and frequency resources. In CF-ISAC networks, AP clustering can be performed from the user point of view~\cite{Cao2023}, or from the target point of view~\cite{Buzzi2024}.  In the former scheme, user scheduling is performed across different APs, with each communication user served by a cluster of APs, while all APs collaborate on the sensing task. The latter introduces a target-centric approach in which the surveillance area is divided into non-overlapping regions. Each area is then sensed by a subset of APs.  The target-centric approach presents an intriguing research direction for supporting multi-target sensing in a distributed manner. However, forming AP clusters that account for network dynamics and managing collaboration among overlapping APs in the sensing function remain challenging tasks that require further investigation.

%%%%%%%%%%%%%%%%%%%%%%%%%%%%%%%%%%%%%%%%%%%%%%%%%%%%%%
\section{Key challenges and Open Research Directions}\label{sec_future}
%%%%%%%%%%%%%%%%%%%%%%%%%%%%%%%%%%%%%%%%%%%%%%%%%%%%%%%
This section discusses key challenges, open research directions, and future trends of CF-ISAC.

%------------------------------------------------------------
\subsection{Key Challenges in Cell-Free ISAC}
%-----------------------------------------------------------

\subsubsection{Multi-Target Sensing}
{Accurately detecting and tracking multiple targets simultaneously remains a major CF-ISAC  challenge. This complexity arises from the need to differentiate overlapping echoes from distinct targets across distributed APs, which often leads to ambiguities in detection and tracking. While the distributed architecture of CF-ISAC  provides additional spatial diversity, it also increases coordination complexity among APs~{\cite{Buzzi2024}}. A potential solution is to divide the scan area into disjoint regions, with each region monitored by designated APs. To enhance robustness, APs can span multiple regions, increasing redundancy and detection reliability.}

{A critical aspect of this approach is the design of an effective coordination protocol. It must ensure that adjacent radar cells, i.e., disjoint regions, are not scanned simultaneously, thereby minimizing mutual interference and reducing the risk of target misdetection in overlapping areas. Ideally, radar cells scanned concurrently should be spatially separated to mitigate cross-region interference. Techniques such as time-division multiplexing and spatial scheduling can be employed. Additionally, advanced signal processing algorithms, including compressed sensing and sparse recovery, can resolve targets in dense environments, further enhancing multi-target detection capabilities~{\cite{Buzzi2024}}.}

\subsubsection{Synchronization and Calibration}
{Precise synchronization and calibration are essential for coherent joint operation across distributed APs~{\cite{Demir2021book, Ngo2017, Zhang2019cellfree}}. Synchronization ensures timing, frequency, and phase alignment among APs, which is critical for coherent transmission, high data rates, and accurate target detection in CF-ISAC~{\cite{Demir2021book, Ngo2017, Zhang2019cellfree}}. Any mismatch leads to signal misalignment, resulting in inter-user interference, degraded communication quality, and reduced sensing accuracy. In particular, phase misalignment impairs joint beamforming, leading to beam misdirection and reduced QoS and radar resolution. Hardware imperfections, oscillator drifts, and propagation delays make synchronization challenging~{\cite{Cheng2024}}.}

{Furthermore, array calibration, which ensures consistent amplitude and phase references across distributed APs, is equally important~{\cite{Rogalin2014, Shepard2012}}. Calibration must compensate for hardware drifts, oscillator mismatches, and environment-dependent propagation effects. Practical techniques include over-the-air reference signaling, mutual coupling measurements, and self-calibration via reciprocal channel estimation~{\cite{Rogalin2014, Shepard2012}}. These methods, however, introduce signaling overhead and rely on precise timing support,  typically provided through GPS or fronthaul-based synchronization~{\cite{Emil2017, Vieira2021}}. As the network scales, maintaining calibration accuracy becomes increasingly difficult because clock drift, phase noise, and jitter accumulate across APs, leading to distortions in coherent beamforming and echo processing.}

{To mitigate these issues, CF-ISAC systems can employ disciplined local oscillators, fronthaul-assisted phase tracking, or hybrid synchronization frameworks that combine hardware-based stabilization (e.g., temperature-compensated oscillators) with software-based correction (e.g., adaptive phase feedback)~{\cite{Cheng2024}}. Such hybrid approaches will be critical for large-scale CF-ISAC deployments, where synchronization, calibration, and hardware impairments collectively determine sensing precision and communication reliability~{\cite{Cheng2024}}. However, as the number of spatially distributed APs increases, achieving and maintaining precise synchronization and calibration becomes more complex. Non-ideal clock behaviors, such as phase noise and jitter, introduce additional errors that distort processed echo signals, degrading detection accuracy and overall system performance~{\cite{Cheng2024}}.}

\subsubsection{Interference Management}
{In CF-ISAC networks, multiple forms of interference arise due to the joint operation of communication and sensing over shared spectral and spatial resources~{\cite{Niu2024}}. The primary interference sources include~{\cite{Niu2024}}:}
\begin{itemize}
    \item {\textit{Intra-user (multiuser) interference:} Signals transmitted by different APs or users sharing the same frequency and time resources can interfere with one another, degrading communication throughput.}
    \item {\textit{Inter-sensing interference:} Echoes or reflections from multiple targets may overlap or interfere, reducing sensing resolution and target detection accuracy.}
    \item {\textit{Sensing-to-communication interference:} Reflected sensing signals or echoes can act as interference to communication receivers operating in the same band.}
    \item {\textit{Communication-to-sensing interference:} Strong user or AP transmission signals may mask weak target echoes, thereby deteriorating sensing performance.}
    \item {\textit{Fronthaul and hardware-induced interference:} Imperfect synchronization, hardware impairments (e.g., phase noise, nonlinearities), and limited fronthaul capacity can introduce residual interference across distributed APs.}
\end{itemize}
{Effective interference management is thus a critical challenge in CF-ISAC systems, where communication and sensing share the same spectral resources~{\cite{Niu2024}}. This dual use can cause co-channel interference, degrading both data throughput and target detection accuracy.}

To address these issues, various techniques have been developed, including adaptive beamforming, interference alignment, and spectrum sharing protocols~{\cite{Niu2024, Meng2024}}. Adaptive beamforming dynamically steers communication and sensing beams to minimize interference, whereas interference alignment techniques exploit spatial dimensions to separate interfering signals. Additionally, learning-based resource allocation strategies, including dynamic spectrum access and power control algorithms, optimize the coexistence of communication and sensing functionalities~{\cite{Meng2024}}. Moreover, hierarchical interference management combines low-level signal-domain coordination (e.g., waveform orthogonalization) with higher-level scheduling and spatial resource control. This multi-layer approach balances instantaneous interference suppression with long-term network fairness and efficiency.

\subsubsection{Fronthaul Capacity, Latency, and EE}
{The fronthaul infrastructure that links distributed APs to the CPU significantly affects the performance of CF-ISAC~{\cite{Masoumi2020}}. Limited bandwidth and high latency can disrupt CSI exchange, coordination, and sensing feedback, leading to outdated beamforming and delayed target detection. These effects are amplified when mobility or dynamic target scenes demand real-time updates~{\cite{Masoumi2020}}.}

{Moreover, high fronthaul signaling overhead directly impacts EE~{\cite{Masoumi2020}}. Frequent CSI reporting, sensing data sharing, and synchronization exchanges increase power consumption, particularly in dense deployments. The EE of CF-ISAC thus involves a trade-off between distributed processing (which reduces fronthaul load but increases local computation) and centralized processing (which improves coordination but consumes more fronthaul bandwidth)~{\cite{Masoumi2020}}.}

{To address these challenges, emerging strategies include edge-assisted CF-ISAC architectures that push partial processing to the APs. Techniques such as fronthaul compression, quantization-aware CSI sharing, and latency-adaptive coordination can substantially reduce signaling load. Additionally, integrating energy-aware scheduling and sleep-mode operation for idle APs can improve EE without compromising sensing coverage. Advanced fronthaul technologies, such as millimeter-wave and optical fiber links, further enhance scalability and real-time operation.}

\subsubsection{Hardware Impairments}
{Hardware imperfections, such as phase noise, nonlinear power amplifiers, I/Q imbalance, and analog-to-digital converter quantization errors~{\cite{9503397}}, pose practical limitations for CF-ISAC~{\cite{Demir2021book}}. These impairments cause a mismatch between transmitted and received signals, leading to spectral regrowth, distortion, and degraded radar image quality.}

{Because CF-ISAC relies on distributed coherent processing, hardware distortions at individual APs can accumulate, severely affecting beamforming accuracy and target localization~{\cite{Demir2021book}}. Compensation methods such as digital predistortion, calibration-based equalization, and hardware-in-the-loop feedback are essential. In particular, adaptive compensation schemes that jointly model and mitigate RF impairments across multiple APs could significantly improve sensing reliability and communication integrity.}

%-----------------------------------------
\subsection{Open Research Directions and Future Trends}
%--------------------------------------------------
\subsubsection{Network-Assisted Cell-Free ISAC}
In network-assisted CFMM,  the CPU coordinates the operation modes of APs, i.e., UL or DL, virtually realizing FD operation within CF architecture~\cite{Wang:TCOM.2020}. In particular, this centralized coordination enables dynamic assignment of APs to transmit DL signals or receive UL signals/echoes in response to real-time network demands, user distribution, and channel conditions. Compared to conventional CFMM, this can enhance SE and mitigate interference across the network~\cite{Wang:TCOM.2020,Mohammadi:TCOM:2024}.
Network-assisted CF-ISAC is thus particularly promising for meeting both UL and DL communication demands while simultaneously sharing resources with sensing functionality~\cite{Wang:TCOM.2020,Mohammadi:TCOM:2024}. This approach supports flexible multi-static sensing, in which APs can be adaptively assigned to transmit DL sensing signals or receive UL echoes according to network demands. Additionally, this framework facilitates the simultaneous transmission and reception of UL and DL signals~\cite{Zeng2023, zeng2024}.

However, communication and sensing tasks can significantly interfere with each other. For example, UL communication signals at UL-operating APs contain both echo-sensing signals and cross-link interference from DL APs. Similarly, sensing signals received at UL APs are subject to interference from both UL and DL communication signals, degrading sensing performance. Consequently, resource allocation must be optimized to design the UL and DL power-control coefficients and configure AP-mode operations to meet both communication and sensing requirements. One potential solution is to divide the UL APs into two disjoint groups, with one dedicated to sensing and the other serving UL communications \cite{Mohammadi:JSAC:2023}.

%--------------------------------------------------------
\subsubsection{New Antenna Technologies and Cell-Free ISAC}
%---------------------------------------------------------
To fully exploit spatial resources and enhance ISAC performance, fluid (movable) antennas have been introduced~\cite{Zou:WCL:2024,Zhou:WCL:2024,Qin:WCL:2024}. By optimizing their position within a confined region, fluid antennas provide spatial diversity with fewer RF chains and minimal space requirements~\cite{Wang:WL:2024}, making them particularly suitable for user terminals. In CFMM networks, APs can deploy fluid antennas to serve multi-antenna users, introducing additional spatial DoFs that improve ISAC performance. However, major challenges remain, including accurate channel estimation, efficient position optimization, and the high dimensionality of the resulting ISAC-oriented optimization problems. Since fluid antenna systems rely on spatially correlated signals, incorporating artificial intelligence can help exploit hidden correlations and improve system design.

Another emerging antenna technology attracting significant attention is holographic MIMO (HMIMO): a cost-effective, planar wireless structure composed of sub-wavelength metallic or dielectric scatterers capable of manipulating EM waves for specific objectives~\cite{Huang:WCL:2020}. The HMIMO surface can operate as a transmitter, receiver, or reflector, enabling highly reconfigurable wireless environments. In~\cite{Adhikary:IoT:2024}, an AI-based framework integrates HMIMO APs into CF networks to achieve efficient ISAC-enabled beamforming through adaptive power allocation. The framework selectively activates grids within HMIMO APs to meet user demands by solving an optimization problem that maximizes the sensing utility function. This approach improves both communication and sensing SINR, enhances EE, and ensures effective power utilization.

%-----------------------------------
\subsubsection{Cell-free ISAC and Near-Field}
%-----------------------------
The previously underutilized upper mid-band spectrum ($7-24$ GHz)~\cite{Zhang:MCOM:2025,chaves2024coverage, Azar2024}, also known as FR3, has received increasing interest from both academia and industry because of the significance of extremely large mMIMO (XL-MIMO) antenna arrays. When either the physical aperture of the antenna array is large or the wavelength is short, the Fraunhofer distance, defined as $2D_{\text{array}}^2/\lambda$ , where  $D_{\text{array}}$ is the aperture size and $\lambda$ is the wavelength, can become substantial \cite{Azar2024}. As a result, near-field propagation conditions may arise when serving the user at shorter distances. While current $5$G systems operate under far-field conditions, near-field propagation is expected to be both inevitable and exploitable in $6$G systems, offering new opportunities for spatial multiplexing, high-resolution sensing, and user localization \cite{Zhang:MCOM:2025,chaves2024coverage, Azar2024, Diluka2026, Diluka2026WBNF}.

Near-field operation fundamentally changes how wireless signals propagate and energy (beamforming) is directed. In contrast to far-field beamforming, which produces planar-wave beams that expand conically with distance, near-field beamfocusing enables energy to be concentrated at a specific spatial point. This is achieved because near-field spherical wavefronts introduce a distance-dependent phase curvature, causing the focal point to dissipate beyond a certain range~\cite{Zhang:MCOM:2025,chaves2024coverage,Azar2024}. As a result, the beam pattern resembles a spotlight, providing energy confinement in both the angular and spatial domains. This dual dependency enables XL-MIMO systems to distinguish users not only by direction but also by their range, enriching the multi-user MIMO channel structure and enhancing sum capacity~\cite{Zhang:MCOM:2023,Cui:MCOM:2023}. Moreover, by exploiting the spherical wavefronts captured in the near-field region, an antenna array can achieve high-resolution three-dimensional localization and sensing capabilities~\cite{Azar2024}. These features are particularly valuable at mmWave and sub-THz frequencies, where shorter wavelengths make near-field effects more pronounced. However, the near-field region's extent--characterized by the Fraunhofer distance--remains limited to only a few tens of meters for a typical $6$G BS operating in the upper mid-band~\cite{Cui:MCOM:2023,Lei:WC:2025}.

Recent studies have begun to extend near-field beamfocusing concepts to CFMM systems. For example, the authors in~\cite{bjornson2024enabling} show that coordinating multiple subarrays separated by only a few meters can emulate near-field focusing effects without requiring a large continuous aperture. This idea aligns with the CFMM architecture, in which widely distributed APs collaboratively reconstruct the spherical wavefront structure to improve beam control, sensing, and localization. In \cite{Xie2025}, a delay-based beamfocusing scheme for near-field CFMM using OFDM modulation is developed by introducing an angle-based approach to leverage sparse array configurations. By combining these techniques, it demonstrates that joint delay-angle-based focusing can form a distinct, controllable focal region, with improved performance achieved by increasing the number of subcarriers and expanding the array aperture. Reference \cite{Xie2024} proposes a modified LoS steering model to capture asynchronous reception effects in near-field CFMM systems. This work shows that distributed APs inherently establish a composite near-field environment through multiple far-field planar waves, thereby reducing spatial correlation and mitigating multi-user interference. These studies provide early evidence that CF architectures can naturally exploit near-field effects without the need for excessively large continuous arrays.

Despite these promising developments, practical realization remains challenging. Achieving coherent near-field coordination across distributed APs requires precise synchronization, accurate calibration, and low-latency fronthaul, which may increase energy consumption and system complexity. Consequently, EE, synchronization accuracy, and scalability become critical design considerations for deploying near-field CF-ISAC systems in 6G.

%--------------------------------------------------------
\subsubsection{Consolidation of Complementary Technologies into Cell-Free ISAC}
%---------------------------------------------------------
To enhance CF-ISAC performance, integrating complementary technologies has gained traction. Notably,~\cite{Dong:IoT:2024} proposed a NOMA-aided CF-ISAC architecture to improve connectivity and efficiency. Focusing on a single-radar sensing scenario, the study jointly optimizes user pairing and beamforming to maximize the minimum achievable communication rate while meeting sensing requirements. Future research could extend this work by incorporating statistical CSI-based designs that account for channel estimation or exploring multi-target scenarios with both instantaneous and statistical approaches.

In~\cite{Abdelaziz:TCOM:2024}, an FD  CFMM system enhanced by a single RIS was investigated. The RIS was used to maximize the weighted sum of the radar and communication SINRs. To achieve this, a joint optimization framework was proposed that encompasses the design of radar and communication receive beamformers, UL transmission powers, DL sensing beamformers, and RIS reflection coefficients. The system's performance can be further enhanced by deploying multiple RIS within the network to expand coverage across a larger area. Additionally, integrating advanced technologies such as beyond-diagonal RIS and stacked metasurfaces presents a compelling avenue for future research and development.

The utilization of multiple UAVs in a CFMM architecture for ISAC systems with dedicated sensing signals has been explored in~\cite{Flores:ICC:2024}. Specifically, three deployment scenarios for the UAVs were considered: mobile UAVs, tethered UAVs, and fixed UAVs. For all scenarios, a transmit precoder that jointly optimizes the sensing and communication requirements, subjected to power constraints, was designed.

%--------------------------------------------------------
\subsubsection{Machine Learning-Based Techniques for CF-ISAC}
%---------------------------------------------------------
Unifying sensing and communication in CFMM systems enhances performance, reduces spectrum congestion, and lowers costs by sharing hardware resources. However, these benefits depend on the optimal design of ISAC waveforms, precoders, detectors, and estimators. Achieving optimal designs requires complex multi-objective optimization due to inherent trade-offs between sensing and communication. Moreover, the unified designs and real-time implementation of CF-ISAC based on classical model-based analytical techniques may often become sub-optimal due to model inadequacy or incompleteness. On the other hand, Pareto-optimal algorithms and solutions for multi-objective optimization problems that apply to CF-ISAC may be entirely unknown in some cases. They may also be prohibitively computationally expensive or practically infeasible to implement.

ML-based techniques have emerged as efficient, data-driven alternatives to classical model-based signal processing and optimization \cite{Yu2022, Hu2021, Dai2020,Vaezi2026}. Deep learning, in particular, is well-suited for CF-ISAC, where multiple conflicting objectives must be jointly optimized \cite{Demirhan2024ML}. For instance, deep learning can help design unified beamformers that balance sensing and communication goals \cite{Demirhan2024ML}.

These benefits accrue because data-driven CF-ISAC techniques extract features from high-dimensional multimodal data, leveraging the embedded low-dimensional structures for model learning \cite{Vaezi2026,Cheng2024a}. The multimodal sensory information in CFMM channels can enhance model-driven CF-ISAC designs and optimizations. Retaining 3D data structures improves the accuracy and generalization of learning-based models, even with limited datasets. Deep learning can also aid in jointly estimating communication and sensing channel parameters, including CSI acquisition, data decoding, and target localization through delay, Doppler, and angular-domain parameter estimation.

Deep learning enhances adaptivity and enables intelligent network features in CF-ISAC. Applications include autonomous vehicles, localization, human activity detection, environmental monitoring, object tracking, smart surveillance, and wide-area imaging.

%===================================
\section{Conclusion}
%=============================
This article explored the emerging CF-ISAC paradigm, which integrates CF and ISAC to enhance SE, EE, and sensing in future wireless networks. It began with a review of the fundamentals of CF and ISAC and their integration into CF-ISAC, highlighting its unique characteristics. The benefits of multi-static sensing and key features of CF-ISAC systems were then examined. Current developments were categorized into performance analysis, resource allocation, security, and user- and target-centric designs, with a comprehensive literature survey and case studies. Finally, the article outlined key challenges, open research directions, and emerging trends, offering valuable insights for future advancements.

%===================================
\appendices
%===================================

\section{Performance Analysis in Cell-Free ISAC}\label{apx_perfomance}

\subsection{Channel Model} 
For the system setup in Fig.~\ref{fig_SystemModelPerfomance}, block flat-fading channel models are considered. During each fading block, $\q{h}_{mk} \in \mathbb{C}^{L \times 1}$, $\q{g}_{mt}^{\mt{d}} \in \mathbb{C}^{L \times 1}$, and $\q{g}_{nt}^{\mt{u}} \in \mathbb{C}^{L \times 1}$ are the channel vectors from the $m$-th DL AP to the $k$-th user, the $m$-th DL AP to the $t$-th target, and the $t$-th target to the $n$-th UL AP, respectively. Moreover, $\q{F}_{mn} \in \mathbb{C}^{L\times L}$ represents the channel between the $m$-th DL AP and the $n$-th UL AP. All channels are assumed to be independent quasi-static Rayleigh fading, which remains constant during the coherence interval. A unified representation of all channels is given as 
\begin{eqnarray}\label{eqn_chnl_model}
    \q{a} = \zeta_{\q{a}}^{1/2} \tilde{\q{a}}, 
\end{eqnarray}
where $\q{a} \in \{ \q{h}_{mk}, \q{g}_{mt}^{\mt{d}}, \q{g}_{mt}^{\mt{u}}\}$, ${\zeta}_{\q{a}}$ accounts for the large-scale path loss and shadowing, and $\tilde{\q{a}} \sim \mathcal{CN}\left(\q{0},\q{I}_L \right)$ captures the small-scale Rayleigh fading, which is static during one coherence interval. With time-division duplexing for both channel estimation and data transmission, CSI can be estimated using orthogonal pilots {\cite{Demir2021book}}, yielding highly accurate estimates. Thus, perfect CSI availability is assumed. 

\subsection{Transmission Model} 
The transmitted signal at the $m$-th DL AP, $\q{x}_{m} \in \mathbb{C}^{L \times 1}$, is given as
\begin{align}
    \q{x}_m = \sum\nolimits_{k = 1}^{K} \q{w}_{mk} q_k + \sum\nolimits_{t=1}^{T} \q{s}_{mt},
\end{align}
where $q_k \in \mathbb{C}$ represents the intended data symbol for the $k$-th user with unit power, i.e., $\E{\vert q_i\vert^2}=1$, and $\q{w}_{mk} \in \mathbb{C}^{L\times 1}$ is the $m$-th DL AP transmit beamforming vector for the $k$-th user, and $\q{s}_{mt} \in \mathbb{C}^{L\times 1}$ is the dedicated sensing signal at the $m$-th DL AP for the $t$-th target \cite{Zhenyao2023}. It is also assumed that $q_k$ and $\q{s}_{mk}$ are independent of each other \cite{Zhenyao2023}.

The received signal at the $k$-th user is given as
\begin{align} \label{eqn_rx_DL_k}
    y_k &= \sum\nolimits_{m=1}^{M} \q{h}_{mk}^{\rm{H}} \q{x}_m + z_k \nonumber \\
    &= \underbrace{\sum\nolimits_{m=1}^{M}  \q{h}_{mk}^{\rm{H}} \q{w}_{mk} q_k}_{\text{Desired signal }} + \underbrace{\sum\nolimits_{i\neq k}^{K} \sum\nolimits_{m=1}^{M} \q{h}_{mk}^{\rm{H}} \q{w}_{mi} q_i }_{\text{Multi-user interference}}  \nonumber \\
   &+ \underbrace{\sum\nolimits_{t=1}^{T} \sum\nolimits_{m=1}^{M} \q{h}_{mk}^{\rm{H}} \q{s}_{mt} }_{\text{Sensing signal interference }}  + \underbrace{z_k}_{\text{AWGN}},
\end{align}
where $z_k \sim \mathcal{CN}(0,\sigma^2)$ is the AWGN at the $k$-th user. 

The UL APs use the reflected signals from targets, i.e., target echoes, to extract target state information \cite{Zhenyao2023}. The received signal at the $n$-th UL AP, i.e., $\q{y}_n \in \mathbb{C}^{L\times 1}$, is given as
\begin{align}\label{eqn_rx_AP_n}
    \q{y}_n &= \underbrace{\sum\nolimits_{m=1}^{M} \q{F}_{mn} \q{x}_m}_{\text{Direct-link interference}} \nonumber \\
    &+ \underbrace{\sum\nolimits_{t=1}^{T} \q{g}_{nt}^{\mt{u}} \alpha_t \sum\nolimits_{m=1}^{M} (\q{g}_{mt}^{\mt{d}})^{\rm{H}} \q{x}_m}_{\text{Target echoes}}  + \underbrace{\q{z}_n}_{\text{AWGN}},
\end{align}
where $\q{z}_n \sim \mathcal{CN}(\q{0},\sigma^2 \q{I}_L)$ is the AWGN at the $n$-th UL AP and $\alpha_t \in \mathbb{C}$ is the complex amplitude of the $t$-th target reflection, accounting for the round-trip path loss and the RCS of the target \cite{Liu2022}. Specifically, path loss accounts for signal attenuation with distance, whereas RCS determines the amount of power reflected toward the radar receiver, depending on the target's size, shape, and materials. Additionally, the UL APs are assumed to employ clutter rejection techniques to minimize interference from reflected clutter in the surrounding environment \cite{Mark2010RadarBook}. 

Since the APs are connected to the CPU via backhaul links, it is assumed that the direct-link interference (DLI), i.e., inter-AP interference, is known to the UL APs. Thus, the UL APs remove DLI before applying the sensing combiner, $\q{u}_{nt} \in \mathbb{C}^{L\times 1}$ to process the sensing information from the $t$-th target. The post-processed signal for obtaining the $t$-th target's sensing information at the $n$-th UL AP is given as
\begin{align}\label{eqn_rx_Sens_t}
    {y}_{nt} &= \q{u}_{nt}^{\rm{H}} \left(\q{y}_{n}-\sum\nolimits_{m=1}^{M} \q{F}_{mn} \q{x}_m \right)\nonumber\\
    &= \underbrace{\alpha_t \q{u}_{nt}^{\rm{H}} \q{g}_{nt}^{\mt{u}} \sum_{m=1}^{M} (\q{g}_{mt}^{\mt{d}})^{\rm{H}} \left( \sum_{i=1}^{K} \q{w}_{mi} q_i + \sum_{l=1}^{T} \q{s}_{ml}\right)}_{\text{$t$-th target's desired reflection}} \nonumber \\
    &+ \underbrace{\sum_{j\neq t}^{T} \alpha_j \q{u}_{nt}^{\rm{H}} \q{g}_{nj}^{\mt{u}} \sum_{m=1}^{M} (\q{g}_{mj}^{\mt{d}})^{\rm{H}} \left( \sum_{i=1}^{K} \q{w}_{mi} q_i + \sum_{j=1}^{T} \q{s}_{mj}\right)}_{\text{Multi-target interference reflections}} \nonumber \\
    &+ \q{u}_{nt}^{\rm{H}} \q{z}_n. 
\end{align}

\subsection{Communication SE} 
From \eqref{eqn_rx_DL_k}, communication SINR of the $k$-th user is obtained as
\begin{align}\label{eqn_user_sinr_k}
    \mt{SINR}_k^{\rm{Com}} = \frac{{\rm{DS}}_k}{\sum_{i\neq k}^{K} {\rm{MUI}}_{ki} +  \sum_{t=1}^{T} {\rm{SSI}}_{kt} + \sigma^2},
\end{align}
where 
\begin{subequations}
\begin{align}
    {\rm{DS}}_k &= \E{\Big\vert \sum\nolimits_{m=1}^{M}  \q{h}_{mk}^{\rm{H}} \q{w}_{mk} \Big\vert^2}, \\
    {\rm{MUI}}_{ki} &= \E{\Big\vert \sum\nolimits_{m=1}^{M}  \q{h}_{mk}^{\rm{H}} \q{w}_{mi} \Big\vert^2}, \\
    {\rm{SSI}}_{kt} &= \E{\Big\vert \sum\nolimits_{m=1}^{M}  \q{h}_{mk}^{\rm{H}} \q{s}_{mt} \Big\vert^2}.
\end{align}
\end{subequations}
To derive the closed-form solution, it is assumed that the DL APs adopt MRT beamforming for both communication and sensing, i.e., $\q{w}_{mk} = \q{h}_{mk}$ and $\q{s}_{mt} = \q{g}_{mt}^{\mt{d}}$. To this end, by evaluating the expectation terms in \eqref{eqn_user_sinr_k}, the closed-form solution of the SINR at the $k$-th user is given in \eqref{eqn_SINR_userk_close}.
\begin{figure*}
\begin{align}\label{eqn_SINR_userk_close}
    \mt{SINR}_k^{\rm{Com}} = \frac{L(L+1)\sum_{m=1}^{M} \zeta_{\q{h}_{mk}}^2 + L^2 \sum_{m=1}^{M} \sum_{m'\neq m}^{M} \zeta_{\q{h}_{mk}} \zeta_{\q{h}_{m'k}} }{ L\sum_{i\neq k}^{K} \sum_{m=1}^{M} \zeta_{\q{h}_{mk}} \zeta_{\q{h}_{mi}} + L \sum_{t=1}^{T} \sum_{m=1}^{M} \zeta_{\q{h}_{mk}} \zeta_{\q{g}_{mt}^{\mt{d}}} + \sigma^2}
\end{align}
 \hrulefill
\end{figure*}
Thus, the SE of the $k$-th user is given as
\begin{align}\label{eqn_DL_rate}
    \mathcal{S}_k^{\rm{Com}} = \log_2 \left(1+ \mt{SINR}_k^{\rm{Com}} \right).
\end{align}

\subsection{Sensing SE} 
The sensing SE is used to evaluate sensing performance (Section~\ref{sec_sensing_SE}). The UL APs perform sensing by utilizing the targets' echoes. From  \eqref{eqn_rx_Sens_t}, the sensing SE of the $t$-th target at the $n$-th UL AP is obtained as 
\begin{align}
    \mathcal{S}_{nt}^{\rm{Sen}} \approx \log_2 \left(1+ \mt{SINR}_{nt}^{\rm{Sen}} \right),
\end{align}
where the $t$-th target's sensing SINR an the $n$-th UL AP is given as
\begin{align}\label{eqn_target_sinr_t}
    \mt{SINR}_{nt}^{\rm{Sen}} = \frac{\vert \alpha_t \vert^2 {\rm{TDS}}_{nt}}{\sum_{j\neq t}^{T} \vert \alpha_j\vert^2 {\rm{MTI}}_{n,tj} + \sigma^2 \E{\Vert \q{u}_{nt} \Vert^2}},
\end{align}
where
\begin{subequations}
\begin{align}
    {\rm{TDS}}_{nt} &\!= \! \E{\! \Bigg\vert \q{u}_{nt}^{\rm{H}} \q{g}_{nt}^{\mt{u}} \! \sum_{m=1}^{M} \!(\q{g}_{mt}^{\mt{d}})^{\rm{H}} \! \left(  \sum_{i=1}^{K} \q{w}_{mi} \! + \! \sum_{l=1}^{T} \q{s}_{ml} \!\right) \!\Bigg\vert^2 \!}, \\
    {\rm{MTI}}_{n,tj} &=\! \E{\! \left\vert \q{u}_{nt}^{\rm{H}} \q{g}_{nj}^{\mt{u}} \! \sum_{m=1}^{M} \!(\q{g}_{mj}^{\mt{d}})^{\rm{H}}\! \left( \!\sum_{i=1}^{K} \q{w}_{mi} \! +\! \sum_{j=1}^{T} \q{s}_{mj} \!\right) \!  \right\vert^2 \!}.
\end{align}
\end{subequations}
By assuming the $n$-th UL AP employ the MRC to extract the $t$-th target's state information, i.e., $\q{u}_{nt} = \q{g}_{nt}^{\mt{u}}$, the closed-form expression of the sensing SINR, $\mt{SINR}_{nt}^{\rm{Sen}}$, is given in \eqref{eqn_SINR_targett_close}.   
\begin{figure*}
\begin{align}\label{eqn_SINR_targett_close}
    \mt{SINR}_{nt}^{\rm{Sen}} = \frac{\vert \alpha_t \vert^2 L(L+1) \zeta_{\q{g}_{nt}^{\mt{u}}}^2 \!\! \sum_{m=1}^{M} \! \left( L \!\sum_{i=1}^{K} \! \zeta_{\q{g}_{mt}^{\mt{d}}} \zeta_{\q{h}_{mi}} \!+\! L(L+1) \zeta_{\q{g}_{mt}^{\mt{d}}}^2 \!+\! L^2 \! \sum_{m'\neq m}^{M} \! \zeta_{\q{g}_{mt}^{\mt{d}}} \zeta_{\q{g}_{m't}^{\mt{d}}} \!+\! L \!\sum_{l\neq t}^{T} \zeta_{\q{g}_{mt}^{\mt{d}}} \zeta_{\q{g}_{ml}^{\mt{d}}} \! \right) }{\sum\limits_{j\neq t}^T \vert \alpha_j\vert^2 L \zeta_{\q{g}_{nt}^{\mt{u}}} \zeta_{\q{g}_{nj}^{\mt{u}}} \!\! \sum\limits_{m=1}^{M}\! \left( L \!\sum\limits_{i=1}^{K} \! \zeta_{\q{g}_{mt}^{\mt{d}}} \zeta_{\q{h}_{mi}} \!+\! L(L+1) \zeta_{\q{g}_{mj}^{\mt{d}}}^2 \!+\! L^2 \! \sum\limits_{m'\neq m}^{M} \! \zeta_{\q{g}_{mj}^{\mt{d}}} \zeta_{\q{g}_{m'j}^{\mt{d}}} \!+\! L \!\sum\limits_{l\neq t}^{T} \zeta_{\q{g}_{mj}^{\mt{d}}} \zeta_{\q{g}_{ml}^{\mt{d}}} \! \right) + L \sigma^2 \zeta_{\q{g}_{nt}^{\mt{u}}} }
\end{align}
\hrulefill
\end{figure*}

\balance

\bibliographystyle{IEEEtran}
\bibliography{IEEEabrv,ref}

\begin{thebibliography}{100}
\providecommand{\url}[1]{#1}
\csname url@samestyle\endcsname
\providecommand{\newblock}{\relax}
\providecommand{\bibinfo}[2]{#2}
\providecommand{\BIBentrySTDinterwordspacing}{\spaceskip=0pt\relax}
\providecommand{\BIBentryALTinterwordstretchfactor}{4}
\providecommand{\BIBentryALTinterwordspacing}{\spaceskip=\fontdimen2\font plus
\BIBentryALTinterwordstretchfactor\fontdimen3\font minus
  \fontdimen4\font\relax}
\providecommand{\BIBforeignlanguage}[2]{{%
\expandafter\ifx\csname l@#1\endcsname\relax
\typeout{** WARNING: IEEEtran.bst: No hyphenation pattern has been}%
\typeout{** loaded for the language `#1'. Using the pattern for}%
\typeout{** the default language instead.}%
\else
\language=\csname l@#1\endcsname
\fi
#2}}
\providecommand{\BIBdecl}{\relax}
\BIBdecl

\bibitem{Liu2022ISAC}
A.~Liu \emph{et~al.}, ``A survey on fundamental limits of integrated sensing
  and communication,'' \emph{{IEEE} Commun. Surveys Tuts.}, vol.~24, no.~2, pp.
  994--1034, 2nd Quart. 2022.

\bibitem{Wang2022ISAC}
J.~Wang \emph{et~al.}, ``Integrated sensing and communication: Enabling
  techniques, applications, tools and data sets, standardization, and future
  directions,'' \emph{{IEEE} Internet Things J.}, vol.~9, no.~23, pp.
  23\,416--23\,440, Dec. 2022.

\bibitem{Zhang2022}
J.~A. Zhang \emph{et~al.}, ``Enabling joint communication and radar sensing in
  mobile networks — {A} survey,'' \emph{{IEEE} Commun. Surveys Tuts.},
  vol.~24, no.~1, pp. 306--345, 1st Quart. 2022.

\bibitem{Azar2024}
A.~Hakimi, D.~Galappaththige, and C.~Tellambura, ``A roadmap for {NF-ISAC} in
  {6G}: A comprehensive overview and tutorial,'' \emph{Entropy}, vol.~26,
  no.~9, Sept. 2024.

\bibitem{liu2023integratedbook}
F.~Liu, C.~Masouros, and Y.~Eldar, Eds.,
  \emph{\BIBforeignlanguage{English}{Integrated Sensing and
  Communications}}.\hskip 1em plus 0.5em minus 0.4em\relax Springer Singapore,
  Jul. 2023.

\bibitem{Mao2023}
W.~Mao \emph{et~al.}, ``Beamforming design in cell-free massive {MIMO}
  integrated sensing and communication systems,'' in \emph{Proc. IEEE Global
  Commun. Conf.}, Dec. 2023, pp. 546--551.

\bibitem{Demirhan2023}
U.~Demirhan and A.~Alkhateeb, ``Cell-free joint sensing and communication
  {MIMO}: A max-min fair beamforming approach,'' in \emph{Proc. IEEE Asilomar
  Conf. Signals, Syst., Comput.}, Oct. 2023, pp. 381--386.

\bibitem{Huang2022Coordinated}
Y.~Huang, Y.~Fang, X.~Li, and J.~Xu, ``Coordinated power control for network
  integrated sensing and communication,'' \emph{{IEEE} Trans. Veh. Technol.},
  vol.~71, no.~12, pp. 13\,361--13\,365, Dec. 2022.

\bibitem{Cao2023Design}
Y.~Cao and Q.-Y. Yu, ``Design and performance analyses of {V-OFDM} integrated
  signal for cell-free massive {MIMO} joint communication and radar system,''
  \emph{{IEEE} Syst. J.}, vol.~17, no.~4, pp. 5943--5954, Dec. 2023.

\bibitem{Wang2023}
B.~Wang, L.~Xu, Z.~Cheng, and Z.~He, ``Semi-distributed hybrid beamforming
  design for cooperative cell-free dual-function radar-communication
  networks,'' in \emph{Proc. IEEE Int. Conf.Acoustics, Speech, and Signal
  Process. Workshops}, Jun. 2023, pp. 1--5.

\bibitem{Sakhnini2022Uplink}
A.~Sakhnini, A.~Bourdoux, M.~Guenach, H.~Sahli, and S.~Pollin, ``Uplink payload
  power control in cell-free communication and radar networks,'' in \emph{Proc.
  IEEE Global Commun. Conf.}, Dec. 2022, pp. 5111--5116.

\bibitem{Silva2023}
I.~W.~G. Da~Silva, D.~P.~M. Osorio, and M.~Juntti, ``Multi-static {ISAC} in
  cell-free massive {MIMO}: Precoder design and privacy assessment,'' in
  \emph{Proc. IEEE Globecom Workshops}, Dec. 2023, pp. 461--466.

\bibitem{Behdad2022}
Z.~Behdad, {\"O}.~T. Demir, K.~W. Sung, E.~Bj{\"o}rnson, and C.~Cavdar, ``Power
  allocation for joint communication and sensing in cell-free massive {MIMO},''
  in \emph{Proc. IEEE Global Commun. Conf.}, Dec. 2022, pp. 4081--4086.

\bibitem{Behdad2024Interplay}
Z.~Behdad, {\"O}.~T. Demir, K.~W. Sung, and C.~Cavdar, ``Interplay between
  sensing and communication in cell-free massive {MIMO} with {URLLC} users,''
  in \emph{Proc. IEEE Wireless Commun. Netw. Conf.}, Apr. 2024, pp. 1--6.

\bibitem{Gesbert2010}
D.~Gesbert \emph{et~al.}, ``Multi-cell {MIMO} cooperative networks: A new look
  at interference,'' \emph{{IEEE} J. Sel. Areas Commun.}, vol.~28, no.~9, pp.
  1380--1408, Dec. 2010.

\bibitem{Dahrouj2010}
H.~Dahrouj and W.~Yu, ``Coordinated beamforming for the multicell multi-antenna
  wireless system,'' \emph{{IEEE} Trans. Wireless Commun.}, vol.~9, no.~5, pp.
  1748--1759, May 2010.

\bibitem{Jun2015}
J.~Wu, Z.~Zhang, Y.~Hong, and Y.~Wen, ``Cloud radio access network {(C-RAN)}: A
  primer,'' \emph{{IEEE} Netw.}, vol.~29, no.~1, pp. 35--41, Jan. 2015.

\bibitem{Ngo2017}
H.~Q. Ngo, A.~Ashikhmin, H.~Yang, E.~G. Larsson, and T.~L. Marzetta,
  ``Cell-free massive {MIMO} versus small cells,'' \emph{{IEEE} Trans. Wireless
  Commun.}, vol.~16, no.~3, pp. 1834--1850, Mar. 2017.

\bibitem{Fishler2004}
E.~Fishler \emph{et~al.}, ``{MIMO} radar: An idea whose time has come,'' in
  \emph{Proc. IEEE Radar Conf.}, Aug. 2004, pp. 71--78.

\bibitem{Haimovich2008}
A.~M. Haimovich, R.~S. Blum, and L.~J. Cimini, ``{MIMO} radar with widely
  separated antennas,'' \emph{{IEEE} Signal Process. Mag.}, vol.~25, no.~1, pp.
  116--129, Jan. 2008.

\bibitem{richards2005fundamentals}
M.~A. Richards, \emph{Fundamentals Of Radar Signal Processing}.\hskip 1em plus
  0.5em minus 0.4em\relax McGraw-Hill Education (India) Pvt Limited, 2005.

\bibitem{Demir2021book}
{\"O}.~Demir, E.~Bj{\"o}rnson, and L.~Sanguinetti, \emph{Foundations of
  User-Centric Cell-Free Massive {MIMO}}, ser. Foundations and trends in signal
  processing.\hskip 1em plus 0.5em minus 0.4em\relax Now Publishers, 2021.

\bibitem{Ammar2022}
H.~A. Ammar, R.~Adve, S.~Shahbazpanahi, G.~Boudreau, and K.~V. Srinivas,
  ``User-centric cell-free massive {MIMO} networks: A survey of opportunities,
  challenges and solutions,'' \emph{{IEEE} Commun. Surveys Tuts.}, vol.~24,
  no.~1, pp. 611--652, 1st Quart. 2022.

\bibitem{Giovanni2018}
G.~Interdonato, E.~Bj{\"{o}}rnson, H.~Q. Ngo, P.~K. Frenger, and E.~G. Larsson,
  ``Ubiquitous cell-free massive {MIMO} communications,'' \emph{EURASIP J.
  Wireless Commun. Netw.}, vol. 2019, pp. 1687--1499, Aug. 2019.

\bibitem{Zhang2020}
J.~Zhang \emph{et~al.}, ``Prospective multiple antenna technologies for beyond
  {5G},'' \emph{{IEEE} J. Sel. Areas Commun.}, vol.~38, no.~8, pp. 1637--1660,
  Aug. 2020.

\bibitem{Elhoushy2022}
S.~Elhoushy, M.~Ibrahim, and W.~Hamouda, ``Cell-free massive {MIMO}: A
  survey,'' \emph{{IEEE} Commun. Surveys Tuts.}, vol.~24, no.~1, pp. 492--523,
  1st Quart. 2022.

\bibitem{Zhang2019cellfree}
J.~Zhang \emph{et~al.}, ``Cell-free massive {MIMO}: A new next-generation
  paradigm,'' \emph{{IEEE} Access}, vol.~7, pp. 99\,878--99\,888, Aug. 2019.

\bibitem{Shuaifei2022}
S.~Chen, J.~Zhang, J.~Zhang, E.~Bj\"{o}rnson, and B.~Ai, ``A survey on
  user-centric cell-free massive {MIMO} systems,'' \emph{Digital Commun.
  Netw.}, vol.~8, no.~5, pp. 695--719, Dec. 2022.

\bibitem{Kassam2023}
J.~Kassam, D.~Castanheira, A.~a. Silva, R.~Dinis, and A.~Gameiro, ``A review on
  cell-free massive {MIMO} systems,'' \emph{Electronics}, vol.~12, no.~4, p.
  1001, Feb. 2023.

\bibitem{Mohammadi2024}
M.~Mohammadi, Z.~Mobini, H.~Quoc~Ngo, and M.~Matthaiou, ``Next-generation
  multiple access with cell-free massive {MIMO},'' \emph{Proc. {IEEE}}, vol.
  112, no.~9, pp. 1372--1420, Sept. 2024.

\bibitem{Diluka2019}
D.~Galappaththige and G.~Amarasuriya, ``Cell-free massive {MIMO} with underlay
  spectrum-sharing,'' in \emph{Proc. IEEE Int. Conf. Commun.}, May 2019, pp.
  1--7.

\bibitem{Galappaththige2021}
D.~Galappaththige, R.~Shrestha, and G.~A. Aruma~Baduge, ``Exploiting cell-free
  massive {MIMO} for enabling simultaneous wireless information and power
  transfer,'' \emph{IEEE Trans. Green Commun. Netw.}, vol.~5, no.~3, pp.
  1541--1557, Sept. 2021.

\bibitem{Diluka2020}
D.~Galappaththige and G.~Amarasuriya, ``{NOMA}-aided cell-free massive {MIMO}
  with underlay spectrum-sharing,'' in \emph{Proc. IEEE Int. Conf. Commun.},
  Jun. 2020, pp. 1--6.

\bibitem{Galappaththige2024}
D.~Galappaththige and C.~Tellambura, ``Sum rate maximization for
  {RSMA}-assisted {CF} {mMIMO} networks with {SWIPT} users,'' \emph{{IEEE}
  Wireless Commun. Lett.}, vol.~13, no.~5, pp. 1300--1304, May 2024.

\bibitem{Galappaththige2021Cellfree}
D.~Galappaththige, D.~Kudathanthirige, and G.~Amarasuriya, ``Performance
  analysis of {IRS}-assisted cell-free communication,'' in \emph{Proc. IEEE
  Global Commun. Conf.}, Dec. 2021, pp. 1--6.

\bibitem{Diluka2021}
D.~Galappaththige and G.~A.~A. Baduge, ``Exploiting underlay spectrum sharing
  in cell-free massive {MIMO} systems,'' \emph{{IEEE} Trans. Commun.}, vol.~69,
  no.~11, pp. 7470--7488, Nov. 2021.

\bibitem{Emil2017}
E.~Bj{\"o}rnson, J.~Hoydis, and L.~Sanguinetti, \emph{Massive {MIMO} Networks:
  Spectral, Energy, and Hardware Efficiency}, 2017, vol.~11.

\bibitem{Marzetta2016book}
T.~L. Marzetta, E.~G. Larsson, H.~Yang, and H.~Q. Ngo, \emph{Fundamentals of
  Massive {MIMO}}.\hskip 1em plus 0.5em minus 0.4em\relax Cambridge University
  Press, 2016.

\bibitem{Ngo:TGCN:2018}
H.~Q. Ngo, L.-N. Tran, T.~Q. Duong, M.~Matthaiou, and E.~G. Larsson, ``On the
  total energy efficiency of cell-free massive {MIMO},'' \emph{IEEE Trans.
  Green Commu. and Networking}, vol.~2, no.~1, pp. 25--39, Mar., 2018.

\bibitem{Wang2020}
D.~Wang, F.~Rezaei, and C.~Tellambura, ``Performance analysis and resource
  allocations for a {WPCN} with a new nonlinear energy harvester model,''
  \emph{IEEE Open J. Commun. Soc.}, vol.~1, pp. 1403--1424, Sept. 2020.

\bibitem{5429879}
S.~Atapattu, C.~Tellambura, and H.~Jiang, ``Energy detection of primary signals
  over $\eta-\mu$ fading channels,'' in \emph{Proc. 2009 Int. Conf. Ind. Inf.
  Syst.}, Dec. 2009, pp. 118--122.

\bibitem{5208031}
S.~P. Herath, N.~Rajatheva, and C.~Tellambura, ``Unified approach for energy
  detection of unknown deterministic signal in cognitive radio over fading
  channels,'' in \emph{IEEE Int. Conf. Commun.}, Jun. 2009, pp. 1--5.

\bibitem{6987540}
S.~Atapattu, C.~Tellambura, H.~Jiang, and N.~Rajatheva, ``Unified analysis of
  low-{SNR} energy detection and threshold selection,'' \emph{{IEEE} Trans.
  Veh. Technol.}, vol.~64, no.~11, pp. 5006--5019, Nov. 2015.

\bibitem{Diluka2024CFBiBC}
D.~Galappaththige, F.~Rezaei, C.~Tellambura, and A.~Maaref, ``Cell-free
  bistatic backscatter communication: Channel estimation, optimization, and
  performance analysis,'' \emph{{IEEE} Trans. Commun.}, vol.~72, no.~10, pp.
  6617--6632, Oct. 2024.

\bibitem{Emil:TCOM:2020}
E.~Björnson and L.~Sanguinetti, ``Scalable cell-free massive {MIMO} systems,''
  \emph{{IEEE} Trans. Commun.}, vol.~68, no.~7, pp. 4247--4261, Jul. 2020.

\bibitem{Parida2023}
P.~Parida and H.~S. Dhillon, ``Cell-free massive {MIMO} with finite fronthaul
  capacity: A stochastic geometry perspective,'' \emph{{IEEE} Trans. Wireless
  Commun.}, vol.~22, no.~3, pp. 1555--1572, Mar. 2023.

\bibitem{Elhoshy2016}
S.~Elhoshy \emph{et~al.}, ``A dimensioning framework for indoor {DAS} {LTE}
  networks,'' in \emph{Proc. Int. Conf. Sel. Topics Mobile Wireless Netw.},
  Apr. 2016, pp. 1--8.

\bibitem{Irmer2011}
R.~Irmer \emph{et~al.}, ``Coordinated multipoint: Concepts, performance, and
  field trial results,'' \emph{{IEEE} Commun. Mag.}, vol.~49, no.~2, pp.
  102--111, Feb. 2011.

\bibitem{Venkatesan2007}
S.~Venkatesan, A.~Lozano, and R.~Valenzuela, ``Network {MIMO}: Overcoming
  intercell interference in indoor wireless systems,'' in \emph{Proc. IEEE
  Asilomar Conf. Signals, Syst., Comput.}, Nov. 2007, pp. 83--87.

\bibitem{Simeone2008}
O.~Simeone, O.~Somekh, H.~Vincent~Poor, and S.~Shamai, ``Distributed {MIMO} in
  multi-cell wireless systems via finite-capacity links,'' in \emph{Proc. 3rd
  Int. Symp. Commun., Control Signal Process.}, Mar. 2008, pp. 203--206.

\bibitem{intel2021fronthaul}
\BIBentryALTinterwordspacing
{Intel Corporation}, ``Exploring {5G} fronthaul network architecture:
  Intelligence splits and connectivity,'' Intel Corporation, Tech. Rep., 2021.
  [Online]. Available:
  \url{https://www.intel.com/content/dam/www/public/us/en/documents/white-papers/exploring-5g-fronthaul-network-architecture-white-paper.pdf}
\BIBentrySTDinterwordspacing

\bibitem{Peng2015}
M.~Peng, Y.~Li, Z.~Zhao, and C.~Wang, ``System architecture and key
  technologies for {5G} heterogeneous cloud radio access networks,''
  \emph{{IEEE} Netw.}, vol.~29, no.~2, pp. 6--14, Apr. 2015.

\bibitem{Mao2017}
Y.~Mao, C.~You, J.~Zhang, K.~Huang, and K.~B. Letaief, ``A survey on mobile
  edge computing: The communication perspective,'' \emph{{IEEE} Commun. Surveys
  Tuts.}, vol.~19, no.~4, pp. 2322--2358, 4th Quart. 2017.

\bibitem{Larsen2018}
L.~M.~P. Larsen, M.~S. Berger, and H.~L. Christiansen, ``Fronthaul for
  cloud-{RAN} enabling network slicing in {5G} mobile networks,''
  \emph{Wireless Commun. Mobile Comput.}, vol. 2018, no.~1, p. 4860212, Aug.
  2018.

\bibitem{Chen2018ChannelHardening}
Z.~Chen and E.~Björnson, ``Channel hardening and favorable propagation in
  cell-free massive {MIMO} with stochastic geometry,'' \emph{{IEEE} Trans.
  Commun.}, vol.~66, no.~11, pp. 5205--5219, Nov. 2018.

\bibitem{Polegre2020}
A.~A. Polegre, F.~Riera-Palou, G.~Femenias, and A.~G. Armada, ``Channel
  hardening in cell-free and user-centric massive {MIMO} networks with
  spatially correlated {Ricean} fading,'' \emph{IEEE Access}, vol.~8, pp.
  139\,827--139\,845, Aug. 2020.

\bibitem{papoulis2002probability}
A.~Papoulis and S.~Pillai, \emph{Probability, Random Variables, and Stochastic
  Processes}, ser. McGraw-Hill series in electrical and computer
  engineering.\hskip 1em plus 0.5em minus 0.4em\relax McGraw-Hill, 2002.

\bibitem{Ngo2017EE}
H.~Q. Ngo, L.-N. Tran, T.~Q. Duong, M.~Matthaiou, and E.~G. Larsson, ``Energy
  efficiency optimization for cell-free massive {MIMO},'' in \emph{Proc. IEEE
  Int. Workshop Signal Process. Adv. Wireless Commun.}, Jul. 2017, pp. 1--5.

\bibitem{Papazafeiropoulos2020}
A.~Papazafeiropoulos, P.~Kourtessis, M.~D. Renzo, S.~Chatzinotas, and J.~M.
  Senior, ``Performance analysis of cell-free massive {MIMO} systems: A
  stochastic geometry approach,'' \emph{{IEEE} Trans. Veh. Technol.}, vol.~69,
  no.~4, pp. 3523--3537, Apr. 2020.

\bibitem{Yang2018}
H.~Yang and T.~L. Marzetta, ``Energy efficiency of massive {MIMO}: Cell-free
  vs. cellular,'' in \emph{Proc. IEEE Veh. Technol. Conf.}, Jun. 2018, pp.
  1--5.

\bibitem{Mark2010RadarBook}
M.~A. Richards, J.~A. Scheer, and W.~A. Holm, Eds., \emph{Principles of Modern
  Radar: Basic principles}, ser. Radar, Sonar and Navigation.\hskip 1em plus
  0.5em minus 0.4em\relax Institution of Engineering and Technology, 2010.

\bibitem{knott2004radar}
E.~Knott, J.~Schaeffer, and M.~Tulley, \emph{Radar Cross Section}, ser. Radar,
  Sonar and Navigation Series.\hskip 1em plus 0.5em minus 0.4em\relax
  Institution of Engineering and Technology, 2004.

\bibitem{Rezende2002RCS}
M.~C. Rezende, I.~M. Martin, M.~A.~S. Miacci, and E.~L. Nohara, ``Radar cross
  section measurements {(8-12 GHz)} of magnetic and dielectric microwave
  absorbing thin sheets,'' in \emph{Proc. SBMO/IEEE MTT-S Int. Microw.
  Optoelectronics Conf.}, Dec. 2002, pp. 439--443.

\bibitem{skolnik2001introduction}
M.~Skolnik, \emph{Introduction to Radar Systems}, ser. Electrical engineering
  series.\hskip 1em plus 0.5em minus 0.4em\relax McGraw-Hill, 2001.

\bibitem{Sayama2001}
S.~Sayama and H.~Sekine, ``Weibull, log-{Weibull} and {$K$}-distributed ground
  clutter modeling analyzed by {AIC},'' \emph{{IEEE} Trans. Aerosp. Electron.
  Syst.}, vol.~37, no.~3, pp. 1108--1113, Apr. 2001.

\bibitem{levanon2004radar}
N.~Levanon and E.~Mozeson, \emph{Radar Signals}, ser. IEEE Press.\hskip 1em
  plus 0.5em minus 0.4em\relax Wiley, 2004.

\bibitem{Blunt2016}
S.~D. Blunt and E.~L. Mokole, ``Overview of radar waveform diversity,''
  \emph{{IEEE} Aerosp. Electron. Syst. Mag.}, vol.~31, no.~11, pp. 2--42, Jul.
  2016.

\bibitem{Liu2023a}
F.~Liu, Y.~Xiong, K.~Wan, T.~X. Han, and G.~Caire, ``Deterministic-random
  tradeoff of integrated sensing and communications in {Gaussian} channels: A
  rate-distortion perspective,'' in \emph{Proc. IEEE Int. Symp. Inf. Theory},
  Jun. 2023, pp. 2326--2331.

\bibitem{Ahmadipour2024}
M.~Ahmadipour, M.~Kobayashi, M.~Wigger, and G.~Caire, ``An
  information-theoretic approach to joint sensing and communication,''
  \emph{{IEEE} Trans. Inf. Theory}, vol.~70, no.~2, pp. 1124--1146, Feb. 2024.

\bibitem{Liao2025}
Z.~Liao \emph{et~al.}, ``Pulse shaping for random {ISAC} signals: The ambiguity
  function between symbols matters,'' \emph{{IEEE} Trans. Wireless Commun.},
  vol.~24, no.~4, pp. 2832--2846, Apr. 2025.

\bibitem{Liu2025}
F.~Liu \emph{et~al.}, ``Uncovering the iceberg in the sea: Fundamentals of
  pulse shaping and modulation design for random {ISAC} signals,'' \emph{{IEEE}
  Trans. Signal Process.}, vol.~73, pp. 2511--2526, Jul. 2025.

\bibitem{Vandendorpe2025}
L.~Vandendorpe, L.~Defraigne, G.~Thiran, T.~Pairon, and C.~Craeye,
  ``Positioning and transmission in cell-free networks: Ambiguity function, and
  {MRC/MRT} array gains,'' in \emph{Proc. IEEE Int. Conf.Acoustics, Speech, and
  Signal Process. Workshops}, Apr. 2025, pp. 1--5.

\bibitem{li2009mimoradar}
J.~Li, J.~Li, and P.~Stoica, \emph{{MIMO} Radar Signal Processing}, ser. IEEE
  Press.\hskip 1em plus 0.5em minus 0.4em\relax Wiley, 2009.

\bibitem{Mohammadi2023}
M.~Mohammadi, Z.~Mobini, D.~Galappaththige, and C.~Tellambura, ``A
  comprehensive survey on full-duplex communication: Current solutions, future
  trends, and open issues,'' \emph{{IEEE} Commun. Surveys Tuts.}, vol.~25,
  no.~4, pp. 2190--2244, 4th Quart. 2023.

\bibitem{Diluka2024CFFD}
D.~Galappaththige, M.~Mohammadi, H.~Q. Ngo, M.~Matthaiou, and C.~Tellambura,
  ``Cell-free full-duplex communication -- {An} overview,'' \emph{{IEEE} Trans.
  Commun.}, pp. 1--1, 2025.

\bibitem{Xiong2022}
Y.~Xiong, F.~Liu, Y.~Cui, W.~Yuan, and T.~X. Han, ``Flowing the information
  from {Shannon} to {Fisher}: Towards the fundamental tradeoff in {ISAC},'' in
  \emph{Proc. IEEE Global Commun. Conf.}, Dec. 2022, pp. 5601--5606.

\bibitem{Ma2020}
D.~Ma, N.~Shlezinger, T.~Huang, Y.~Liu, and Y.~C. Eldar, ``Joint
  radar-communication strategies for autonomous vehicles: Combining two key
  automotive technologies,'' \emph{{IEEE} Signal Process. Mag.}, vol.~37,
  no.~4, pp. 85--97, Jun. 2020.

\bibitem{Ma2021}
------, ``{FRaC}: {FMCW}-based joint radar-communications system via index
  modulation,'' \emph{{IEEE} J. Sel. Topics Signal Process.}, vol.~15, no.~6,
  pp. 1348--1364, Nov. 2021.

\bibitem{Chen2022}
L.~Chen, Z.~Wang, Y.~Du, Y.~Chen, and F.~R. Yu, ``Generalized transceiver
  beamforming for {DFRC} with {MIMO} radar and {MU-MIMO} communication,''
  \emph{{IEEE} J. Sel. Areas Commun.}, vol.~40, no.~6, pp. 1795--1808, Jun.
  2022.

\bibitem{Liu2020}
X.~Liu \emph{et~al.}, ``Joint transmit beamforming for multiuser {MIMO}
  communications and {MIMO} radar,'' \emph{{IEEE} Trans. Signal Process.},
  vol.~68, pp. 3929--3944, Jul. 2020.

\bibitem{Bell1993}
M.~Bell, ``Information theory and radar waveform design,'' \emph{{IEEE} Trans.
  Inf. Theory}, vol.~39, no.~5, pp. 1578--1597, Sept. 1993.

\bibitem{Tang2010}
B.~Tang, J.~Tang, and Y.~Peng, ``{MIMO} radar waveform design in colored noise
  based on information theory,'' \emph{{IEEE} Trans. Signal Process.}, vol.~58,
  no.~9, pp. 4684--4697, Sept. 2010.

\bibitem{Zhang2021Radar}
J.~A. Zhang \emph{et~al.}, ``An overview of signal processing techniques for
  joint communication and radar sensing,'' \emph{{IEEE} J. Sel. Topics Signal
  Process.}, vol.~15, no.~6, pp. 1295--1315, Nov. 2021.

\bibitem{Kay1998}
S.~M. Kay, \emph{Fundamentals of Statistical Signal Processing, Vol. {I}:
  {E}stimation Theory}.\hskip 1em plus 0.5em minus 0.4em\relax Englewood
  Cliffs, NJ, USA: Prentice Hall, 1998.

\bibitem{He2022}
Z.~He, W.~Xu, H.~Shen, Y.~Huang, and H.~Xiao, ``Energy efficient beamforming
  optimization for integrated sensing and communication,'' \emph{{IEEE}
  Wireless Commun. Lett.}, vol.~11, no.~7, pp. 1374--1378, Jul. 2022.

\bibitem{Stoica2007}
P.~Stoica, J.~Li, and Y.~Xie, ``On probing signal design for {MIMO} radar,''
  \emph{{IEEE} Trans. Signal Process.}, vol.~55, no.~8, pp. 4151--4161, Jul.
  2007.

\bibitem{Cui2014}
G.~Cui, H.~Li, and M.~Rangaswamy, ``{MIMO} radar waveform design with constant
  modulus and similarity constraints,'' \emph{{IEEE} Trans. Signal Process.},
  vol.~62, no.~2, pp. 343--353, Jan. 2014.

\bibitem{Hua2023}
H.~Hua, J.~Xu, and T.~X. Han, ``Optimal transmit beamforming for integrated
  sensing and communication,'' \emph{{IEEE} Trans. Veh. Technol.}, vol.~72,
  no.~8, pp. 10\,588--10\,603, Mar. 2023.

\bibitem{Tang2019}
B.~Tang and J.~Li, ``Spectrally constrained {MIMO} radar waveform design based
  on mutual information,'' \emph{{IEEE} Trans. Signal Process.}, vol.~67,
  no.~3, pp. 821--834, Feb. 2019.

\bibitem{Zhenyao2023}
Z.~He \emph{et~al.}, ``Full-duplex communication for {ISAC}: Joint beamforming
  and power optimization,'' \emph{{IEEE} J. Sel. Areas Commun.}, vol.~41,
  no.~9, pp. 2920--2936, Sept. 2023.

\bibitem{Ouyang2022}
C.~Ouyang, Y.~Liu, and H.~Yang, ``Performance of downlink and uplink integrated
  sensing and communications {(ISAC)} systems,'' \emph{{IEEE} Wireless Commun.
  Lett.}, vol.~11, no.~9, pp. 1850--1854, Sept. 2022.

\bibitem{Diluka2023}
D.~Galappaththige, C.~Tellambura, and A.~Maaref, ``Integrated sensing and
  backscatter communication,'' \emph{{IEEE} Wireless Commun. Lett.}, vol.~12,
  no.~12, pp. 2043--2047, Dec. 2023.

\bibitem{Bekkerman:TSP:2006}
I.~Bekkerman and J.~Tabrikian, ``Target detection and localization using {MIMO}
  radars and sonars,'' \emph{{IEEE} Trans. Signal Process.}, vol.~54, no.~10,
  pp. 3873--3883, Sept. 2006.

\bibitem{dan2024}
Q.~Dan, H.~Lei, K.-H. Park, G.~Pan, and M.-S. Alouini, ``Beamforming design for
  joint target sensing and proactive eavesdropping,'' \emph{arXiv}, 2024.

\bibitem{Li2007}
J.~Li and P.~Stoica, ``{MIMO} radar with colocated antennas,'' \emph{{IEEE}
  Signal Process. Mag.}, vol.~24, no.~5, pp. 106--114, Sept. 2007.

\bibitem{Liu2020Radar}
F.~Liu, C.~Masouros, A.~P. Petropulu, H.~Griffiths, and L.~Hanzo, ``Joint radar
  and communication design: Applications, state-of-the-art, and the road
  ahead,'' \emph{{IEEE} Trans. Commun.}, vol.~68, no.~6, pp. 3834--3862, Jun.
  2020.

\bibitem{Rivetti:WCNC:2024}
S.~Rivetti, E.~Björnson, and M.~Skoglund, ``Secure spatial signal design for
  {ISAC} in a cell-free {MIMO} network,'' in \emph{Proc. IEEE Wireless Commun.
  Netw. Conf.}, Apr. 2024, pp. 01--06.

\bibitem{galappaththige2024RSMA}
D.~Galappaththige, S.~Zargari, C.~Tellambura, and G.~Y. Li, ``Optimization of
  rate-splitting multiple access with integrated sensing and backscatter
  communication,'' \emph{{IEEE} Trans. Veh. Technol.}, pp. 1--16, 2025.

\bibitem{zargari2024ISABC}
S.~Zargari, D.~Galappaththige, and C.~Tellambura, ``Transmit power-efficient
  beamforming design for integrated sensing and backscatter communication,''
  \emph{{IEEE} Open J. Commun. Soc.}, vol.~6, pp. 775--792, Jan. 2025.

\bibitem{Diluka2026}
D.~Galappaththige, C.~Tellambura, and S.~Herath, ``Wideband cognitive radio for
  joint communication and sensing: Optimization of subcarrier allocation and
  beamforming,'' \emph{{IEEE} Trans. on Cogn. Commun. Netw.}, vol.~12, pp.
  1694--1709, Jan. 2026.

\bibitem{Diluka2026WBNF}
D.~Galappaththige and C.~Tellambura, ``Wideband {NF-ISAC}: Subcarrier
  allocation for sensing and beamforming,'' \emph{{IEEE} Wireless Commun.
  Lett.}, vol.~15, pp. 2164--2168, 2026.

\bibitem{zargari2024CFISAC}
S.~Zargari, D.~Galappaththige, C.~Tellambura, and G.~Y. Li, ``Downlink
  beamforming for cell-free {ISAC}: A fast complex oblique manifold approach,''
  \emph{{IEEE} Trans. Wireless Commun.}, 2025.

\bibitem{IMT2023}
\BIBentryALTinterwordspacing
``{The ITU-R Framework for IMT-2030},'' Jul. 2023, accessed: 5-15-2024.
  [Online]. Available:
  \url{https://www.itu.int/en/ITU-R/study-groups/rsg5/rwp5d/imt-2030/Documents/IMT-2030%20Framework_WP%205D%20Management%20Team.pdf}
\BIBentrySTDinterwordspacing

\bibitem{3GPPISAC2024}
\BIBentryALTinterwordspacing
``{3GPP TR} 22.837, {Feasibility} study on integrated sensing and
  communication, {V}.19.3.0 {R}el. 19,'' Apr. 2024. [Online]. Available:
  \url{https://portal.3gpp.org/desktopmodules/Specifications/SpecificationDeta
  ils.aspx?specificationId=4044}
\BIBentrySTDinterwordspacing

\bibitem{Zeng2019}
T.~Zeng, O.~Semiari, W.~Saad, and M.~Bennis, ``Joint communication and control
  for wireless autonomous vehicular platoon systems,'' \emph{{IEEE} Trans.
  Commun.}, vol.~67, no.~11, pp. 7907--7922, Nov. 2019.

\bibitem{Milanes2012}
V.~Milanes \emph{et~al.}, ``An intelligent {V2I}-based traffic management
  system,'' \emph{{IEEE} Trans. Intell. Transp. Syst.}, vol.~13, no.~1, pp.
  49--58, Mar. 2012.

\bibitem{Yongsen2019}
Y.~Ma, G.~Zhou, and S.~Wang, ``{WiFi} sensing with channel state information:
  {A} survey,'' \emph{ACM Comput. Surv.}, vol.~52, no.~3, Jun. 2019.

\bibitem{Popovski2019}
P.~Popovski \emph{et~al.}, ``Wireless access in ultra-reliable low-latency
  communication ({URLLC}),'' \emph{{IEEE} Trans. Commun.}, vol.~67, no.~8, pp.
  5783--5801, Aug. 2019.

\bibitem{Philip2021}
N.~Y. Philip, J.~J. P.~C. Rodrigues, H.~Wang, S.~J. Fong, and J.~Chen,
  ``{Internet of Things} for in-home health monitoring systems: Current
  advances, challenges and future directions,'' \emph{{IEEE} J. Sel. Areas
  Commun.}, vol.~39, no.~2, pp. 300--310, Feb. 2021.

\bibitem{Chander2023}
C.~Prakash, L.~P. Singh, A.~Gupta, and S.~K. Lohan, ``Advancements in smart
  farming: A comprehensive review of {IoT}, wireless communication, sensors,
  and hardware for agricultural automation,'' \emph{Sensors and Actuators A:
  Physical}, vol. 362, p. 114605, Nov. 2023.

\bibitem{zargari2024riemannian}
S.~Zargari, D.~Galappaththige, C.~Tellambura, and H.~Vincent~Poor, ``A
  {Riemannian} manifold approach to constrained resource allocation in
  {ISAC},'' \emph{{IEEE} Trans. Commun.}, vol.~73, no.~5, pp. 3655--3670, May
  2025.

\bibitem{Rahman2020}
M.~L. Rahman, J.~A. Zhang, X.~Huang, Y.~J. Guo, and R.~W. Heath, ``Framework
  for a perceptive mobile network using joint communication and radar
  sensing,'' \emph{{IEEE} Trans. Aerosp. Electron. Syst.}, vol.~56, no.~3, pp.
  1926--1941, Jun. 2020.

\bibitem{Diluka2024NF}
D.~Galappaththige, S.~Zargari, C.~Tellambura, and G.~Y. Li, ``Near-field
  {ISAC}: Beamforming for multi-target detection,'' \emph{{IEEE} Wireless
  Commun. Lett.}, vol.~13, no.~7, pp. 1938--1942, Jul. 2024.

\bibitem{Diluka2025NF}
------, ``Low-complexity multi-target detection in {ELAA} {ISAC},''
  \emph{{IEEE} Commun. Lett.}, vol.~29, no.~3, pp. 620--624, Mar. 2025.

\bibitem{Ouyang2022Uplink}
C.~Ouyang, Y.~Liu, and H.~Yang, ``On the performance of uplink {ISAC}
  systems,'' \emph{{IEEE} Commun. Lett.}, vol.~26, no.~8, pp. 1769--1773, Aug.
  2022.

\bibitem{Zhiqing:Net:2024}
Z.~Wei \emph{et~al.}, ``Integrated sensing and communication enabled multiple
  base stations cooperative sensing towards {6G},'' \emph{IEEE Network},
  vol.~38, no.~4, pp. 207--215, July 2024.

\bibitem{Xu:COMMG:2024}
D.~Xu \emph{et~al.}, ``Interference mitigation for network-level {ISAC}: An
  optimization perspective,'' \emph{{IEEE} Commun. Mag.}, vol.~62, no.~9, pp.
  28--34, Sept. 2024.

\bibitem{Guo:WC:2025}
H.~Guo \emph{et~al.}, ``Integrated communication, localization, and sensing in
  {6G D-MIMO} networks,'' \emph{{IEEE} Wireless Commun.}, vol.~32, no.~2, pp.
  214--221, Apr. 2025.

\bibitem{Meng:WC:2025}
K.~Meng, C.~Masouros, A.~P. Petropulu, and L.~Hanzo, ``Cooperative {ISAC}
  networks: Opportunities and challenges,'' \emph{{IEEE} Wireless Commun.},
  vol.~32, no.~3, pp. 212--219, June 2025.

\bibitem{Wang:WC:2025}
X.~Wang \emph{et~al.}, ``Cooperative integrated sensing and communication in
  {6G}: From operators perspective,'' \emph{{IEEE} Wireless Commun.}, vol.~32,
  no.~1, pp. 52--59, February 2025.

\bibitem{Meng:WC:2024}
K.~Meng, C.~Masouros, A.~P. Petropulu, and L.~Hanzo, ``Cooperative {ISAC}
  networks: Opportunities and challenges,'' \emph{IEEE Wireless Commun.}, pp.
  1--8, 2024.

\bibitem{Kaitao:TWC:2025}
K.~Meng, K.~Han, C.~Masouros, and L.~Hanzo, ``Network-level {ISAC}: An
  analytical study of antenna topologies ranging from massive to cell-free
  {MIMO},'' \emph{{IEEE} Trans. Wireless Commun.}, pp. 1--1, 2025.

\bibitem{Li:TSP:1993}
J.~Li and R.~Compton, ``Maximum likelihood angle estimation for signals with
  known waveforms,'' \emph{{IEEE} Trans. Signal Process.}, vol.~41, no.~9, pp.
  2850--2862, Sept. 1993.

\bibitem{Li:TWC:2024}
R.~Li, Z.~Xiao, and Y.~Zeng, ``Toward seamless sensing coverage for cellular
  multi-static integrated sensing and communication,'' \emph{{IEEE} Trans.
  Wireless Commun.}, vol.~23, no.~6, pp. 5363--5376, June 2024.

\bibitem{Meng:TWC:2024}
K.~Meng, C.~Masouros, G.~Chen, and F.~Liu, ``Network-level integrated sensing
  and communication: Interference management and {BS} coordination using
  stochastic geometry,'' \emph{{IEEE} Trans. Wireless Commun.}, vol.~23,
  no.~12, pp. 19\,365--19\,381, Dec. 2024.

\bibitem{Ismail2024}
M.~I. Ismail, A.~M. Shaheen, M.~M. Fouda, and A.~S. Alwakeel, ``{RIS}-assisted
  integrated sensing and communication systems: Joint reflection and
  beamforming design,'' \emph{{IEEE} Open J. Commun. Soc.}, vol.~5, pp.
  908--927, Feb. 2024.

\bibitem{Yang2024}
X.~Yang, Z.~Wei, Y.~Liu, H.~Wu, and Z.~Feng, ``{RIS}-assisted cooperative
  multicell {ISAC} systems: A multi-user and multi-target case,'' \emph{{IEEE}
  Trans. Wireless Commun.}, vol.~23, no.~8, pp. 8683--8699, Aug. 2024.

\bibitem{Luo2023}
H.~Luo, R.~Liu, M.~Li, and Q.~Liu, ``{RIS}-aided integrated sensing and
  communication: Joint beamforming and reflection design,'' \emph{{IEEE} Trans.
  Veh. Technol.}, vol.~72, no.~7, pp. 9626--9630, Jul. 2023.

\bibitem{Chen2024}
J.~Chen, K.~Wu, J.~Niu, and Y.~Li, ``Joint active and passive beamforming in
  {RIS}-assisted secure {ISAC} systems,'' \emph{Sensors}, vol.~24, no.~1, Jan.
  2024.

\bibitem{Rezaei2023Coding}
F.~Rezaei, D.~Galappaththige, C.~Tellambura, and S.~Herath, ``Coding techniques
  for backscatter communications - {A} contemporary survey,'' \emph{{IEEE}
  Commun. Surveys Tuts.}, pp. 1020--1058, 2th Quart. 2023.

\bibitem{Diluka2022}
D.~Galappaththige, F.~Rezaei, C.~Tellambura, and S.~Herath, ``Link budget
  analysis for backscatter-based passive {IoT},'' \emph{{IEEE} Access},
  vol.~10, pp. 128\,890--128\,922, Dec. 2022.

\bibitem{rezaei2023timespread}
F.~Rezaei, D.~Galappaththige, C.~Tellambura, and A.~Maaref, ``Time-spread
  pilot-based channel estimation for backscatter networks,'' \emph{{IEEE}
  Trans. Commun.}, vol.~72, no.~1, pp. 434--449, Jan. 2024.

\bibitem{Galappaththige2023SR}
D.~Galappaththige, F.~Rezaei, C.~Tellambura, and S.~Herath, ``Beamforming
  designs for enabling symbiotic backcom multiple access under imperfect
  {CSI},'' \emph{{IEEE} Access}, vol.~11, pp. 89\,986--90\,005, Aug. 2023.

\bibitem{Galappaththige2023RIS}
------, ``{RIS}-empowered ambient backscatter communication systems,''
  \emph{{IEEE} Wireless Commun. Lett.}, vol.~12, no.~1, pp. 173--177, Jan.
  2023.

\bibitem{Rezaei2024NOMA}
F.~Rezaei, D.~Galappaththige, C.~Tellambura, and S.~Herath, ``{NOMA}-assisted
  symbiotic backscatter: Novel beamforming designs under imperfect {SIC},''
  \emph{{IEEE} Trans. Veh. Technol.}, vol.~73, no.~5, pp. 6829--6844, May 2024.

\bibitem{Galappaththige2024passive}
D.~Galappaththige, F.~Rezaei, C.~Tellambura, and S.~Herath, ``Optimizing
  passive tag performance with reconfigurable intelligent surfaces in bistatic
  backscatter networks,'' \emph{{IEEE} Trans. Veh. Technol.}, vol.~73, no.~9,
  pp. 12\,917--12\,933, Sept. 2024.

\bibitem{Diluka2025ISABC}
D.~Galappaththige, S.~Zargari, and C.~Tellambura, ``Dual function of sensing
  and backscatter communication in cellular networks,'' \emph{{IEEE} Internet
  Things M.}, vol.~8, no.~3, pp. 64--71, May 2025.

\bibitem{Jiang2024}
Y.~Jiang, Q.~Wu, W.~Chen, and K.~Meng, ``{UAV}-enabled integrated sensing and
  communication: Tracking design and optimization,'' \emph{{IEEE} Commun.
  Lett.}, vol.~28, no.~5, pp. 1024--1028, May 2024.

\bibitem{Orikumhi2022}
I.~Orikumhi, J.~Bae, and S.~Kim, ``{UAV}-assisted integrated sensing and
  communications for user blockage prediction,'' in \emph{Proc. 13th Int. Conf.
  Inf. Commun. Technol. Converg.}, Oct. 2022, pp. 471--473.

\bibitem{Song2025}
Y.~Song \emph{et~al.}, ``An overview of cellular {ISAC} for low-altitude {UAV}:
  New opportunities and challenges,'' \emph{{IEEE} Commun. Mag.}, pp. 1--8,
  2025.

\bibitem{Zhao:TWc:2025}
C.~Zhao \emph{et~al.}, ``Networked isac-based {UAV} tracking and handover
  toward low-altitude economy,'' \emph{{IEEE} Trans. Wireless Commun.},
  vol.~24, no.~9, pp. 7670--7685, Sept. 2025.

\bibitem{Cheng:TCOM:2025}
G.~Cheng, X.~Song, Z.~Lyu, and J.~Xu, ``Networked {ISAC} for low-altitude
  economy: Coordinated transmit beamforming and {UAV} trajectory design,''
  \emph{{IEEE} Trans. Commun.}, vol.~73, no.~8, pp. 5832--5847, Aug. 2025.

\bibitem{Shaoqiang:TWC:2025}
S.~Yan, H.~Luo, P.~Yang, J.~Zhao, and F.~Gao, ``{UAV} trajectory monitoring for
  integrated sensing and communications system,'' \emph{{IEEE} Trans. Wireless
  Commun.}, pp. 1--1, 2025.

\bibitem{Zhang2022Holographic}
H.~Zhang \emph{et~al.}, ``Holographic integrated sensing and communication,''
  \emph{{IEEE} J. Sel. Areas Commun.}, vol.~40, no.~7, pp. 2114--2130, Jul.
  2022.

\bibitem{Adhikary2024}
A.~Adhikary \emph{et~al.}, ``Holographic {MIMO} with integrated sensing and
  communication for energy-efficient cell-free {6G} networks,'' \emph{{IEEE}
  Internet Things J.}, vol.~11, no.~19, pp. 30\,617--30\,635, Oct. 2024.

\bibitem{Gavras2023}
I.~Gavras, M.~A. Islam, B.~Smida, and G.~C. Alexandropoulos, ``Full duplex
  holographic {MIMO} for near-field integrated sensing and communications,'' in
  \emph{Proc. IEEE 31st European Signal Process. Conf.}, Sept. 2023, pp.
  700--704.

\bibitem{Zhu:COMMG:2024}
L.~Zhu, W.~Ma, and R.~Zhang, ``Movable antennas for wireless communication:
  Opportunities and challenges,'' \emph{{IEEE} Commun. Mag.}, vol.~62, no.~6,
  pp. 114--120, Jun. 2024.

\bibitem{Wong:TWC:2021}
K.-K. Wong, A.~Shojaeifard, K.-F. Tong, and Y.~Zhang, ``Fluid antenna
  systems,'' \emph{{IEEE} Trans. Wireless Commun.}, vol.~20, no.~3, pp.
  1950--1962, March 2021.

\bibitem{Shao:Tut:2025}
X.~Shao \emph{et~al.}, ``A tutorial on six-dimensional movable antenna for {6G}
  networks: Synergizing positionable and rotatable antennas,'' \emph{{IEEE}
  Commun. Surveys Tuts.}, pp. 1--1, 3rd Quart. 2025.

\bibitem{Yuanwei:MCOM:2025}
Y.~Liu \emph{et~al.}, ``Pinching-antenna systems: Architecture designs,
  opportunities, and outlook,'' \emph{{IEEE} Commun. Mag.}, pp. 1--7, 2025.

\bibitem{Zhendong:WC:2025}
Z.~Li \emph{et~al.}, ``Movable antennas enabled {ISAC} systems: Fundamentals,
  opportunities, and future directions,'' \emph{{IEEE} Wireless Commun.}, pp.
  1--8, 2025.

\bibitem{Lyu:TWC:2025}
W.~Lyu \emph{et~al.}, ``Movable antenna enabled integrated sensing and
  communication,'' \emph{{IEEE} Trans. Wireless Commun.}, vol.~24, no.~4, pp.
  2862--2875, Apr. 2025.

\bibitem{Jiang:TWC:2025}
C.~Jiang \emph{et~al.}, ``Movable antenna-assisted integrated sensing and
  communication systems,'' \emph{{IEEE} Trans. Wireless Commun.}, vol.~24,
  no.~8, pp. 6397--6412, Aug. 2025.

\bibitem{Guo:TWC:2025}
Y.~Guo \emph{et~al.}, ``Movable antenna enhanced networked integrated sensing
  and communication system,'' \emph{{IEEE} Trans. Wireless Commun.}, pp. 1--1,
  2025.

\bibitem{Ding:TWC:2025}
J.~Ding, Z.~Zhou, X.~Shao, B.~Jiao, and R.~Zhang, ``Movable antenna-aided
  near-field integrated sensing and communication,'' \emph{{IEEE} Trans.
  Wireless Commun.}, pp. 1--1, 2025.

\bibitem{Sun:IoT:2025}
Y.~Sun, H.~Xu, C.~Ouyang, and H.~Yang, ``Rotatable and movable antenna-enabled
  near-field integrated sensing and communication,'' \emph{{IEEE} Internet
  Things J.}, vol.~12, no.~21, pp. 45\,119--45\,132, Nov. 2025.

\bibitem{Yunhui:WCL:2025}
Y.~Qin, Y.~Fu, and H.~Zhang, ``Joint antenna position and transmit power
  optimization for pinching antenna-assisted {ISAC} systems,'' \emph{{IEEE}
  Wireless Commun. Lett.}, pp. 1--1, 2025.

\bibitem{Weihao:TWC:2025}
W.~Mao \emph{et~al.}, ``Multi-waveguide pinching antennas for {ISAC},''
  \emph{{IEEE} Trans. Wireless Commun.}, pp. 1--1, 2025.

\bibitem{Diluka2025CFBook}
D.~Galappaththige and C.~Tellambura, ``Cell-free integrated sensing and
  communication,'' \emph{Foundations and Trends in Networking}, vol.~16, no.
  1-2, pp. 1--170, 02 2025.

\bibitem{Mohammadi:JSAC:2023}
M.~Mohammadi, T.~T. Vu, H.~Q. Ngo, and M.~Matthaiou, ``Network-assisted
  full-duplex cell-free massive {MIMO}: Spectral and energy efficiencies,''
  \emph{{IEEE} J. Sel. Areas Commun.}, vol.~41, no.~9, pp. 2833--2851, Sept.
  2023.

\bibitem{Fishler2006}
E.~Fishler \emph{et~al.}, ``Spatial diversity in radars—{Models} and
  detection performance,'' \emph{{IEEE} Trans. Signal Process.}, vol.~54,
  no.~3, pp. 823--838, Mar. 2006.

\bibitem{Yuanyuan2024}
Y.~Dong \emph{et~al.}, ``Cell-free {ISAC} massive {MIMO} systems with
  capacity-constrained fronthaul links,'' \emph{Digital Signal Processing},
  vol. 145, p. 104341, Jan. 2024.

\bibitem{demirhan2024cellfree}
U.~Demirhan and A.~Alkhateeb, ``Cell-free {ISAC} {MIMO} systems: Joint sensing
  and communication beamforming,'' \emph{arXiv}, 2024.

\bibitem{Elfiatoure2023}
M.~Elfiatoure, M.~Mohammadi, H.~Q. Ngo, and M.~Matthaiou, ``Cell-free massive
  {MIMO} for {ISAC}: Access point operation mode selection and power control,''
  in \emph{Proc. IEEE Global Commun. Conf.}, Dec. 2023, pp. 104--109.

\bibitem{Mao2024}
W.~Mao \emph{et~al.}, ``Communication-sensing region for cell-free massive
  {MIMO} {ISAC} systems,'' \emph{{IEEE} Trans. Wireless Commun.}, vol.~23,
  no.~9, pp. 12\,396--12\,411, Sept. 2024.

\bibitem{Liu2024}
S.~Liu, R.~Liu, Z.~Lu, M.~Li, and Q.~Liu, ``Cooperative cell-free {ISAC}
  networks: Joint {BS} mode selection and beamforming design,'' in \emph{Proc.
  IEEE Wireless Commun. Netw. Conf.}, Apr. 2024, pp. 1--6.

\bibitem{elfiatoure2024multiple}
M.~Elfiatoure, M.~Mohammadi, H.~Q. Ngo, H.~Shin, and M.~Matthaiou,
  ``Multiple-target detection in cell-free massive {MIMO}-assisted {ISAC},''
  \emph{{IEEE} Trans. Wireless Commun.}, vol.~24, no.~5, pp. 4283--4298, May
  2025.

\bibitem{Moerman2022}
A.~Moerman \emph{et~al.}, ``Beyond {5G} without obstacles: {mmWave}-over-fiber
  distributed antenna systems,'' \emph{{IEEE} Commun. Mag.}, vol.~60, no.~1,
  pp. 27--33, Jan. 2022.

\bibitem{You2010}
X.-H. You \emph{et~al.}, ``Cooperative distributed antenna systems for mobile
  communications [coordinated and distributed {MIMO}],'' \emph{{IEEE} Wireless
  Commun.}, vol.~17, no.~3, pp. 35--43, Jun. 2010.

\bibitem{1199275}
C.~Tellambura, A.~Annamalai, and V.~Bhargava, ``Closed form and infinite series
  solutions for the {MGF} of a dual-diversity selection combiner output in
  bivariate {Nakagami} fading,'' \emph{{IEEE} Trans. Commun.}, vol.~51, no.~4,
  pp. 539--542, Apr. 2003.

\bibitem{1556835}
Y.~Chen and C.~Tellambura, ``Infinite series representations of the trivariate
  and quadrivariate {R}ayleigh distribution and their applications,''
  \emph{{IEEE} Trans. Commun.}, vol.~53, no.~12, pp. 2092--2101, Dec. 2005.

\bibitem{545899}
C.~Tellambura, ``Evaluation of the exact union bound for trellis-coded
  modulations over fading channels,'' \emph{{IEEE} Trans. Commun.}, vol.~44,
  no.~12, pp. 1693--1699, Dec. 1996.

\bibitem{4694096}
T.~Nechiporenko, P.~Kalansuriya, and C.~Tellambura, ``Performance of optimum
  switching adaptive $m$ -{QAM} for amplify-and-forward relays,'' \emph{{IEEE}
  Trans. Veh. Commun.}, vol.~58, no.~5, pp. 2258--2268, Jun. 2009.

\bibitem{Sakhnini2022}
A.~Sakhnini, M.~Guenach, A.~Bourdoux, H.~Sahli, and S.~Pollin, ``A target
  detection analysis in cell-free massive {MIMO} joint communication and radar
  systems,'' in \emph{Proc. IEEE Int. Conf. Commun.}, May 2022, pp. 2567--2572.

\bibitem{Behdad2024}
Z.~Behdad, {\"O}.~T. Demir, K.~W. Sung, E.~Björnson, and C.~Cavdar,
  ``Multi-static target detection and power allocation for integrated sensing
  and communication in cell-free massive {MIMO},'' \emph{{IEEE} Trans. Wireless
  Commun.}, vol.~23, no.~9, pp. 11\,580--11\,596, Sept. 2024.

\bibitem{Kulathunga2025}
R.~Kulathunga, J.~Dassanayake, and G.~Amarasuriya, ``Cell-free massive
  {MIMO}-aided {ISAC},'' in \emph{Proc. IEEE Int. Conf. Commun.}, Montreal,
  Quebec, Canada, Jun. 2025, pp. 1--6, accepted.

\bibitem{3GPP2010}
``{3GPP TR} 36.814, further advancements for {E-UTRA} physical layer aspects,
  {V}.9.0.0 {R}el. 9,'' Mar. 2010. Available Online:
  \url{https://portal.3gpp.org/desktopmodules/Specifications/SpecificationDetails.aspx?specificationId=2493}.

\bibitem{behdad2024Joint}
Z.~Behdad, {\"O}.~T. Demir, K.~W. Sung, and C.~Cavdar, ``Joint processing and
  transmission energy optimization for {ISAC} in cell-free massive {MIMO} with
  {URLLC},'' \emph{arXiv}, 2024.

\bibitem{Mobini:SPAWC:2025}
Z.~Mobini, M.~Mohammadi, J.~He, H.~Q. Ngo, and M.~Matthaiou, ``Cell-free
  massive {MIMO}-assisted {ISAC} with beam scanning,'' in \emph{Proc. IEEE Int.
  Workshop Signal Process. Adv. Wireless Commun.}, Jul. 2025, pp. 1--5.

\bibitem{Wenrui:TWC:2025}
W.~Li, M.~Li, M.-M. Zhao, and A.~Liu, ``Transmit beamforming optimization for
  cell-free integrated sensing and communication systems,'' \emph{{IEEE} Trans.
  Wireless Commun.}, pp. 1--1, 2025.

\bibitem{Qu2024}
K.~Qu, J.~Ye, X.~Li, and S.~Guo, ``Privacy and security in ubiquitous
  integrated sensing and communication: Threats, challenges and future
  directions,'' \emph{{IEEE} Internet Things Mag.}, vol.~7, no.~4, pp. 52--58,
  Jul. 2024.

\bibitem{Zhu:TuT:2024}
X.~Zhu \emph{et~al.}, ``Enabling intelligent connectivity: A survey of secure
  {ISAC} in {6G} networks,'' \emph{{IEEE} Commun. Surveys Tuts.}, vol.~27,
  no.~2, pp. 748--781, 2nd Quart. 2025.

\bibitem{Nasir2024}
A.~A. Nasir, ``Joint users’ secrecy rate and target’s sensing {SNR}
  maximization for a secure cell-free {ISAC} system,'' \emph{{IEEE} Commun.
  Lett.}, vol.~28, no.~7, pp. 1549--1553, Jul. 2024.

\bibitem{Ren2024}
Z.~Ren, J.~Xu, L.~Qiu, and D.~W.~K. Ng, ``Secure cell-free integrated sensing
  and communication in the presence of information and sensing eavesdroppers,''
  \emph{{IEEE} J. Sel. Areas Commun.}, vol.~42, no.~11, pp. 3217--3231, Nov.
  2024.

\bibitem{Cao2023}
Y.~Cao and Q.-Y. Yu, ``Joint resource allocation for user-centric cell-free
  integrated sensing and communication systems,'' \emph{{IEEE} Commun. Lett.},
  vol.~27, no.~9, pp. 2338--2342, Sept. 2023.

\bibitem{Buzzi2024}
S.~Buzzi, C.~D’Andrea, and S.~Liesegang, ``Scalability and implementation
  aspects of cell-free massive {MIMO} for {ISAC},'' in \emph{Proc. IEEE Int.
  Symp. Wireless Commun. Syst.}, Jun. 2024, pp. 1--6.

\bibitem{Cheng2024}
G.~Cheng, Y.~Fang, J.~Xu, and D.~W.~K. Ng, ``Optimal coordinated transmit
  beamforming for networked integrated sensing and communications,''
  \emph{{IEEE} Trans. Wireless Commun.}, vol.~23, no.~8, pp. 8200--8214, Aug.
  2024.

\bibitem{Rogalin2014}
R.~Rogalin \emph{et~al.}, ``Scalable synchronization and reciprocity
  calibration for distributed multiuser {MIMO},'' \emph{{IEEE} Trans. Wireless
  Commun.}, vol.~13, no.~4, pp. 1815--1831, Apr. 2014.

\bibitem{Shepard2012}
C.~Shepard \emph{et~al.}, ``Argos: practical many-antenna base stations,'' in
  \emph{Proc. the 18th Annual Int. Conf. Mobile Comput. Netw.}, ser. Mobicom
  '12.\hskip 1em plus 0.5em minus 0.4em\relax New York, NY, USA: Association
  for Computing Machinery, Aug. 2012, p. 53–64.

\bibitem{Vieira2021}
J.~Vieira and E.~G. Larsson, ``Reciprocity calibration of distributed massive
  {MIMO} access points for coherent operation,'' in \emph{Proc. IEEE Annual
  Int. Symp. Personal, Indoor and Mobile Radio Commun.}, Sept. 2021, pp.
  783--787.

\bibitem{Niu2024}
Y.~Niu, Z.~Wei, L.~Wang, H.~Wu, and Z.~Feng, ``Interference management for
  integrated sensing and communication systems: A survey,'' \emph{{IEEE}
  Internet Things J.}, vol.~12, no.~7, pp. 8110--8134, Apr. 2025.

\bibitem{Meng2024}
K.~Meng, C.~Masouros, G.~Chen, and F.~Liu, ``Network-level integrated sensing
  and communication: Interference management and {BS} coordination using
  stochastic geometry,'' \emph{{IEEE} Trans. Wireless Commun.}, vol.~23,
  no.~12, pp. 19\,365--19\,381, Dec. 2024.

\bibitem{Masoumi2020}
H.~Masoumi and M.~J. Emadi, ``Performance analysis of cell-free massive {MIMO}
  system with limited fronthaul capacity and hardware impairments,''
  \emph{{IEEE} Trans. Wireless Commun.}, vol.~19, no.~2, pp. 1038--1053, Feb.
  2020.

\bibitem{9503397}
A.~Mohammadian and C.~Tellambura, ``{RF} impairments in wireless transceivers:
  Phase noise, {CFO}, and {IQ} imbalance – a survey,'' \emph{IEEE Access},
  vol.~9, pp. 111\,718--111\,791, 2021.

\bibitem{Wang:TCOM.2020}
D.~Wang \emph{et~al.}, ``Performance of network-assisted full-duplex for
  cell-free massive {MIMO},'' \emph{IEEE Trans. Commun.}, vol.~68, no.~3, pp.
  1464--1478, Mar. 2020.

\bibitem{Mohammadi:TCOM:2024}
M.~Mohammadi, Z.~Mobini, H.~Q. Ngo, and M.~Matthaiou, ``Ten years of research
  advances in full-duplex massive {MIMO},'' \emph{{IEEE} Trans. Commun.},
  vol.~73, no.~3, pp. 1756--1786, Mar. 2025.

\bibitem{Zeng2023}
F.~Zeng \emph{et~al.}, ``Integrated sensing and communication for
  network-assisted full-duplex cell-free distributed massive {MIMO} systems,''
  \emph{arXiv}, 2023.

\bibitem{zeng2024}
------, ``Multi-static {ISAC} based on network-assisted full-duplex cell-free
  networks: Performance analysis and duplex mode optimization,'' \emph{arXiv},
  2024.

\bibitem{Zou:WCL:2024}
Q.~Zou, Z.~Behdad, O.~T. Demir, and C.~Cavdar, ``Distributed versus centralized
  sensing in cell-free massive {MIMO},'' \emph{{IEEE} Wireless Commun. Lett.},
  vol.~13, no.~12, pp. 3345--3349, Dec 2024.

\bibitem{Zhou:WCL:2024}
L.~Zhou, J.~Yao, M.~Jin, T.~Wu, and K.-K. Wong, ``Fluid antenna-assisted {ISAC}
  systems,'' \emph{{IEEE} Wireless Commun. Lett.}, Dec. 2024.

\bibitem{Qin:WCL:2024}
H.~Qin \emph{et~al.}, ``{Cramér-Rao} bound minimization for movable
  antenna-assisted multiuser integrated sensing and communications,''
  \emph{{IEEE} Wireless Commun. Lett.}, Dec. 2024.

\bibitem{Wang:WL:2024}
C.~Wang \emph{et~al.}, ``{AI}-empowered fluid antenna systems: Opportunities,
  challenges, and future directions,'' \emph{{IEEE} Wireless Commun.}, vol.~31,
  no.~5, pp. 34--41, Oct. 2024.

\bibitem{Huang:WCL:2020}
C.~Huang \emph{et~al.}, ``Holographic {MIMO} surfaces for {6G} wireless
  networks: Opportunities, challenges, and trends,'' \emph{{IEEE} Wireless
  Commun.}, vol.~27, no.~5, pp. 118--125, Oct. 2020.

\bibitem{Adhikary:IoT:2024}
A.~Adhikary \emph{et~al.}, ``Holographic {MIMO} with integrated sensing and
  communication for energy-efficient cell-free {6G} networks,'' \emph{{IEEE}
  Internet Things J.}, vol.~11, no.~19, pp. 30\,617--30\,635, Oct. 2024.

\bibitem{Zhang:MCOM:2025}
J.~Zhang, H.~Miao, P.~Tang, L.~Tian, and G.~Liu, ``New mid-band for {6G}:
  Several considerations from the channel propagation characteristics
  perspective,'' \emph{{IEEE} Commun. Mag.}, vol.~63, no.~1, pp. 175--180, Jan.
  2025.

\bibitem{chaves2024coverage}
F.~Chaves \emph{et~al.}, ``Coverage evaluation of {7--15 GHz} bands from
  existing sites,'' \emph{Nokia White Paper}, 2024.

\bibitem{Zhang:MCOM:2023}
H.~Zhang, N.~Shlezinger, F.~Guidi, D.~Dardari, and Y.~C. Eldar, ``{6G} wireless
  communications: From far-field beam steering to near-field beam focusing,''
  \emph{IEEE Communications Magazine}, vol.~61, no.~4, pp. 72--77, Apr. 2023.

\bibitem{Cui:MCOM:2023}
M.~Cui, Z.~Wu, Y.~Lu, X.~Wei, and L.~Dai, ``Near-field {MIMO} communications
  for {6G}: Fundamentals, challenges, potentials, and future directions,''
  \emph{{IEEE} Commun. Mag.}, vol.~61, no.~1, pp. 40--46, Jan. 2023.

\bibitem{Lei:WC:2025}
H.~Lei, J.~Zhang, Z.~Wang, B.~Ai, and E.~Bjornson, ``Near-field user
  localization and channel estimation for {XL-MIMO} systems: Fundamentals,
  recent advances, and outlooks,'' \emph{{IEEE} Wireless Commun.}, pp. 1--9,
  2025.

\bibitem{bjornson2024enabling}
E.~Bj{\"o}rnson \emph{et~al.}, ``Enabling {6G} performance in the upper
  mid-band by transitioning from massive to gigantic {MIMO},'' \emph{arXiv
  preprint arXiv:2407.05630}, 2024.

\bibitem{Xie2025}
M.~Xie, X.~Yu, J.~Xu, B.-H. Soong, and C.~Yuen, ``Beam focusing for near-field
  cell-free massive {MIMO},'' \emph{{IEEE} Wireless Commun. Lett.}, vol.~14,
  no.~8, pp. 2491--2495, Aug. 2025.

\bibitem{Xie2024}
M.~Xie, X.~Yu, X.~Dang, B.-H. Soong, and C.~Yuen, ``Near-field cell-free
  massive {MIMO} with channel aging: A modified {LoS} model,'' \emph{{IEEE}
  Wireless Commun. Lett.}, vol.~13, no.~12, pp. 3553--3557, Dec. 2024.

\bibitem{Dong:IoT:2024}
Y.~Dong, Z.~Yang, H.~Wang, N.~Hao, and H.~Li, ``Joint user pairing and
  beamforming design for {NOMA}-aided {CFMM-ISAC} systems,'' \emph{{IEEE}
  Internet Things J.}, vol.~12, no.~6, pp. 6749--6763, Mar. 2025.

\bibitem{Abdelaziz:TCOM:2024}
A.~Abdelaziz~Salem, M.~A. Albreem, K.~Alnajjar, S.~Abdallah, and M.~Saad,
  ``Integrated cooperative sensing and communication for {RIS}-enabled
  full-duplex cell-free {MIMO} systems,'' \emph{{IEEE} Trans. Commun.},
  vol.~73, no.~6, pp. 3804--3819, Jun. 2025.

\bibitem{Flores:ICC:2024}
X.~A. Flores~Cabezas, I.~W. da~Silva, and M.~Juntti, ``Performance of
  {UAV}-based cell-free {mMIMO ISAC} networks: Tethered vs. mobile,'' in
  \emph{Proc. IEEE Int. Conf. Commun.}, Jun. 2024, pp. 3549--3554.

\bibitem{Yu2022}
W.~Yu, F.~Sohrabi, and T.~Jiang, ``Role of deep learning in wireless
  communications,'' \emph{IEEE BITS the Info. Theory Mag.}, vol.~2, no.~2, pp.
  56--72, Nov. 2022.

\bibitem{Hu2021}
S.~Hu, X.~Chen, W.~Ni, E.~Hossain, and X.~Wang, ``Distributed machine learning
  for wireless communication networks: Techniques, architectures, and
  applications,'' \emph{{IEEE} Commun. Surveys Tuts.}, vol.~23, no.~3, pp.
  1458--1493, 3rd Quart. 2021.

\bibitem{Dai2020}
L.~Dai, R.~Jiao, F.~Adachi, H.~V. Poor, and L.~Hanzo, ``Deep learning for
  wireless communications: An emerging interdisciplinary paradigm,''
  \emph{{IEEE} Wireless Commun.}, vol.~27, no.~4, pp. 133--139, Aug. 2020.

\bibitem{Vaezi2026}
M.~Vaezi, G.~A. {Aruma Baduge}, E.~Ollila, and S.~A. Vorobyov, ``A tutorial on
  {AI}-empowered integrated sensing and communications,'' \emph{{IEEE} Commun.
  Surveys Tuts.}, vol.~28, pp. 4980--5013, 2026.

\bibitem{Demirhan2024ML}
U.~Demirhan and A.~Alkhateeb, ``Learning beamforming in cell-free massive {MIMO
  ISAC} systems,'' in \emph{Proc. IEEE Int. Workshop Signal Process. Adv.
  Wireless Commun.}, Sept. 2024, pp. 326--330.

\bibitem{Cheng2024a}
X.~Cheng \emph{et~al.}, ``Intelligent multi-modal sensing-communication
  integration: Synesthesia of machines,'' \emph{{IEEE} Commun. Surveys Tuts.},
  vol.~26, no.~1, pp. 258--301, 1st Quart. 2024.

\bibitem{Liu2022}
F.~Liu, Y.-F. Liu, A.~Li, C.~Masouros, and Y.~C. Eldar, ``Cram\'{e}r-{Rao}
  bound optimization for joint radar-communication beamforming,'' \emph{{IEEE}
  Trans. Signal Process.}, vol.~70, pp. 240--253, Jan. 2022.

\end{thebibliography}

\end{document}